\documentclass[11pt]{utSASreport}
\usepackage{subfigure}    
\usepackage{url}          
\usepackage{amsfonts,amstext,amsmath,amssymb,amscd}
\usepackage[dvips]{graphicx}
\usepackage{fancyhdr}     
\usepackage{verbatim}     
\usepackage{color}        

\usepackage[english]{babel}
\usepackage{geometry}
\usepackage{a4wide}
\usepackage{epsfig}

\usepackage{graphicx}

\usepackage{layout}

\title{\bf Incorporating a Volatility Smile into the Markov-Functional Model\footnote{This is the author's thesis for the Master of Science in Applied Mathematics with specialization in Financial Engineering. The thesis was under the supervision of A. Bagchi, M.H. Vellekoop and D. Kandhai and was defended on 22 December 2006. The project was executed from May 2006 to November 2006 at the ABN AMRO Bank in Amsterdam.}}

\author{Feijia Wang}

\sasreportdate{December 22, 2006}

\sasperiodofwork{01/05/2006}{30/11/2006}

\sascommittee{\Bagchi\\ \Vellekoop\\ \Kandhai\\ }

\begin{document}

\maketitle
\pagenumbering{roman} 



\chapter*{Preface $\&$ Acknowledgement} \addcontentsline{toc}{chapter}{Preface $\&$ Acknowledgement}
This thesis is a detailed report of my 7-month internship at the
Product $\&$ Transaction Analysis group of ABN AMRO in Amsterdam.
\\ \\
I would like to take this opportunity to thank all the people who
have given help related to this project. First of all, I would like
to thank my university supervisors Prof. Bagchi and Michel Vellekoop
for giving comments on my work and all other contributions. Without
Prof. Bagchi's initial effort, anything related to this project
would not have happened. I also want to thank Marije Elkenbracht and
Lukas Phaf from ABN AMRO for offering me such a nice internship.
Here I do owe a major acknowledgement to my company supervisor Drona
Kandhai, who took care of every aspect of the project in detail and
had the time lines well under control so that the project could be
finished without significant delays. Thanks also to Rutger Pijls,
Bert-Jan Nauta, Alex Zilber and Artem Tsvetkov for their useful
opinions on parts of my work, and other colleagues at ABN for
creating a harmonious atmosphere for work. Last but not least, I
would like to thank all the teachers of the MSc programme in
Financial Engineering of the University of Twente for educating me
in both math and finance.


\tableofcontents \addcontentsline{toc}{chapter}{Table of Contents}


\clearemptydoublepage
\pagenumbering{arabic}

\chapter{Introduction}\label{chapter:intro}
Interest rate models evolved from short rate models, which model the
instantaneous rate implied from the yield curve, to market models
that are based on LIBOR/swap rates. A nice property of short rate
models is that they are based on low-dimensional Markov processes.
This allows for analytical valuation or the use of tree/PDE based
approaches. But on the other hand, it has the difficulty of
calibration to caps/floors or swaptions. Market models are more
intuitive as LIBOR/swap rates are something that exists in reality.
They can be also easily calibrated to market instruments. However,
due to the large dimensionality which is inherent to these models,
the only tractable approach is to apply Monte-Carlo simulation.
Markov-Functional ({\bf MF}) models contain the nice properties from
both these two classes of models. Only a low-dimensional Markov
process $X_t$ is tracked such that the value of exotic derivatives
can be computed efficiently on a lattice. Meanwhile MF models can
still be calibrated to caps/floors or swaptions in a relatively easy
way.
\\ \\
The major question in MF models is how to go from $X_t$'s
stochasticity to the distributions of LIBOR/swap rates. The original
MF models map $X_t$ to the lognormal distribution of the underlying,
and thus volatility smile is not taken into account. A natural
extension of this model is a mapping to another distribution that is
consistent with volatility smile. The objective of this project is
to study the effect of volatility smile on the values and hedging
performance of co-terminal Bermudan swaptions in the
Markov-Functional model. We focus on Bermudan swaptions because they
are one of the most liquid American-style interest rate derivatives.
A convenient choice that can fit to the static volatility smile and
satisfy the arbitrage-free condition is the Uncertain Volatility
Displaced Diffusion ({\bf UVDD}) model. This model can generate both
the effects of skew and smile, as has already been demonstrated in
Abouchoukr \cite{Abouchoukr 2003}. However, it is not clear whether
its hedging performance is good or not. The fact that different
models can calibrate to today's smile but disagree on the hedging
performance has been discussed in the literature \cite{GS
2005}\cite{Henrard 2005}. In this report, we present in detail the
performance of the Markov-Functional model with UVDD digital mapping
in terms of pricing and hedging of Bermudan swaptions.
\\ \\
The rest of this report is organized as follows. Chapter
\ref{chapter:MF} focuses on explaining the original
Markov-Functional models in every aspect. Chapter
\ref{chapter:smile} studies the effect of incorporating volatility
smile for pricing. Chapter \ref{chapter:dynamics} investigates the
future smile and smile dynamics of the extended MF model. Chapter
\ref{chapter:calibration} presents the calibration results of the
UVDD model. Chapter \ref{chapter:Hedging} reports the details of the
hedging simulations. Finally, Chapter \ref{chapter:conclusions}
summarizes the main conclusions of this study and some suggestions
for future research.

\chapter{Markov-Functional Models} \label{chapter:MF}

\section{Quick Review of Interest Rate Models} \label{sec:review}
The first generation of interest rate models was a family of short
rate models whose governing SDE is specified under a martingale
measure Q. These short rate models share the general form:
\begin{equation}
dr(t)=\mu(t,r(t))dt+\sigma(t)r^{\beta}(t)dW^{Q}(t),
\end{equation}
where $\beta$ ranges from 0 to 1 and $W^{Q}$ is Brownian motion
under $Q$. In practice only $\beta$ values of $0$, $\frac{1}{2}$ and
$1$ are typically used, which correspond to, for example, the
Hull-White model, the Cox-Ingersoll-Ross model and the
Black-Karasinski model. Specifying r as the solution of a SDE allows
us to use Markov process theory, and thus we may work within a PDE
framework. If the term structure $\{D(t,T);0\leq t \leq T,
T>0\}$\footnote{$D(t,T)$ denotes the value at time $t$ of a discount
bond maturing at $T$.} has the form
\begin{equation}
D(t,T)=e^{a(t,T)+b(t,T)r(t)},
\end{equation}
where $a(t,T)$ and $b(t,T)$ are deterministic functions, then the
model is said to process an {\bf affine term structure} (ATS)
\cite{Bjork 2003}. Hence the yield\footnote{The continuously
compounded yield from $t$ to $T$ is defined as
$y(t,T)=\log\frac{1}{D(t,T)}$.} $y(t,T)$ from t to T has the form:
\begin{equation} \label{eq:ATS}
y(t,r(t);T)=-a(t,T)-b(t,T)r(t).
\end{equation}
This makes it particularly convenient to obtain analytical formulas
for the values of bonds and derivatives on bonds. However, the
obviously very unrealistic fact that Equation \ref{eq:ATS} implies
is that all yields are perfectly correlated, as short rate is the
only source of risk, \emph{i.e.},
\begin{equation}
\rho(y(t,r(t);T_{1}),y(t,r(t);T_{2}))=\rho(r(t),r(t))=1.
\end{equation}
\\
Instead of specifying a much more complicated short rate model, for
example a two-factor or even multi-factor short rate model,
Heath-Jarrow-Morton \cite{HJM 1992} chose to model the entire
forward rate curve as their (infinite dimensional) state variable.
The HJM approach to interest rate modelling is a general framework
for analysis rather than a specific model, like, for example, the
Hull-White model. In this framework, the forward rate can be
specified directly under a martingale measure Q as
\begin{equation}
df(t,T)=\mu(t,T)dt+\sigma(t,T)dW^{Q}(t),
\end{equation}
where $\sigma(t,T)$ is a d-dimensional row-vector and $dW^{Q}(t)$ is
a d-dimensional column-vector. By the choice of volatilities
$\sigma(t,T)$, the drift parameters $\mu(t,T)$ are determined by the
arbitrage-free principle\footnote{For proof, we refer to Chapter 23
of Bjork \cite{Bjork 2003}.}.
\begin{equation}
\mu(t,T)=\sigma(t,T)\int_{t}^{T}\sigma(t,T)^T ds,
\end{equation}
where in the formula $^T$ denotes transpose. Then we implicitly
observe today's forward rate structure $\{f^{\star}(0,T);T\geq 0\}$
from the market so that we can integrate to get the whole spectrum
of the forward rates.
\begin{equation} \label{eq:HJM}
f(t,T)=f^{\star}(0,T)+\int_{0}^{t}\mu(s,T)ds+\int_{0}^{t}\sigma(s,T)dW^Q(s),
\end{equation}
Using the results obtained from Equation \ref{eq:HJM}, we can
compute the prices of bonds and derivatives on bond\footnote{For
details, we refer to Bjork \cite{Bjork 2003}.}.
\\ \\
Short rate and forward rate models are mimicking the modelling of
equity/currency derivatives, whose underlying dynamics has the
following form
\begin{equation} \label{eq:equity}
dV(t)=\mbox{\{drift\}}dt+\mbox{\{diffusion\}}dW(t),
\end{equation}
where $V$ can function as either the spot or forward value of the
underlying. However, the interest rate we observe in reality, like
LIBOR or swap rate, always carries a tenor from overnight to years.
Then it's by intuition more suitable to model the interest rate
dynamics by carrying a tenor parameter as well, \emph{i.e.},
\begin{equation} \label{eq:interest}
dV(t;\alpha)=\mbox{\{drift\}}dt+\mbox{\{diffusion\}}dW(t),
\end{equation}
where $\alpha$ is the tenor length. Comparing Equation
\ref{eq:equity} and Equation \ref{eq:interest}, we see that
short/forward rate models are dealing with interest rates with an
\emph{infinitesimal} tenor length.
\\ \\
{\bf Remark: From now on we will use the notation defined in
Appendix \ref{sec:notation}.}
\\ \\
A historic breakthrough came from Brace-Gatarek-Musiela \cite{BGM
1997} and Jamshidian \cite{Jamshidian 1997}, whose approach was to
directly model discrete market rates such as forward LIBOR rates in
the LIBOR market models or forward swap rates in the swap market
models. For LIBOR market models, let's look at Equation
\ref{eq:libor}. If we choose $D_{n+1}(t)$ as the numeraire, it can
be proved that the LIBOR process $L_{n}(t)$ is a martingale under
the forward measure $Q^{n+1}$.\footnote{For proof, we refer to
Chapter 25 of Bjork \cite{Bjork 2003}.} If we then further assume
the LIBOR rate $L_{n}(t)$ to be lognormally distributed under its
forward measure, \emph{i.e.},
\begin{equation}
dL_{n}(t)=L_{n}(t)\sigma_{n}(t)dW^{n+1}_t,
\end{equation}
where $\sigma_n(t)$ is a d-dimensional row-vector and $dW^{n+1}_t$
is a d-dimensional column-vector, we can transform all the LIBOR
process $L_{n}(t)$ to the terminal measure $Q^{N+1}$ by application
of Girsanov theorem,\footnote{For derivation of Equation
\ref{eq:libor_process}, we refer to Chapter 25 of Bjork \cite{Bjork
2003}.}
\begin{equation} \label{eq:libor_process}
dL_{n}(t)=-L_{n}(t)(\sum_{k=n+1}^{N}\frac{\alpha_{k}L_{k}(t)}{1+\alpha_{k}L_{k}(t)}
\sigma_{n}(t)\sigma_{k}(t)^T)dt +L_{n}(t)\sigma_{n}(t)dW^{N+1}_t.
\end{equation}
The valuation and risk management of interest rate derivatives by
means of LIBOR market models then resort to multi-dimensional
Monte-Carlo simulation. A similar line is followed by swap market
models, in which PVBP, $P_{n}(t)$, is chosen to be the numeraire.
Hence, the forward swap process $S_{n}(t)$ is a martingale under the
forward measure $Q^{n,N+1}$.\footnote{For the definitions of
$P_{n}(t)$, $S_{n}(t)$ and $Q^{n,N+1}$, please refer to Appendix
\ref{sec:notation}.} What's worth mentioning is that the terminal
measure in swap market models, $Q^{N,N+1}$, coincides with that of
LIBOR market model $Q^{N+1}$ as their numeraires just differ by a
constant $\alpha_{N}$. This can be further explained by the fact
that division by a certain numeraire determines a certain measure,
which is the rule of allocating probabilities. So division by an
extra constant won't matter for the distributions of random
variables.
\\ \\
Market models are more intuitive than short rate models as
LIBOR/swap rates exist in reality. They can be also easily
calibrated to market instruments. However, due to the large
dimensionality which is inherent to these models, the only tractable
approach is to apply Monte-Carlo simulation. Hunt-Kennedy-Pelsser
\cite{HKP 2000} proposed a general class of Markov-Functional
interest rate models, which contain nice properties from both these
two classes of models. In Markov-Functional models, we would only
have to track a low-dimensional process $X$ which is Markovian in
some martingale measure, usually the terminal measure $Q^{N+1}$,
\begin{equation}
dX(t)=\tau(t)dW^{N+1}_t,
\end{equation}
where $\tau(t)$ can be either a deterministic or a stochastic
process as long as $X(t)$ retains the Markov property\footnote{For
details of the Markov property, please refer to Chapter 7 of
Oksendal \cite{Oksendal 2000}.}. For each terminal time point $T_n$,
the random variable $X(T_n)$, which has no financial interpretation
at all, is mapped to the terminal LIBOR rate $L_n(T_n)$ or swap rate
$S_n(T_n)$. The former leads to the LIBOR MF models and the latter
leads to the swap MF models.\footnote{In this report, we focus on
the swap MF models rather than the LIBOR MF models, both of which
nevertheless work in the same fashion.} Each of these state
variables is originally modelled in market models by a stand-alone
process $L_n(t)$ or $S_n(t)$. In such a setting, we can avoid
Monte-Carlo simulations, which reduces the computing time
significantly in comparison with market models for the same task
\cite{PP 2005}. Because of the freedom to choose the functional
forms of state variables, MF models retain the advantage of accurate
calibration to relevant market prices. Besides, MF models are
capable of controlling the state transition to some extent thanks to
the freedom to choose the volatility process $\tau(t)$. We will
discuss these aspects in more detail in the following sections.

\section{Markov-Functional Interest Rate Models}
This section explains the details of Markov-Functional models and is
based on Hunt-Kennedy-Pelsser \cite{HKP 2000}, Pelsser \cite{Pelsser
2000} and Regenmortel \cite{Regenmortel 2003}.

\subsection{Assumptions of MF Model} \label{sec:assumption}
\begin{itemize}
\item \emph{\bf Assumption 1} The state of the economy at time t is
entirely described via some low-dimensional Markov process, which
will be denoted by $X(t)$. A convenient and typical choice of the
process has the following form
\begin{equation} \label{eq:MF}
dX(t)=\tau(t)dW^{N+1}_t,
\end{equation}
where $\tau(t)$ is a deterministic function. Thus this corresponds
to a one-factor MF model. Actually, throughout this report, we stick
to the {\bf one-dimensional} MF model.
\\ \\
To be more concrete, we assume that the numeraire discount bond
$D_{N+1}(t,X(t))$ is a function of $X(t)$. This implies that
$D_{N+1}$ is totally determined by $X(t)$. By applying the
martingale property it can be shown that every discount bond
$D_{k}(t,X(t))$, for all $k\leq N+1$, is a function of $X(t)$:
\begin{equation} \label{eq:martingale}
\frac{D_{k}(t,X(t))}{D_{N+1}(t,X(t))} =
\mathbb{E}_t^{N+1}[\frac{D_{k}(T_k,X(T_k))}{D_{N+1}(T_k,X(T_k))}]
=\mathbb{E}_t^{N+1}[\frac{1}{D_{N+1}(T_k,X(T_k))}].
\end{equation}

Note
$\mathbb{E}_t^{N+1}(\cdot)=\mathbb{E}^{N+1}(\;\cdot\;|\mathcal{F}_t^X)$,
where $\mathcal{F}_t^X$ is the information generated by $X$ on
$[0,t]$.
\\ \\
Conditional on $X(t)=x_{t}$ the random variable $X(s)$ follows, for
$s\geq t$, a normal probability distribution with mean $x_{t}$ and
variance $\int_{s}^{t}\tau^{2}(u)du$.\footnote{For derivation,
please refer to Appendix \ref{sec:derivation1}.} The probability
density function of $X(s)$ given $X(t)=x_{t}$ is denoted by
$\phi(X(s)|X(t)=x_{t})$ and can be expressed as
\begin{equation} \label{eq:density}
\phi(X(s)|X(t)=x_{t})=\frac{exp(-\frac{1}{2}\frac{(X(s)-x_{t})^2}{\int_{t}^{s}\tau^{2}(u)du})}
{\sqrt{2\pi\int_{t}^{s}\tau^{2}(u)du}}.
\end{equation}

\item \emph{\bf Assumption 2} The terminal swap rate $S_{n}(T_n,x)$, for all $n=1,\ldots,N$,
is a strictly monotonically increasing function of $x$.
\end{itemize}

\subsection{What is Modelled in MF?} \label{section:mf_model}
\begin{figure}[h!]
\centering
\includegraphics [width=\textwidth,height=55mm]{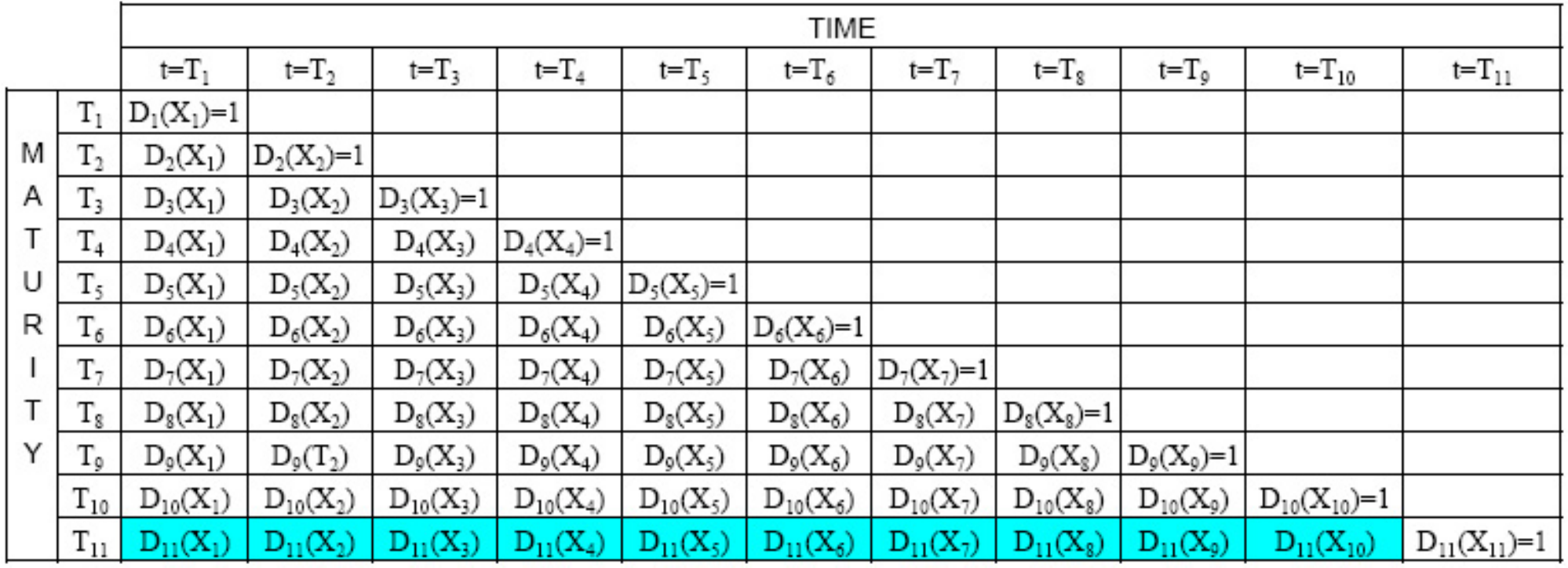}
\centering \caption{State variables we are interested in (N=10).}
\label{fig:to_model}
\end{figure}

{\bf Remark: From now on we are applying the simplified notation
defined in Appendix \ref{sec:notation2}.}
\\ \\
An interest rate model should be able to describe the distribution
of the future yield curve, whose fundamental quantities are discount
bonds. For pricing Bermudan swaptions, it is more convenient to use
a swap Markov-functional model that is calibrated to the underlying
European swaptions. Roughly speaking, by the relationship (see
Equation \ref{eq:swap} and \ref{eq:PVBP})
\begin{equation} \label{eq:terminal_swaprate}
S_n(X_n) = \frac{D_n(X_n)-D_{N+1}(X_n)}{P_n(X_n)}
=\frac{1-D_{N+1}(X_n)}{\sum_{k=n+1}^{N+1}\alpha_{k-1}D_k(X_n)},
\end{equation}
we should determine $D_k(X_n)$'s functional form, shown in Figure
\ref{fig:to_model}, such that $S_n(X_n)$ fits its market
distribution. Actually we only need to determine the functional form
of the numeraire discount bond $D_{N+1}(X_n)$, the shadowed state
variables in Figure \ref{fig:to_model}, as functional forms of all
other discount bonds can be determined by Equation
\ref{eq:martingale} and \ref{eq:density},
\begin{eqnarray}
D_{k}(X_n)&=&D_{N+1}(X_n)\mathbb{E}_{T_{n}}^{N+1}[\frac{1}{D_{N+1}(X_k)}] \nonumber\\
&=&D_{N+1}(X_n)\int_{-\infty}^{\infty}\frac{1}{D_{N+1}(y)}\phi(y|X_n)dy,
\end{eqnarray}
where $\phi$ denotes the probability density function of $X_k$
conditional on $X_n=x_n$.
\\ \\

In conclusion, given a specified $X(t)$ process, we determine the
functional forms of $D_{N+1}(X_n)$ such that the model is
calibrated to the market prices of European swaptions.

\subsection{Black-Scholes Digital Mapping} \label{sec:BS_mapping}
Let's illustrate the mapping from $X_n$ to $D_{N+1}(X_n)$ assuming
the terminal swap rate $S_n(T_n)$ is lognormally distributed and
thus smile is not taken into account. We conduct the mapping via
Digital swaptions\footnote{For details of Digital swaption, please
refer to Appendix \ref{sec:notation}} for the sake of its relatively
simple payoff. This is why it's called a \emph{"Digital Mapping"}.
Because of the lognormal assumption above, the digital mapping here
is call the "Black-Scholes Digital Mapping".
\\ \\
The functional form of the numeraire discount bond $D_{N+1}(X_n)$
($n=N,\ldots,1$) is determined by following a backward induction
process from $T_N$ to $T_1$.
\\ \\
First the value at time 0 of a Digital Receiver Swaption with strike
$K$ and maturity $T_n$, \emph{i.e.}, $DSN_n(0;K)$, is given
by\footnote{For details, please refer to Appendix
\ref{sec:notation}.}
\begin{equation} \label{eq:BS_formula}
DSN_n(0;K)=P_n(0)\Phi(\frac{\log(K/S_{n}(0))+\frac{1}{2}\bar{\sigma}_n^2
T_n}{\bar{\sigma}_n\sqrt{T_n}}).
\end{equation}
As explained in Appendix \ref{sec:notation} (see Equation
\ref{eq:second_dev}), Digital swaption values across a continuum of
strikes imply the terminal density of the underlying swap rate. In
the Black-Scholes world, this is assumed to be a lognormal
distribution.\footnote{Note this is the only place we should change
in the digital mapping if we want to calibrate the model to the
market smile. More concretely, we use another option pricing model's
formula for Digital swaption to imply the market distribution.}
\\ \\
On the other hand, the option's value can be expressed under the
terminal measure $Q^{N+1}$ as
\begin{equation} \label{eq:risk_neutral}
DSN_n(0;K)=D_{N+1}(0)\mathbb{E}_{0}^{N+1}[I_{\{S_n(X_n)<K\}}
\frac{P_n(X_n)}{D_{N+1}(X_n)}].
\end{equation}
By Assumption 2 in Section \ref{sec:assumption}, we have that
$S_{n}(X_n)$ is a strictly monotonically increasing function of
$X_n$, which implies that we can invert the function to get a
certain $x_n$ such that $\{S_n(X_n)<K\}\Leftrightarrow \{X_n<x_n\}$.
Thus $DSN_n(0;K)$ can be rewritten as
\begin{equation}
DSN_n(0;K)=D_{N+1}(0)\mathbb{E}_{0}^{N+1}[I_{\{X_n<x_n\}}
\frac{P_n(X_n)}{D_{N+1}(X_n)}],
\end{equation}
which we denote by a new symbol $\widetilde{DSN_n(0;x_n)}$ instead
of the original symbol $DSN_n(0;K)$. Applying the martingale
property to $\frac{P_n(X_n)}{D_{N+1}(X_n)}$, we would get
\begin{eqnarray} \label{eq:risk_neutral2}
\widetilde{DSN_n(0;x_n)} &=& \nonumber
D_{N+1}(0)\mathbb{E}_{0}^{N+1}[I_{\{X_n<x_n\}}
{E}_{T_n}^{N+1}[\frac{P_n(X_{n+1})}{D_{N+1}(X_{n+1})}]] \\
&=& D_{N+1}(0)\int_{-\infty}^{x_n}[\int_{-\infty}^{\infty}
\frac{P_n(y)}{D_{N+1}(y)} \phi_1(y|z)dy]\phi_2(z)dz,
\end{eqnarray}
where $\phi_1$ denotes the probability density function of $X_{n+1}$
conditional on $X_n=x_n$ and $\phi_2$ denotes the probability
density function of $X_n$. Note the functional form of
$\frac{P_n(X_{n+1})}{D_{N+1}(X_{n+1})}$ can be determined by
Equation \ref{eq:R_PVBP}. Therefore $\widetilde{DSN_n(0;x_n)}$ can
be evaluated at least numerically for different values of $x_n$
which correspond to different values of K observed in the market.
\\ \\
Equating \ref{eq:BS_formula} and \ref{eq:risk_neutral2}, we get
\begin{equation} \label{eq:mapping}
S_n(x_n)=K=S_n(0)exp(-\frac{1}{2}\bar{\sigma}_n^2 T_n +
\bar{\sigma}_n\sqrt{T_n}\Phi^{-1}(\frac{\widetilde{DSN_n(0;x_n)}}{P_n(0)})).
\end{equation}
As $x_n$ is a certain value of $X_n$, we generalize Equation
\ref{eq:mapping} to get the functional form of $S_n(X_n)$.
\begin{equation} \label{eq:Sn}
S_n(X_n)=S_n(0)exp(-\frac{1}{2}\bar{\sigma}_n^2 T_n +
\bar{\sigma}_n\sqrt{T_n}\Phi^{-1}(\frac{\widetilde{DSN_n(0;X_n)}}{P_n(0)})).
\end{equation}
\\ \\
Then the functional form of $D_{N+1}(X_n)$ can be determined by
rewriting Equation \ref{eq:terminal_swaprate}
\begin{equation}
D_{N+1}(X_n)=\frac{1}{1+S_{n}(X_n)\frac{P_n(X_n)}{D_{N+1}(X_n)}},
\end{equation}
where $\frac{P_n(X_n)}{D_{N+1}(X_n)}$ has already been calculated in
Equation \ref{eq:risk_neutral2}.

\subsection{Numerical Solution} \label{sec:numerical}
\begin{figure}[h!]
\centering
\includegraphics[width=\textwidth]{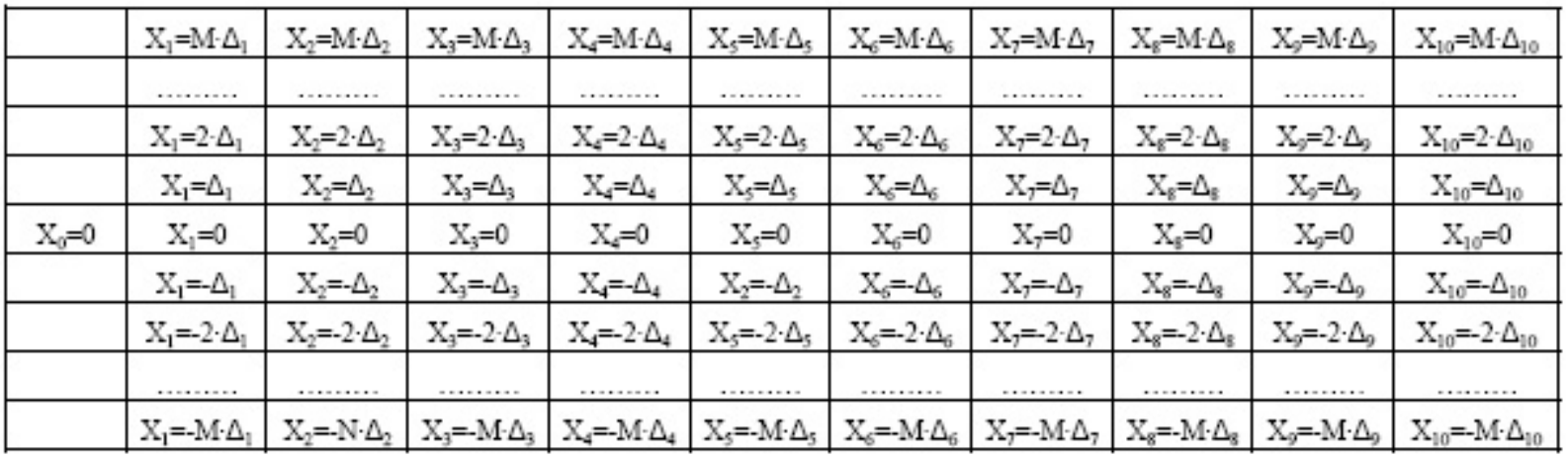}
\centering \caption{Lattice of $X_n$ (N=10).} \label{fig:lattice}
\end{figure}

In practice, the MF model is solved on a lattice. For each floating
reset date $T_n$, we choose $2M+1$ values of $X_n$ from
$-m\times\sigma_{X_n}$ to
$m\times\sigma_{X_n}$\footnote{$\sigma_{X_n}$ denotes the standard
deviation of $X_n$.}, or equivalently $-M\times\Delta_n$ to
$M\times\Delta_n$ with step length $\Delta_n$, see Figure
\ref{fig:lattice}, where
\begin{eqnarray}
M &= & m \times (\mbox{number of steps in the interval length equal
to one $\sigma_{X_n}$}) \nonumber \\
\Delta_n &=& \frac{\sigma_{X_n}}{\mbox{number of steps in the interval
length equal to one $\sigma_{X_n}$}} \\
&=& \frac{\sqrt{\int_{0}^{T_n}\tau^{2}(u)du}}{\mbox{number of steps
in the interval length equal to one $\sigma_{X_n}$}}, \nonumber
\end{eqnarray}
\\

An implementation of the MF model relies heavily on the evaluation
of expectations using numerical integration routines.  The numerical
integration adopted was introduced by Pelsser \cite{Pelsser 1998-c},
which is outlined as follows\footnote{Detailed explanation of the
method can be found in Appendix \ref{chapter:Gaussians}.}:
\begin{itemize}
\item fit a polynomial to the payoff function defined on the grid by
applying Neville's algorithm\footnote{For details of Neville's
algorithm, please refer to Section 3.1 of "Numerical Recipes in C++"
\cite{NR 2002}.};
\item calculate analytically the
integral of the polynomial against the Gaussian distribution.
\end{itemize}

\section{Volatility Function and Terminal Correlation}
\label{sec:MR_all}
\subsection{Volatility Function and Terminal
Correlation}\label{sec:MR}

The prices of Bermudan swaptions depend strongly on the joint
distribution or the terminal correlations of underlying swap rates
$S_n(T_n)$.\footnote{This section is based on Pelsser \cite{Pelsser
1998-a}\cite{Pelsser 2000}.} By applying a first order Taylor
expansion to $\log S_n(X_n)$, we could get the following linear
approximation. It it accurate enough for $X_n$ close to zero, where
a majority of the probability mass concentrates\footnote{You will
see the validity check for this approximation in Section
\ref{sec:assumption_check}.}.
\begin{eqnarray} \label{eq:approximation}
\log S_n(T_n,X_n) &\approx& \log S_n(T_n,x)|_{x=0} +
X_n\frac{\partial \log S_n(T_n,x)}{\partial
x}|_{x=0}  \nonumber \\
&=& \mbox{constant1} + \mbox{constant2} \times X_n.
\end{eqnarray}
\begin{flushleft}
Hence for $n<k$ we approximately have
\end{flushleft}
\begin{equation}
Corr(\log S_n(T_n),\log S_k(T_k))\approx Corr(X_n(T_n),X_k(T_k)).
\end{equation}
The problem in turn transforms to finding the auto-correlation of
the process $X(t)$. Pelsser \cite{Pelsser 2000} got inspired by the
Hull-White model, whose short rate process $r(t)$ follows
\begin{equation} \label{eq:HW}
dr(t)=(\theta(t)-ar(t))dt+\sigma dW(t).
\end{equation}
By some algebra, we derive the auto-correlation structure of the
short rates depending on the mean-reversion parameter $a$ via the
relationship\footnote{For derivation, please refer to Appendix
\ref{sec:derivation2}.}
\begin{equation} \label{eq:HW_MR}
Corr(r(t),r(s))=\begin{cases} \sqrt{\frac{t}{s}}
& \mbox{if $a=0$}\\
\sqrt{\frac{e^{2at}-1}{e^{2as}-1}} & \mbox{if $a\neq 0$}
\end{cases},
\end{equation}
for $t<s$. If we set $X(t)$ process' volatility function in
equation \ref{eq:MF} to be
\begin{equation} \label{eq:MF_vol}
\tau(t)=e^{at},
\end{equation}
we would get an equivalent expression for the auto-correlation of
the process $X(t)$\footnote{For derivation, please refer to Appendix
\ref{sec:derivation2}.}
\begin{equation} \label{eq:MR}
Corr(X(t),X(s)) =\begin{cases} \sqrt{\frac{t}{s}}
& \mbox{if $a=0$}\\
\sqrt{\frac{e^{2at}-1}{e^{2as}-1}} & \mbox{if $a\neq 0$}
\end{cases},
\end{equation}
for $t<s$. Thus, parameter $a$ can be interpreted as the
mean-reversion parameter of the process $X(t)$. We can see from
Equation \ref{eq:MR} that increasing the mean-reversion parameter
$a$ has the effect of reducing the auto-correlation between the
values of $X(T_n)$ for different floating reset dates $T_n$. Thus
increasing the mean-reversion parameter reduces the auto-correlation
between terminal swap rates $S_n(T_n)$.

\subsection{Estimation of the Mean-Reversion Parameter}
\label{sec:MR_estimation}

Because of the ill-liquidity of other exotic interest rate
derivatives that contain the information of terminal correlation of
co-terminal swap rates, we are left with estimating the terminal
correlations by historical data analysis.\footnote{This section is
based on Pelsser \cite{Pelsser 1998-a}.} The correlation of $\log
S_n(T_n)$ and $\log S_k(T_k)$, for $n<k$, is equivalent to the
correlation of their log differences $\log\frac{S_n(T_n)}{S_n(0)}$
and $\log\frac{S_k(T_k)}{S_k(0)}$. This is because $S_n(0)$ and
$S_k(0)$ are known today, division by which is sort of a
normalization. One approach is to estimate the correlation by
analyzing the most recently historical time series\footnote{Note we
are using small letter $s$ to denote a time series of swap rates
because they are market quotes.} of
$\log\frac{s_{t+T_n}(t+T_n)}{s_{t+T_n}(t)}$ and
$\log\frac{s_{t+T_k}(t+T_k)}{s_{t+T_k}(t)}$. However, this method
turns out to give estimates with large standard deviation due to the
long lags needed for the calculation of the difference (see
\cite{Pelsser 1998-a}). Therefore we instead analyze the time series
with shorter lags, \emph{i.e.}, $\log\frac{s_{t+T_n}(t+\Delta
u)}{s_{t+T_n}(t)}$ and $\log\frac{s_{t+T_k}(t+\Delta
u)}{s_{t+T_k}(t)}$, where $\Delta u$ represents the lag size. If
$\Delta u\rightarrow 0$, we are virtually analyzing the time series
with \emph{infinitesimal} lags, \emph{i.e.}, $d\log s_{t+T_n}(t+u)$
and $d\log s_{t+T_k}(t+u)$. If we stick to the lognormal assumption,
\emph{i.e.},
\begin{equation} \label{eq:MR_esti_assumption}
\begin{cases}
dS_{T_n}(u)=\sigma_{n}S_{T_n}(u)dW^{n,N+1}_u\\
dS_{T_k}(u)=\sigma_{k}S_{T_k}(u)dW^{k,N+1}_u
\end{cases},
\end{equation}
by applying It$\hat{o}$'s lemma we would have
\begin{equation}
\begin{cases}
d\log S_{T_n}(u) = \sigma_n dW_u^{n,N+1}
- \frac{1}{2}\sigma_n^2 du \\
d\log S_{T_k}(u) = \sigma_k dW_u^{k,N+1} - \frac{1}{2}\sigma_k^2 du
\end{cases},
\end{equation}
If we apply the Girsanov transformation\footnote{For details of
Girsanov Theorem, please refer to Chapter 11 of Bjork \cite{Bjork
2003}.}, \emph{i.e.}, we set
\begin{equation}
\begin{cases}
dW_u^1 = dW_u^{n,N+1} - \frac{1}{2}\sigma_n du\\
dW_u^2 = dW_u^{k,N+1} - \frac{1}{2}\sigma_k du
\end{cases},
\end{equation}
where $W_u^1$ and $W_u^2$ denote Brownian motions under the new
measure, we would have
\begin{equation}
\begin{cases}
d\log S_{T_n}(u) = \sigma_n dW_u^1\\
d\log S_{T_k}(u) = \sigma_k dW_u^2
\end{cases}.
\end{equation}
Let's denote the instantaneous correlation between $W_u^1$ and
$W_u^2$ by $\rho$, \emph{i.e.},
\begin{equation}
dW_u^1 dW_u^2 = \rho du.
\end{equation}

\begin{flushleft}
Then the correlation we are interested in can be expressed as
\end{flushleft}
\begin{equation} \label{eq:MR_corr_factor}
Corr(\log\frac{S_{T_n}(T_n)}{S_n(0)},\log\frac{S_{T_k}(T_k)}{S_k(0)})
=Corr(W_u^1(T_n),W_u^2(T_k))=\rho\sqrt{\frac{T_n}{T_k}}.
\end{equation}
\\
The problem thus transforms to a historical estimation\footnote{The
instantaneous correlation $\rho$ is the same under the real world
measure and risk-neutral measure, because the dynamics under the two
measures differ only by the drift term.} of the instantaneous
correlation $\rho$. In practice, we could approximately estimate
$\rho$ by choosing a smallest possible lag size $\Delta u$, that is,
one day. How valid this approach is depends on how valid the
lognormal assumption is and how valid the approximation of
instantaneous correlation by historically estimating the correlation
on a daily basis is.

\section{Bermudan Swaption Pricing under Markov-Functional} \label{sec:pricing_general}

In this section, we first discuss the general backward induction
method for valuing an American-style option, and then illustrate the
pricing procedure under MF's framework.

\subsection{American-style Option Pricing in a Discrete Time Model}
Let's express everything in the swap/swaption context. Suppose we
are under some risk-neutral measure Q with numeraire B(t) and the
American swaption is allowed to exercise at any floating reset date
$T_n(n=1,\ldots,N)$. Then the value of the American swaption
$BSN(T_n;K)$\footnote{An American option applied on a set of
discrete time points is literally still a Bermudan option, so we
adopt the notation $BSN$ here.} at time $T_n$ can be computed
backwardly as follows \cite{Vellekoop 2005},
\begin{eqnarray}
BSN(T_N;K) &=& ESN_N(T_N;K)\nonumber\\
\frac{BSN(T_{n-1};K)}{B(T_{n-1})} &=&
max\{\;\frac{ESN_{n-1}(T_{n-1};K)}{B(T_{n-1})}\;, \;
\mathbb{E}_{T_{n-1}}^Q[\frac{BSN(T_n;K)}{B(T_n)}]\;\}
\label{eq:American} \\
BSN(0;K) &=& B(0)\mathbb{E}_{T_0}^Q[\frac{BSN(T_1;K)}{B(T_1)}].
\nonumber
\end{eqnarray}
where the payoff of a European swaption at maturity $T_n$ is
\begin{equation}   \label{eq:European_payoff}
ESN_{n}(T_{n};K)=max\{\;SV(T_n;K)\;,\;0\},
\end{equation}
where $SV(T_n;K)$ is the swap value at time $T_n$ (see Section
\ref{sec:notation} for notation).\footnote{
$\frac{BSN(T_{n};K)}{B(T_{n})}$ is actually a
Q-\emph{supermartingale}, meaning
\begin{equation} \nonumber
\mathbb{E}_{T_n}^Q[\frac{BSN(T_{n+1};K)}{B(T_{n+1})}]\;\leqslant\;
\frac{BSN(T_n;K)}{B(T_n)}.
\end{equation}
}

\subsection{Bermudan Swaption Pricing with the MF Model} \label{sec:pricing}
\begin{figure}[h!]
\centering
\includegraphics[width=120mm, height=45mm]{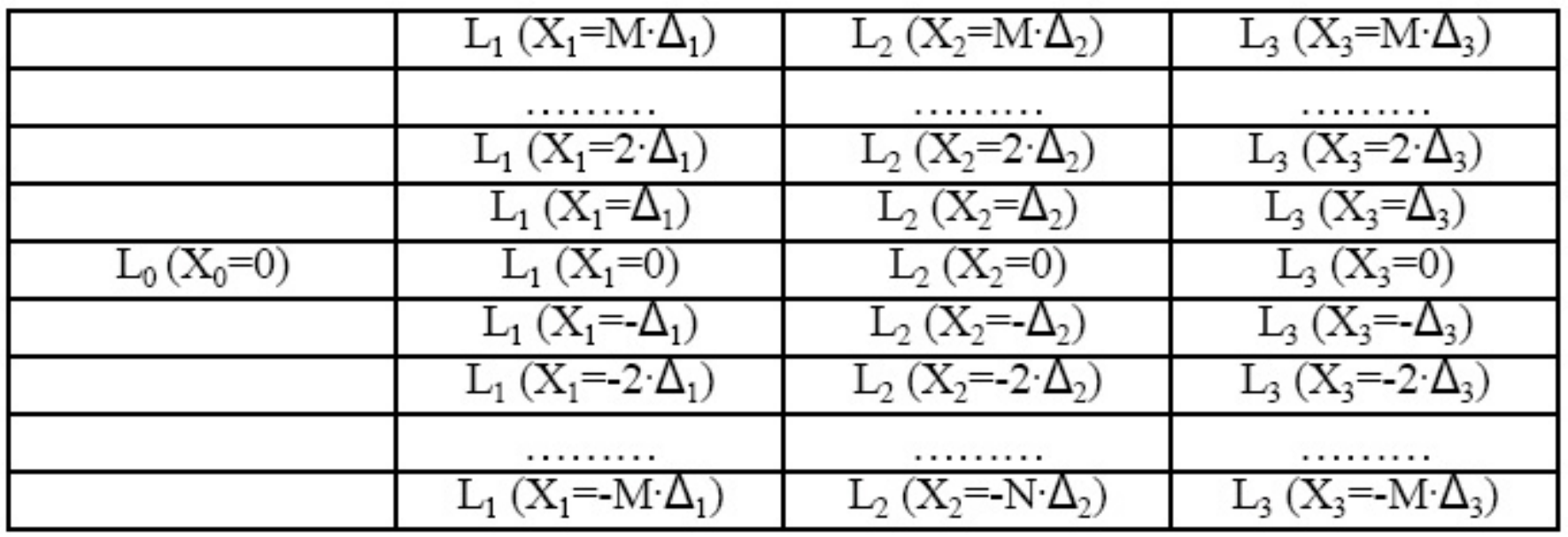}
\centering \caption{LIBOR "tree" by MF model.} \label{fig:LIBOR}
\end{figure}
\begin{figure}[h!]
\centering
\includegraphics[width=120mm, height=45mm]{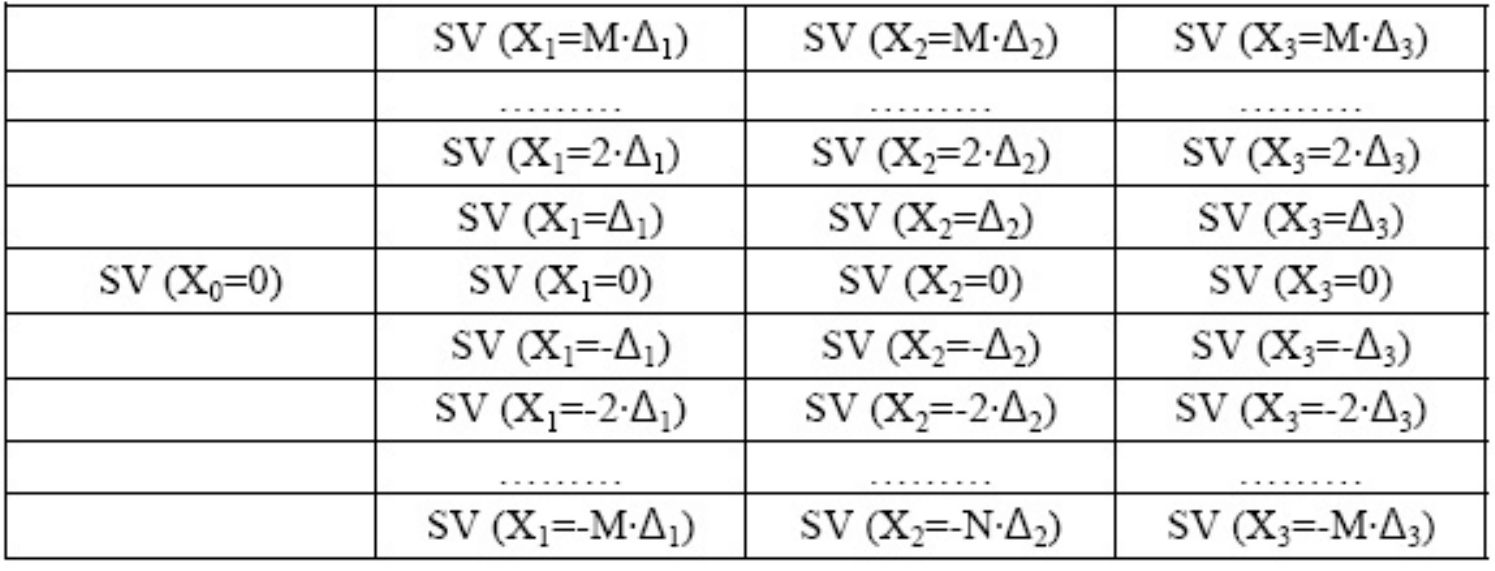}
\centering \caption{Swap value "tree" by MF model.}
\label{fig:swap_value2}
\end{figure}
\begin{figure}[h!]
\includegraphics[width=120mm, height=45mm]{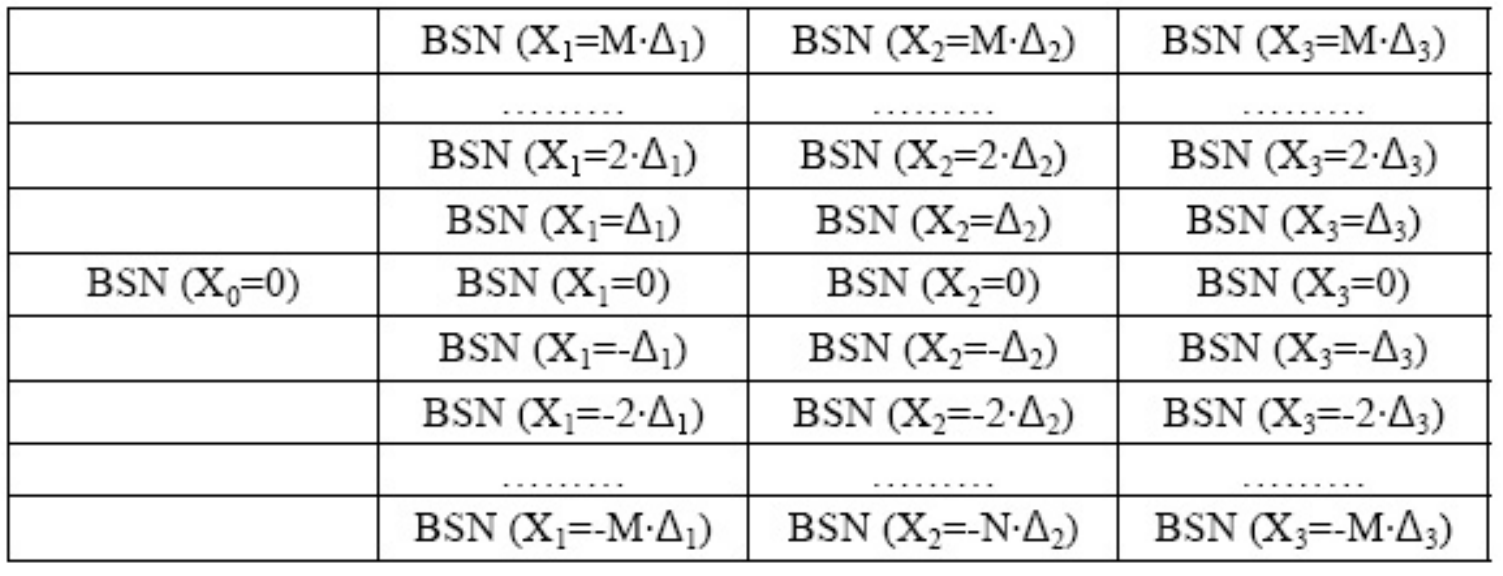}
\centering \caption{Option value "tree" by MF model.}
\label{fig:option_value2} \centering
\end{figure}

Assume N=4 and we have the LIBOR "tree" after the digital
mapping,\footnote{$L_n(X_n)$ is determined by Equation
\ref{eq:libor}.} shown in Figure \ref{fig:LIBOR}, we can compute the
corresponding swap value "tree" and option valuation "tree", shown
in Figure \ref{fig:swap_value2} and \ref{fig:option_value2},
respectively. In other words, we ought to determine the functional
forms of $\frac{SV(X_n;K)}{D_{N+1}(X_n)}$ and
$\frac{BSN(X_n;K)}{D_{N+1}(X_n)}$ so as to find out today's option
value $BSN(0;K)$.\footnote{It's obviously more convenient to get the
functional forms of $\frac{SV(X_n;K)}{D_{N+1}(X_n)}$ and
$\frac{BSN(X_n;K)}{D_{N+1}(X_n)}$ rather than $SV(X_n;K)$ and
$BSN(X_n;K)$.}
\\ \\
The functional form of numeraire-discounted swap value
$\frac{SV(X_n;K)}{D_{N+1}(X_n)}$ can be determined by the backward
induction:
\begin{eqnarray}
\frac{SV(X_{N+1};K)}{D_{N+1}(T_{N+1})} &=& SV(X_{N+1};K) \;\;=\;\;
\varphi[\alpha_N (r(X_N)-K)] \nonumber\\
\frac{SV(X_n;K)}{D_{N+1}(X_n)} &=&
\mathbb{E}_{T_n}^{N+1}[\frac{SV(X_{n+1};K)}{D_{N+1}(X_n)}] +
\frac{\varphi[\alpha_{n-1} (r(X_{n-1})-K)]}{D_{N+1}(X_n)}
 \label{eq:swap_value2}\\
\frac{SV(0;K)}{D_{N+1}(0)} &=&
\mathbb{E}_0^{N+1}[\frac{SV(X_1;K)}{D_{N+1}(X_1)}]\nonumber,
\end{eqnarray}
where $\varphi$ is 1 for a payer swap and -1 for a receiver swap.
Note there is no cash exchange at time $T_1$.
\\ \\
The functional form of numeraire-discounted option value
$\frac{BSN(X_n;K)}{D_{N+1}(X_n)}$ can be determined backwards from
$T_N$ to $T_1$:
\\ \\
If $T_n$ is not an exercise date, we have
\begin{equation} \label{eq:MF_riskneutral}
\frac{BSN(X_n;K)}{D_{N+1}(X_n)}=\mathbb{E}_{T_n}^{N+1}[\frac{BSN(X_{n+1};K)}{D_{N+1}(X_{n+1})}].
\end{equation}
If $T_n$ is an exercise date, we have
\begin{eqnarray}
\frac{BSN(X_n;K)}{D_{N+1}(X_n)}&=&
max\{\;\frac{ESN(X_n;K)}{D_{N+1}(X_n)}\;,\;
\mathbb{E}_{T_n}^{N+1}[\frac{BSN(X_{n+1};K)}{D_{N+1}(X_{n+1})}]
\;\} \nonumber \\
\label{eq:Bermudan} &=&
max\{\;max[\frac{SV(X_n;K)}{D_{N+1}(X_n)},0]\;,\;
\mathbb{E}_{T_n}^{N+1}[\frac{BSN(X_{n+1};K)}{D_{N+1}(X_{n+1})}] \;\}
\nonumber \\
&=& max\{\;\frac{SV(X_n;K)}{D_{N+1}(X_n)}\;,\;
\mathbb{E}_{T_n}^{N+1}[\frac{BSN(X_{n+1};K)}{D_{N+1}(X_{n+1})}]
\;\},
\end{eqnarray}
where $BSN(X_{N+1};K)=0$.
\\ \\
Now we can obtain today's Bermudan swaption value $BSN(0;K)$.
\begin{equation}
BSN(0;K)=D_{N+1}(0)\mathbb{E}_0^{N+1}[\frac{BSN(X_1;K)}{D_{N+1}(X_1)}].
\end{equation}

\chapter{Integration of Volatility Smile} \label{chapter:smile}
\section{Incorporating Volatility Smile into the MF Model}
In the MF model a smile can be incorporated quite naturally. What is
required for this, is a model to obtain swaption prices across a
\emph{continuum} of strikes given a limited number of market quotes.

\subsection{Interpolation of Implied Volatility}
A first approach that might come into one's mind is to keep the
Black-Scholes mapping and interpolate/extrapolate the market quotes
to obtain a continuum of implied volatilities as a function of the
strike, \emph{i.e.}, $\bar{\sigma}_n(K)$. Then, provided that
assumption 2 in Section \ref{sec:assumption} still holds, the only
thing that we need to change in the mapping procedure described in
Section \ref{sec:BS_mapping}, is to replace
$\bar{\sigma}_n$\footnote{In the lognormal case $\bar{\sigma}_n$ is
constant.} in Equation \ref{eq:Sn} with $\bar{\sigma}_n(S_n(X_n))$.
More precisely, we instead solve numerically the following equation
for $S_n(X_n)$,
\begin{equation}
S_n(X_n) = S_n(0)exp\{-\frac{1}{2}\bar{\sigma}_n^2(S_n(X_n))T_n +
\bar{\sigma}_n(S_n(X_n))\sqrt{T_n}\Phi^{-1}(\frac{\widetilde{DSN_n(0;X_n)}}{P_n(0)})\}.
\end{equation}
\\

In Ref. \cite{Johnson 2006} a method is used for interpolation of
the prices corresponding to the intermediate strikes such that the
price of the ATM European swaption is preserved. However, as noted
by Johnson \cite{Johnson 2006} these smoothing methods may not
satisfy the arbitrage-free constraints at all.

\subsection{Uncertain Volatility Displaced Diffusion Model} \label{sec:UVDD}
An alternative approach is to base the digital mapping on an option
pricing model that includes smile. In other words, we use another
distribution rather than the lognormal one to approximate the
terminal density of the swap rate, which allows for a good fit to
the volatility smile observed in the market. In this project, we
will use the Uncertain Volatility Displaced Diffusion model
(hereafter {\bf UVDD}) proposed by Brigo-Mercurio-Rapisarda
\cite{BMR 2003}.

In the following, we will first describe the displaced diffusion
model which is the simplest extension of the lognormal model that
can include skew effects. The description of the UVDD model will
follow afterwards.

\subsubsection{Displaced Diffusion Model (hereafter DD)}
We assume in this setting that the dynamics of the swap rate
$S_n(t)$ under the forward measure $Q^{n,N+1}$ is as follows,
\begin{equation} \label{eq:DD}
dS_{n}(t)=\sigma_{n}(S_{n}(t)+m_n)dW^{n,N+1}_t.
\end{equation}
The parameter $m_n$ is called the displacement coefficient.
Following the same reasoning from Equation \ref{eq:DSN} to
\ref{eq:DSN_BS}, we can derive the closed form solution for the
value of a Digital receiver swaption,
\begin{equation} \label{eq:DD_Digital}
DSN_n(0;K)=P_n(0)\Phi(\frac{\log(\frac{K+m_n}{S_{n}(0)+m_n})+\frac{1}{2}\sigma_n^2
T_n}{\sigma_n\sqrt{T_n}}).
\end{equation}
Similarly, the value of a European swaption is given by
\begin{eqnarray} \label{eq:DD_European}
ESN_n(0;K) &=& \varphi P_n(0)((S_n(0)+m_n)\Phi(\varphi d_+)-(K+m_n)\Phi(\varphi d_-)) \\
d_{\pm} &=&
\frac{\log(\frac{S_{n}(0)+m_n}{K+m_n})\pm\frac{1}{2}\sigma_n^2
T_n}{\sigma_n\sqrt{T_n}} \nonumber,
\end{eqnarray}
where $\varphi$ is 1 for a payer European swaption and -1 for a
receiver one.
\\
\begin{figure}[h!]
\centering
\includegraphics[width=120mm,height=70mm]{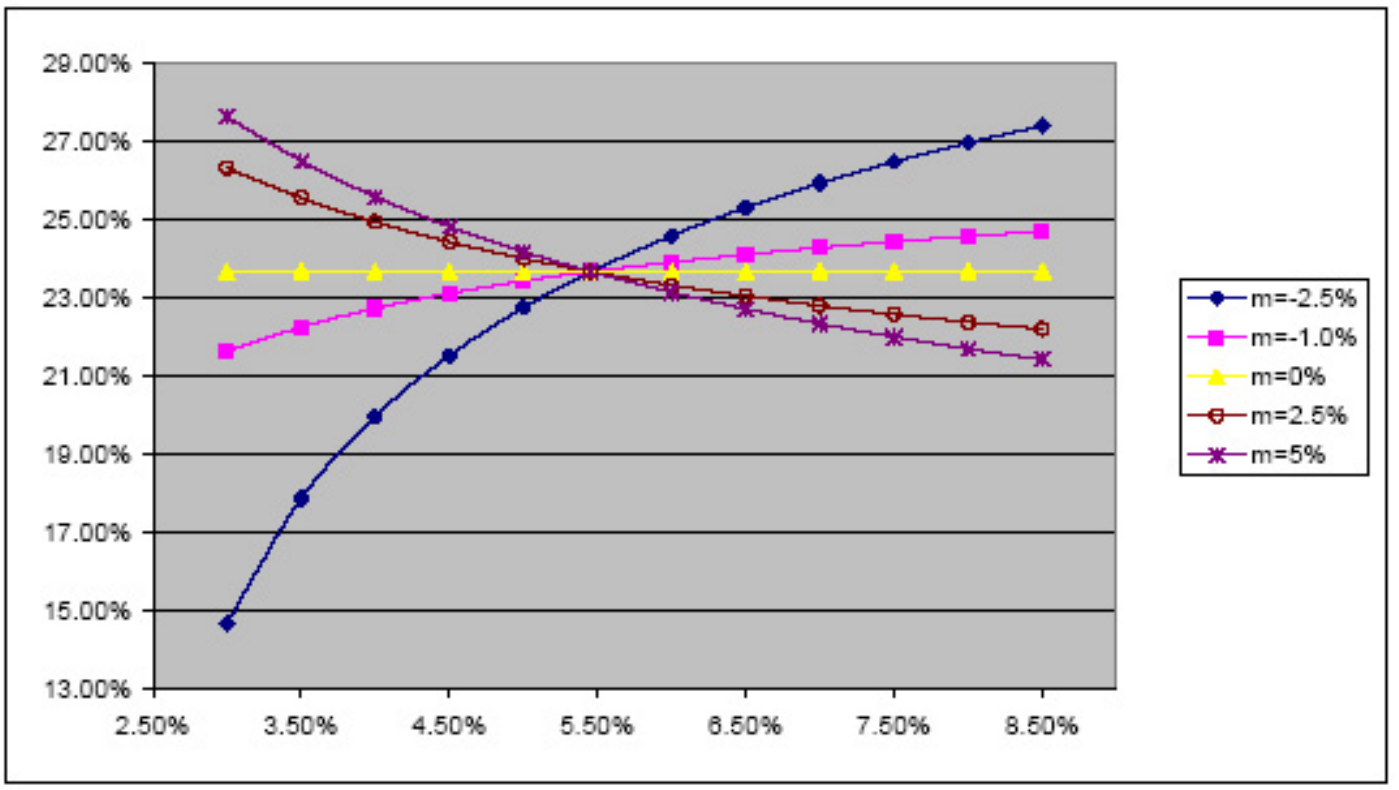}
\centering \caption{Implied skew for various values of the
displacement coefficient.} \label{fig:DD}
\end{figure}

The displacement coefficient can be used to generate an implied
volatilities' skew shape. A positive value of the displacement
coefficient generates a downward sloping skew, while a negative
value generates an upward slopping skew. The latter is unrealistic
and should not be used. We report in Figure \ref{fig:DD} the implied
skew for various values of the displacement coefficient. The case
$m_n = 0\%$ corresponds to the usual lognormal model. We have used
the market data corresponding to Data Set I in Appendix
\ref{sec:DataSet1} and Trade I in Appendix \ref{sec:TestTrades}. The
tested instrument was a European swaption expiring at the fifth
floating reset date, {\em i.e.}, $T_5$, with at-the-money swap rate
of $5.45\%$. The parameter $\sigma_5$ was adjusted such that the
implied ATM volatility was the same for all cases. More precisely,
we determine $\sigma_5$ such that the UVDD ATM price equals the BS
ATM price.

The DD model can only incorporate the volatility skew, but market
data suggest that the volatility quotes of swaption is typically a
smile shape \cite{Hagan 2002}. Thus the DD model is insufficient for
describing the market quotes.

\subsubsection{UVDD Model}
In the UVDD setup, $S_{n}(t)+m_n$ is assumed to have the following
dynamics
\begin{equation} \label{eq:UVDD_whole}
dS_n(t)=\begin{cases} \sigma_n^0(S_n(t)+m_n)dW^{n,N+1}_t \;\;\; t\in[0,\varepsilon]\\
\eta_n(S_n(t)+m_n)dW^{n,N+1}_t \;\;\; t>\varepsilon,
\end{cases}
\end{equation}
where $\sigma_n^0$ is a constant and $\eta_n$ is a random variable
that is independent of $W^{n,N+1}_t$ and can take the following
values,
\begin{equation}
\eta_n=\begin{cases} \sigma_n^1 \;\;\; \mbox{with probability $\lambda_n^1$}\\
\sigma_n^2 \;\;\; \mbox{with probability $\lambda_n^2$} \\
\vdots \\
\sigma_n^M \;\;\; \mbox{with probability $\lambda_n^M$}
\end{cases},
\end{equation}
\\
where $\Sigma_{i=1}^{M}\lambda_n^i=1$. Denoting by $\mathbb{P}$ the
risk neutral probability under the forward measure $Q^{n,N+1}$, we
have
\begin{equation}
\label{eq:UVDD_cdf} \mathbb{P}\{S_n(t)+m_n\leq y\} =
\Sigma_{i=1}^{M} \mathbb{P}\{\{S_n(t)+m_n\leq y\}\cap\{\eta_n =
\sigma_n^i \}\}=\Sigma_{i=1}^{M}\lambda^i
\mathbb{P}\{S_n^i(t)+m_n\leq y|\eta_n = \sigma_n^i \},
\end{equation}
Differentiating Equation \ref{eq:UVDD_cdf} with respect to $y$, we
get the probability density function of $S_{n}(t)+m_n$,
\begin{equation}
p_{n,t}(y)=\frac{\partial}{\partial y}\mathbb{P}\{S_{n}(t)+m_n\leq
y\} =\Sigma_{i=1}^{M}\lambda_n^i\frac{\partial}{\partial
y}\mathbb{P}\{S_{n}^i(t)+m_n\leq y|\eta_n = \sigma_n^i\}
=\Sigma_{i=1}^{M}\lambda_n^i p_{n,t}^i(y),
\end{equation}
where $p_{n,t}^i(y)$ is the density of a displaced lognormal
variable with constant volatilities $\sigma_n^i$. The value of a
Digital receiver swaption can be expressed under $Q^{n,N+1}$ as
follows,
\begin{eqnarray}
DSN_n(0;K) &=& P_n(0)\mathbb{E}_{0}^{n,N+1}[
\frac{P_n(T_n)I_{\{S_n(T_n)<K\}}}{P_n(T_n)}] \nonumber \\
&=& P_n(0)\mathbb{E}_{0}^{n,N+1}[
\frac{P_n(T_n)I_{\{S_n(T_n)+m_n<K+m_n\}}}{P_n(T_n)}]
\nonumber\\
&=& P_n(0)\int_0^{+\infty}I_{\{y<K+m_n\}}
\Sigma_{i=1}^{M}\lambda_n^i p_{n,t}^i(y)dy \nonumber \\
&=& \Sigma_{i=1}^{M}\lambda_n^i
P_n(0)\int_0^{+\infty}I_{\{y<K+m_n\}}
p_{n,t}^i(y)dy \nonumber \\
&=& \Sigma_{i=1}^{M}\lambda_n^i
P_n(0)\mathbb{E}_{0}^{n,N+1}[I_{\{S_n^i(T_n)+m_n<K+m_n\}}]
\nonumber \\
&=& \Sigma_{i=1}^{M}\lambda_n^i
P_n(0)\mathbb{E}_{0}^{n,N+1}[I_{\{S_n^i(T_n)<K\}}].
\end{eqnarray}
\\
Now following once more the same line of reasoning from \ref{eq:DSN}
to \ref{eq:DSN_BS}, we derive a closed form solution for the value
of Digital receiver swaption,
\begin{equation} \label{eq:UVDD_Digital}
DSN_n(0;K)=P_n(0)\Sigma_{i=1}^{M}\lambda_n^i
\Phi(\frac{\log(\frac{K+m_n}{S_{n}(0)+m_n})+\frac{1}{2}(\sigma_n^i)^2
T_n}{\sigma_n^i\sqrt{T_n}}).
\end{equation}
Similarly, the value of a European swaption can be determined
analytically by
\begin{eqnarray} \label{eq:UVDD_European}
ESN_n(0;K) &=& \varphi P_n(0)\Sigma_{i=1}^{M}\lambda_n^i
((S_n(0)+m_n)\Phi(\varphi d_+^i)-(K+m_n)\Phi(\varphi d_-^i)) \\
d_{\pm}^i &=&
\frac{\log(\frac{S_{n}(0)+m_n}{K+m_n})\pm\frac{1}{2}(\sigma_n^i)^2
T_n}{\sigma_n^i\sqrt{T_n}}, \nonumber
\end{eqnarray}
where $\varphi$ is 1 for a payer European swaption and -1 for a
receiver one.
\\ \\
We have performed the tests reported in this chapter with two
components ($M=2$). The model can be expressed in terms of the
following parameters $m_n$, $\sigma_n^1$, $\sigma_n^2$,
$\lambda_n^1$, $\lambda_n^2$. It can be also expressed in terms of
the parameters $m_n$, $\sigma_n$, $\omega_n$, $\lambda_n$ with:
\begin{equation}      \label{eq:2components}
\begin{cases}
\sigma_n^1 = \sigma_n \\
\sigma_n^2 = \omega_n\sigma_n  \\
\lambda_n^1 = \lambda_n \\
\lambda_n^2 = 1-\lambda_n
\end{cases}.
\end{equation}
\\
\begin{figure}[h!]
\includegraphics[width=120mm,height=70mm]{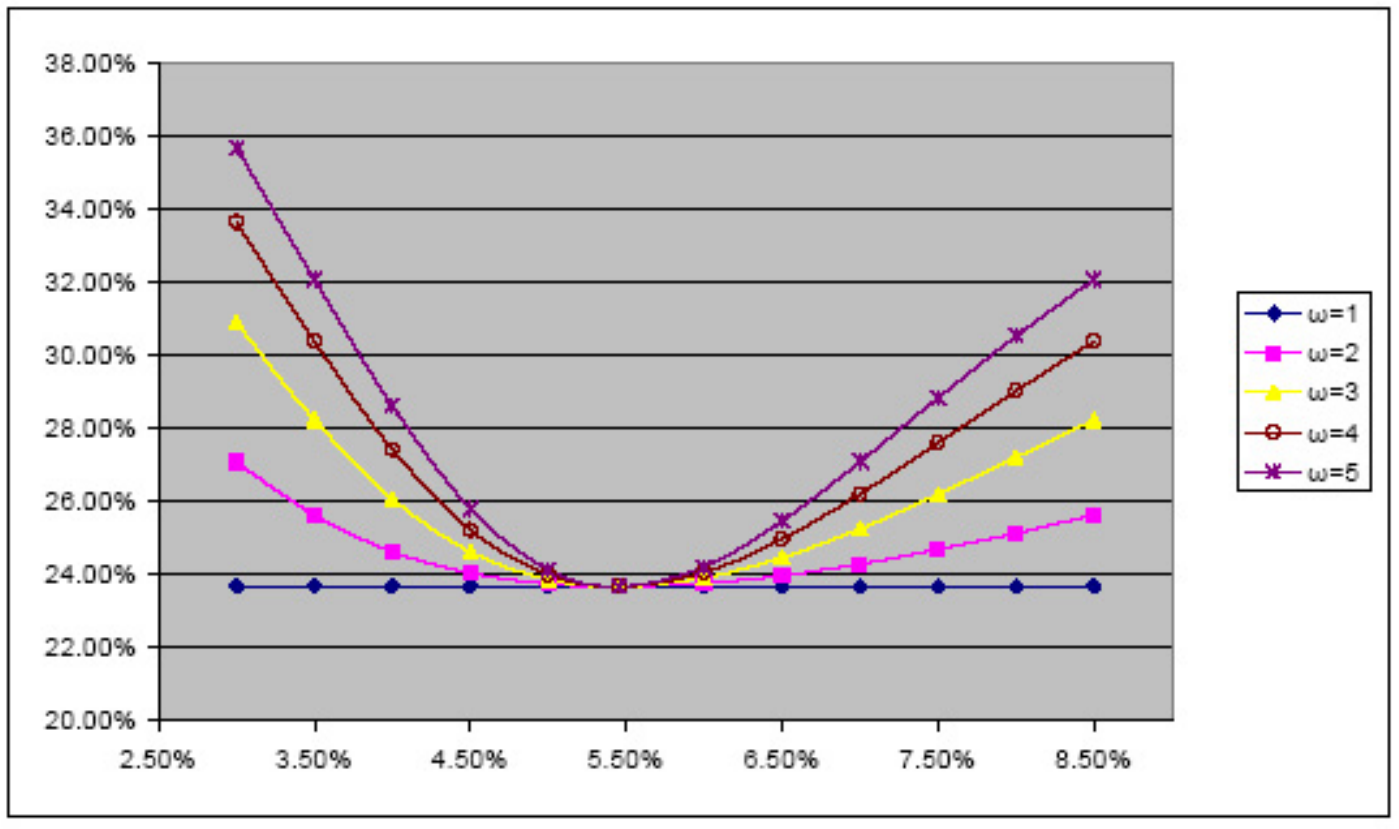}
\centering \caption{Implied smile by UVDD for various values of
$\omega$ ($\lambda=0.75$, $m=0$).} \label{fig:UVDD1} \centering
\end{figure}
We first report in Figure \ref{fig:UVDD1} the shapes of the
volatility smile obtained by setting $\lambda_n$ to 0.75, $m_n$ to 0
and by varying $\omega_n$  from 1 to 5. The case $\omega_n = 1$
reduces to the usual lognormal model. In this test and also the
following one, we use the same data set and trade specification as
was used in DD case described previously. We again adjust the
parameter $\sigma_5$ such that the implied Black volatility
corresponding to the at-the-money strike is the same for all the
cases. We see that a mixture of lognormal components without
displacement produces a symmetric smile centered around the
at-the-money strike. The smile shape is more pronounced for higher
values of $\omega_5$. This is because increasing the value of
$\omega_5$ implies that there are fatter tails (both left side and
right side) in the underlying distribution, and thus that
away-from-the-money swaptions are more underpriced in a lognormal
model.
\begin{figure}[h!]
\centering
\includegraphics[width=120mm,height=70mm]{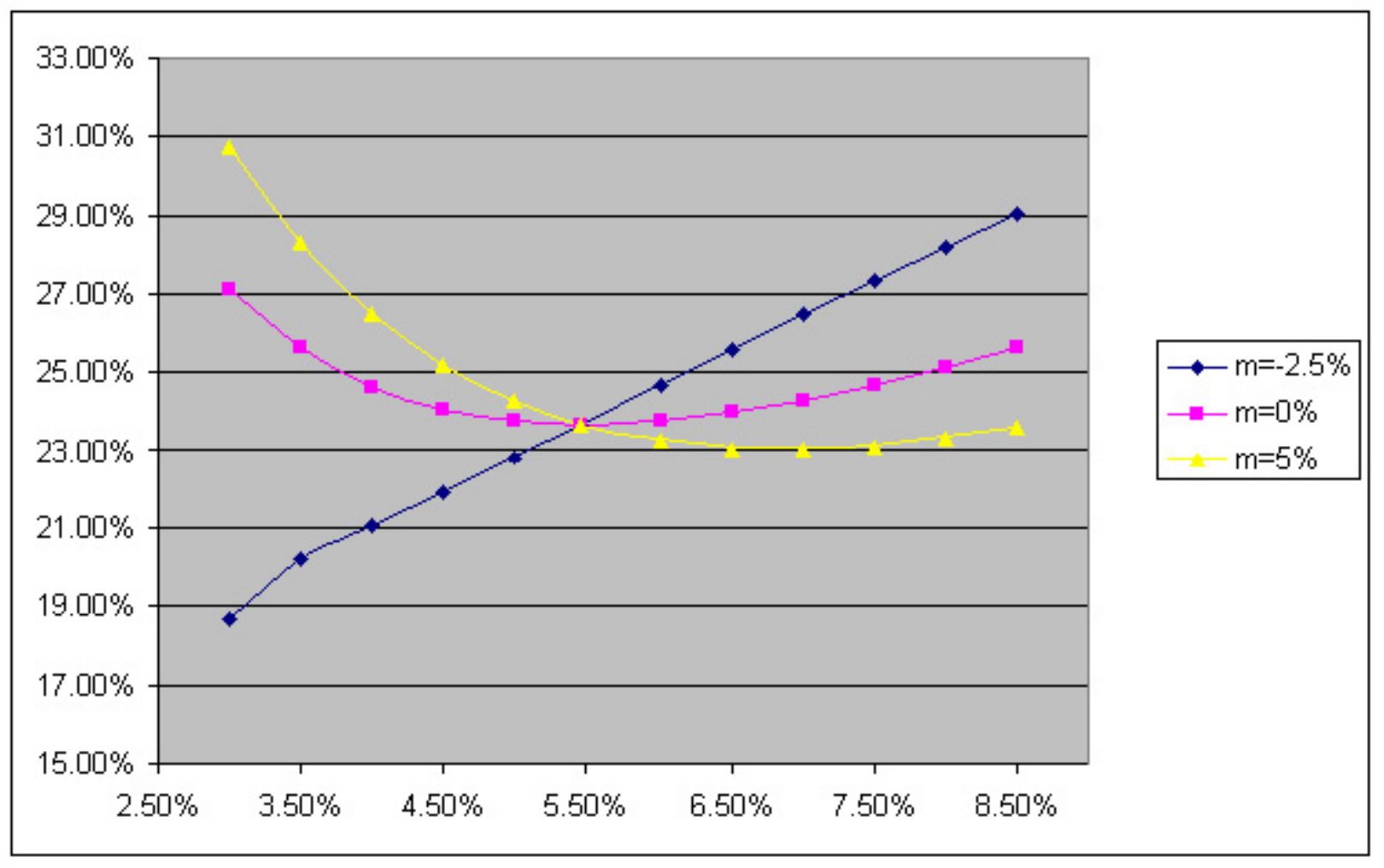}
\centering \caption{Implied smile by UVDD for various values of $m$
($\lambda=0.75$, $\omega=2$).} \label{fig:UVDD2}
\end{figure}
We report in Figure \ref{fig:UVDD2} the shape of the volatility
smile obtained by setting $\lambda_5$ to 0.75, $\omega_5$ to 2 and
by varying $m_5$ from 1 to 5. When $m_5$ is set to zero, a symmetric
smile is produced. Assigning a positive value to this parameter puts
more weight on low strike and less weight on the high strike. The
opposite happens when $m_5$ is set to a negative value. This is
because a higher displacement implies a fatter left tail and a
thinner right tail of the underlying distribution. Therefore, the
UVDD approach allows for combining a symmetric shape of the smile
with upward or downward sloping behavior.

\subsection{UVDD Digital Mapping}

We can perform the UVDD digital mapping by applying a small change
in the original BS mapping. The BS mapping is explained in Section
\ref{sec:BS_mapping}. In the UVDD digital mapping, the analytical
formula for digital swaptions corresponding to the UVDD model
(Equation \ref{eq:UVDD_Digital}) should be used instead of the Black
digital formula. We get the functional form of $S_n(X_n)$ by solving
the following equation numerically with respect to $x_n$,

\begin{eqnarray} \label{eq:UVDD_mapping}
DSN_n(0;K) &=& DSN_n(0;S_n(x_n)) \nonumber \\
&=& P_n(0)\Sigma_{i=1}^{M}\lambda^i
\Phi(\frac{\log(\frac{S_n(x_n)+m_n}{S_{n}(0)+m_n})+\frac{1}{2}(\sigma_n^i)^2
T_n}{\sigma_n^i\sqrt{T_n}}) \nonumber \\
 &=& \widetilde{DSN_n(0;x_n)}.
\end{eqnarray}

Note that $\widetilde{DSN_n(0;x_n)}$ is defined in Equation
\ref{eq:risk_neutral2}. This is a non-linear root-finding problem
for which we resort to the Newton-Raphson method\footnote{For
details of Newton-Raphson method, please refer to Section 9.4 of
"Numerical Recipes in C++" \cite{NR 2002}.}.

\section{Test Results of Different Digital Mappings}
\label{sec:pricing_test}

We have performed a number of tests based on the market data of Data
Set I in Appendix E.1 and using the setting of Trade I in Appendix
E.2. The tests were run for the following digital mappings:
\begin{itemize}
\item case 1: a Black-Scholes mapping;
\item case 2: a Displaced Diffusion mapping with $m_n=2.5\%$;
\item case 3: a Displaced Diffusion mapping with $m_n=5\%$;
\item case 4: a Displaced Diffusion mapping with $m_n=-2.5\%$;
\item case 5: a UVDD mapping with $m_n=0\%$, $\lambda_n=0.75$ and
$\omega_n=2$;
\item case 6: a UVDD mapping with $m_n=0\%$, $\lambda_n=0.75$ and
$\omega_n=5$;
\item case 7: a UVDD mapping with $m_n=2.5\%$, $\lambda_n=0.75$ and
$\omega_n=2$;
\item case 8: a UVDD mapping with $m_n=2.5\%$, $\lambda_n=0.75$ and
$\omega_n=3$.
\end{itemize}
For cases 2 to 8 we adjust the parameter $\sigma_n
(n=1,2,\ldots,10)$ in order to recover the same ATM volatilities
$\bar{\sigma}_n (n=1,2,\ldots,10)$ as for case 1. We first discuss
the results obtained for some consistency checks. Next, some test
results will be shown for the validity of the assumptions made in
the MF model and the convergence of the Markov-Functional model with
respect to the discretization parameters. Finally, we will discuss
the effect of the smile on the value of Bermudan swaptions.

\subsection{Consistency of European Swaption Prices}
\label{sec:European}

To demonstrate the correctness of the implementation, we first
compare the European swaption values ($ESN_n$ for n=1..10) obtained
by the MF model with the analytical formula for each of the eight
cases mentioned above. The strike (fixed coupon rate) is set to
5$\%$. The results are shown in Table \ref{table:case1_European} to
Table \ref{table:case78_European}. The first row refers to the 1
into 10 period swaption and the last row to the 10 into 1 period
swaption. The results clearly show that the MF model reproduces the
values of the underlying Europeans with high accuracy: the relative
error is less than 1 bp.


\begin{table}[h!]
\begin{center}
    \begin{tabular}{r|l}
        \hline
            Analytical values & MF values  \\
        \hline
                        0.00  &  0.00 \\
                        109.10 & 109.10 \\
                        194.40 & 194.40 \\
                        241.31 & 241.31 \\
                        246.96 & 246.96 \\
                        241.18 & 241.18 \\
                        208.48 & 208.48 \\
                        171.98 & 171.98 \\
                        119.22 & 119.22 \\
                        64.15 & 64.15 \\
        \hline
    \end{tabular}
\caption{Case 1 (Black-Scholes mapping).}
\label{table:case1_European}
\end{center}
\end{table}

\begin{table}[h!]
\begin{center}
    \begin{tabular}{r|l|r|l|r|l}
        \hline
            \multicolumn{2}{c|}{Case 2} & \multicolumn{2}{c|}{Case
            3} & \multicolumn{2}{c}{Case 4}\\
        \hline
            Analytical & MF values &
            Analytical & MF values &Analytical & MF values \\
        \hline
            0.00 &   0.00 &   0.00 &   0.00 &   0.00 &   0.00 \\
            107.86 & 107.86 & 107.25 & 107.25  &113.05 & 113.05 \\
            194.79 & 194.79 &194.98 &194.98  &193.26  &193.26 \\
            243.10 & 243.10 &244.01 &244.01 &236.28 &236.28 \\
            249.43 & 249.43 & 250.70 &250.70 & 240.23 &240.23 \\
            244.12 &244.12 &245.67 &245.67 & 233.35 & 233.35 \\
            211.25 & 211.25 &212.72 &212.72 & {\bf 201.21} &{\bf 201.20} \\
            174.52 & 174.52 & 175.88 &175.88 & 165.46 &165.46 \\
            121.07 & 121.07 &122.05 &122.05 & 114.54 & 114.54 \\
            65.21  & 65.21 &65.79 &65.79 &  61.52 &61.52\\
        \hline
    \end{tabular}
\caption{Case 2,3,4 (Displaced Diffusion mapping).}
\label{table:case234_European}
\end{center}
\end{table}

\begin{table}[h!]
\begin{center}
    \begin{tabular}{r|l|r|l}
        \hline
            \multicolumn{2}{c|}{Case 5} & \multicolumn{2}{c}{Case 6}\\
        \hline
            Analytical & MF values &
            Analytical & MF values \\
        \hline
            0.01  &  0.01  &  0.35  &  0.35 \\
            109.55 & 109.55 & 111.91 & 111.91 \\
            194.42 & 194.42 & 194.53 & 194.53 \\
            241.63 & 241.63 & 243.31 & 243.31 \\
            247.51 & 247.51 & {\bf 250.35} & {\bf 250.34} \\
            241.90 & 241.90 & {\bf 245.61} & {\bf 245.60} \\
            {\bf 209.14} & {\bf 209.13} & {\bf 212.49} & {\bf 212.48} \\
            172.62 &172.62 & {\bf 175.89} & {\bf 175.88} \\
            119.68 &119.68 &122.00 &122.00 \\
            64.45 &64.45 &65.95 &65.95 \\
        \hline
    \end{tabular}
\caption{Case 5,6 (UV mapping).} \label{table:case56_European}
\end{center}
\end{table}

\begin{table}[h!]
\begin{center}
    \begin{tabular}{r|l|r|l}
        \hline
            \multicolumn{2}{c|}{Case 7} & \multicolumn{2}{c}{Case 8}\\
        \hline
            Analytical & MF values &
            Analytical & MF values \\
        \hline
            0.01  &  0.01  &  0.06 &   0.06 \\
            108.31 & 108.31 & 109.06 & 109.06 \\
            194.81 & 194.81 & 194.84 & 194.84 \\
            243.42 &243.42  &243.95 & 243.95 \\
            249.97 &249.97 & 250.87 & 250.87 \\
            244.84 &244.84 &246.03  &246.03 \\
            211.91 &211.91 &{\bf 212.99} & {\bf 212.98} \\
            175.17 &175.17 &176.22 &176.22 \\
            121.53 &121.53 &122.28 &122.28 \\
            65.52 &65.52 &66.01 &66.01 \\
        \hline
    \end{tabular}
\caption{Case 7,8 (UVDD mapping with $m=2.5\%$).}
\label{table:case78_European}
\end{center}
\end{table}

\newpage
\subsection{Convergence of the Numerical Algorithm}
\begin{figure}[h!]
\centering
\includegraphics[width=120mm,height=70mm]{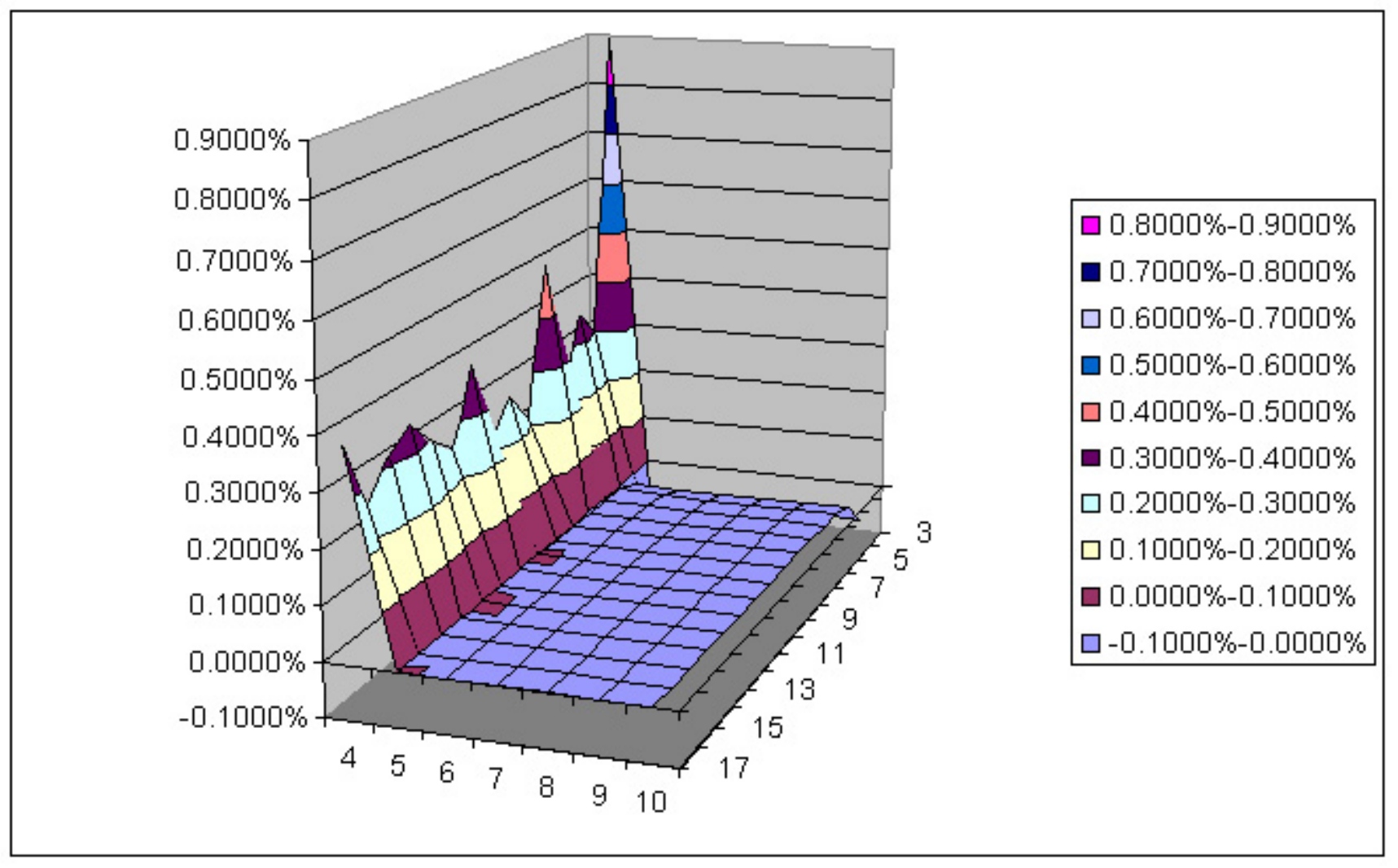}
\centering \caption{Relative pricing errors with respect to "Steps
per Deviation" and "Number of Deviations".} \label{fig:convergence2}
\end{figure}
The MF model is based on a lattice. In this section, we check the
convergence of the numerical integration with respect to the
discretization of the lattice. There are two parameters controlling
the level of discretization, namely,
\begin{itemize}
\item The range of values that $X$ can take expressed in units of
its standard deviation, {\em i.e.} "Number of Deviations";
\item The number of discrete points per standard deviation, {\em
i.e.} "Steps per Deviation".
\end{itemize}
To access the convergence of the method, it is sufficient to look at
the pricing of the European swaptions as this is basically exposed
to the discretization error in the numerical integration. We
consider the swaption which expires at $T_7$ and case 8 of Section
\ref{sec:European}. The analytical value of this European swaption
is 212.986 bp. The relative error is shown in Figure
\ref{fig:convergence2}, where the "Steps per Deviation" ranges from
3 to 17 and the "Number of Deviations" from 4 to 10. A very good
convergence is obtained by using the following setting: "Steps per
Deviation" higher than 5 and "Number of Deviations" higher than 4.


\subsection{Assumption/Approximation Validity Check under UVDD Mapping}
\label{sec:assumption_check}

\begin{figure}[h!]
\centering
\includegraphics[width=120mm,height=70mm]{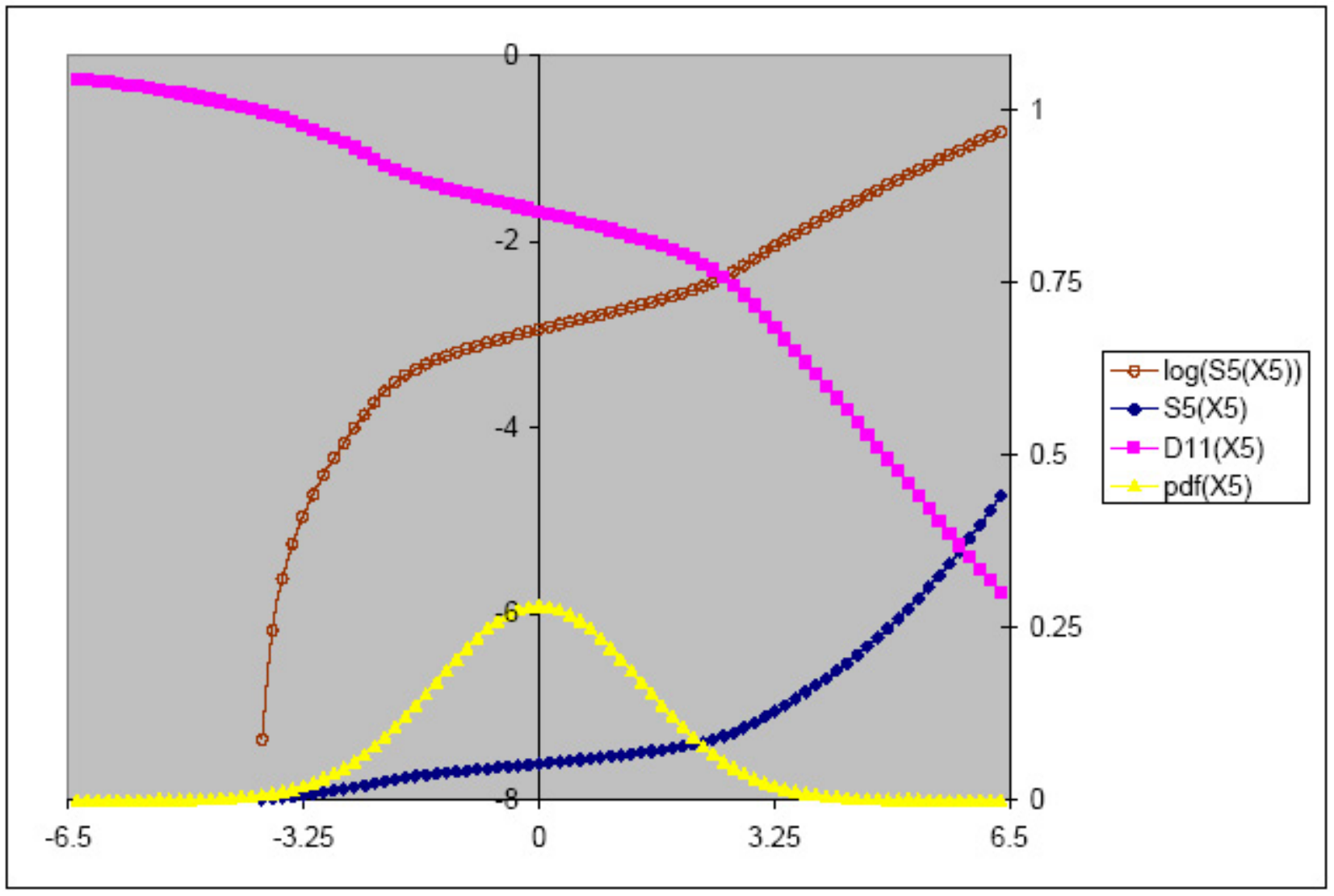}
\centering \caption{Assumption/Approximation validity check for
$T_5$.} \label{fig:assumption2}
\end{figure}

\begin{figure}[h!]
\centering
\includegraphics[width=120mm,height=70mm]{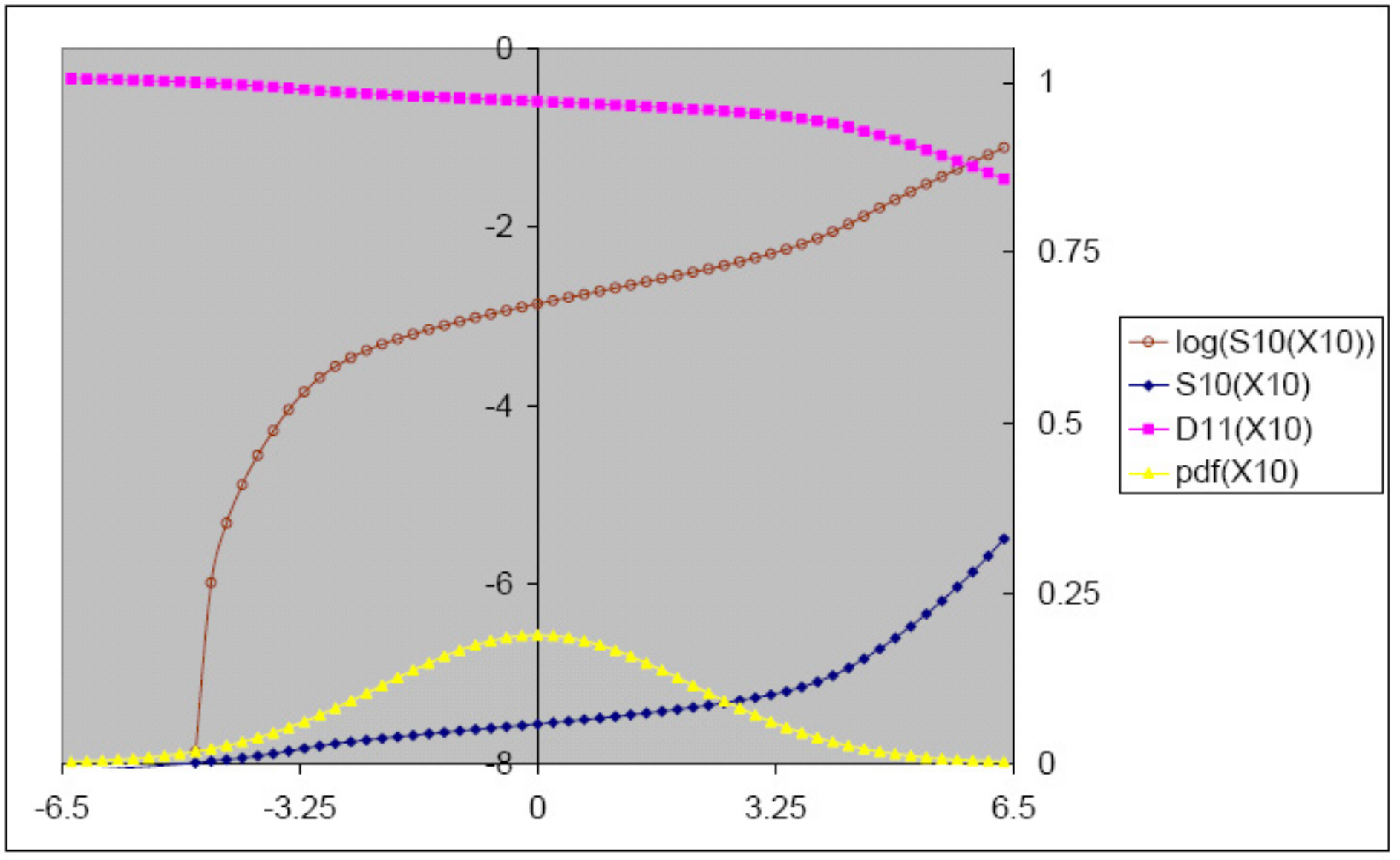}
\centering \caption{Assumption/Approximation validity check for
$T_{10}$.} \label{fig:assumption3}
\end{figure}

Before continuing further with pricing, we would like to check the
validity of the assumptions made in Section \ref{sec:assumption}. To
summarize the following assumptions were made:
\begin{itemize}
\item The numeraire discount bond $D_{N+1}(t,X(t))$ is a function
of $X(t)$;
\item The terminal swap rate $S_{n}(T_n,x)$ is a
strictly monotonically increasing function of $x$.
\end{itemize}
In Figures \ref{fig:assumption2} and \ref{fig:assumption3}, we show
the functional behavior of $D_{N+1}(t,X(t))$ and $S_{n}(T_n,X_n)$,
respectively, for $T_5$ and $T_{10}$ and corresponding to case 8.
Also included in these figures is the probability density function
of $X_n$. The following can be observed from these plots:
\begin{itemize}
\item The numeraire discount bond $D_{11}(X_5)$ and $D_{11}(X_{10})$ are
functions monotonically decreasing in $X_5$ and $X_{10}$,
respectively;
\item The swap rate $S_5(X_5)$ and $S_{10}(X_{10})$, respectively, are functions strictly monotonically increasing in
$X_5$ and $X_{10}$.
\end{itemize}
This demonstrates that the assumptions mentioned above are still
valid for the UVDD mapping.
\\ \\
We would also like to check the linear approximation made in
Equation \ref{eq:approximation} of Section \ref{sec:MR}. Here we
repeat that equation below,
\begin{eqnarray}
\log S_n(T_n,X_n) &\approx& \log S_n(T_n,x)|_{x=0} + X_n
\frac{\partial \log S_n(T_n,x)}{\partial x}|_{x=0}
\nonumber \\
\nonumber &=& \mbox{constant1} + \mbox{constant2} \times X_n.
\end{eqnarray}
We report in Figure \ref{fig:assumption2} and
\ref{fig:assumption3}\footnote{In these two figures only $\log
S_n(X_n)$ for $n=5,10$ is related to the first y-axis in the
middle.} the results for $T_5$ and $T_{10}$, respectively, and using
the setting corresponding to case 8. We see that $\log S_5(X_5)$
$\log S_5(X_{10})$ are linear functions of $X_5$ and $X_{10}$ for
$X_5$ and $X_{10}$ close to zero. It should be noted that the
approximation is even valid for the range of X where the majority of
the probability mass is concentrated. This demonstrates that in the
UVDD mapping the linear approximation is still valid.

\subsection{Effect on Bermudan Swaption Prices}
\label{sec:Bermudan_price}

To test the impact of the shape of the implied volatility smile on
the value of Bermudan swaptions, we consider the following trades
and settings:
\begin{itemize}
\item A Bermudan swaption with the right to exercise at reset
dates $T_5$, $T_6$, $T_7$, $T_8$, $T_9$ and $T_{10}$;
\item All the different cases (case 4 eliminated) with the strike set to $3.5\%$,
$5.5\%$ and $7.5\%$, respectively;
\item Forward par swap rates
$S_5(0)$, $S_6(0)$, $S_7(0)$, $S_8(0)$, $S_9(0)$ and $S_{10}(0)$
were set to $5.45\%$, $5.62\%$, $5.69\%$, $5.83\%$, $5.89\%$ and
$6.06\%$, respectively. The mean-reversion parameter was set to
$0\%$.\footnote{At this point we don't calibrate the mean-reversion
parameter using empirical data. $0\%$ can be seen as a benchmark
mean-reversion level.}
\end{itemize}

The results are reported in Table \ref{table:Bermudan}. Case 1 is
the standard lognormal case.
\\
\begin{table}[h!]
\begin{center}
    \begin{tabular}{c|c c c}
        \hline
            Strike & 3.50\% & 5.50\% & 7.50\% \\
        \hline
            case 1  & 541.00 & 228.45 & 90.11 \\
            case 2 & 548.63 & 228.45 & 82.32  \\
            case 3 & 552.71 & 228.48 & 78.23  \\
            case 5 & 545.76 & 226.27 & 95.52  \\
            case 6 & 567.78 & 223.78 & 126.56 \\
            case 7 & 553.17 & 226.54 & 88.30 \\
            case 8 & 560.57 & 223.97 & 99.17 \\
       \hline
    \end{tabular}
\caption{Effect on the price of Bermudans.} \label{table:Bermudan}
\end{center}
\end{table}

For the displaced diffusion cases, \emph{i.e.}, cases 2-3,
increasing the displacement coefficient leads to an increase in the
implied volatilities corresponding to low strikes and to a decrease
in the implied volatilities corresponding to high strikes. This
results in a fatter left tail and a thinner right tail for the
underlying's distribution. Thus we see an increase in the
displacement coefficient results in an increase in the value of a
deep ITM Bermudan and a decrease in the value of a deep OTM
Bermudan.
\\ \\
For the UVDD cases, \emph{i.e.}, cases 5-6 or 7-8, more pronounced
smiles result in a higher price for a deep ITM/OTM Bermudan
swaption. This can be explained in a similar way as above since a
more pronounced smile leads to fatter tails (both left side and
right side) in the distribution of the underlying. However, for the
near-the-money Bermudans, we see that more pronounced smiles result
in a lower price. This is counter-intuitive and is contrary to the
findings of Abouchoukr \cite{Abouchoukr 2003}. We emphasize that the
tests in that study were based on different market data and trade
specifications. It was found that more pronounced smiles result in a
higher near-the-money Bermudan price. To make sure that this result
is not due to numerical artifacts, we have increased the "Steps per
Deviation" and "Number of Deviations"\footnote{Please refer to
Appendix \ref{sec:TestTrades} for the grid specification.} to 100
and 20, respectively. The findings did not changed. A possible
explanation for this phenomenon is provided in Appendix
\ref{chapter:negative_vega}. It is shown that this behavior is
plausible in the case of a simple example.
\\ \\
In order to analyze the impact of the mean-reversion parameter, we
valued the Bermudan swaption for case 1 and 8 for a wider range of
strikes and two different mean-reversion levels ($0\%$ and $10\%$,
respectively). The results are shown in Table \ref{table:Bermudan2}.
As expected, in both cases, the European value converges to the
underlying swap value as the strike gets lower. Moreover, in both
cases, the Bermudan swaption value converges to its European
counterpart as the strike gets lower. This is because for a payer
Bermudan it becomes more likely to exercise early as the strike gets
lower.

\begin{flushleft}
To summarize the following can be concluded from the results
presented in Table \ref{table:Bermudan2}:
\end{flushleft}
\begin{itemize}

\item Incorporating the volatility smile has a significant impact for the
away-from-the-money Bermudan prices, while it has a relatively
marginal impact for the near-the-money Bermudan prices. This is
inline with the observation that the effect of smile on the
Europeans is more pronounced as one moves away from the ATM level;

\item For payer Bermudans, increasing the mean-reversion level gives
rise to an increase of the overall level of Bermudan prices. This
can be roughly explained as follows. Increasing the mean-reversion
level will reduce the terminal correlations of swap rates $S_n(T_n)$
and $S_k(T_k)$ for $n = 1..N, k > n$ (see Section \ref{sec:MR}). In
our case, if in the future state at $T_5$, $S_5(T_5)$ was below the
strike, a higher mean-reversion level would increase the likelihood
that the economy would transform to a state at a further future time
$T_n$ for $n=6,7,8,9,10$ in which $S_n(T_n)$ would be higher than
the strike, and thus increasing the value of the Bermudan.

Finally, the increase is more pronounced for the near-the-money
Bermudans compared to the away-from-the-money ones. This can be
explained as follows. If the strike is much lower than the ATM
level, the Bermudan will be less sensitive to the terminal
correlations as the likelihood of early exercising at $T_5$ is quite
high. If the strike gets much higher than the ATM strike level, this
effect would become less pronounced too as the likelihood of
postponing the exercise later in future will increase.
\end{itemize}

\begin{table}[h!]
\begin{center}
    \begin{tabular}{c c c c c c c}
        \hline
            Strike & 3.00\% &  3.50\% &  4.00\% &  4.50\% &  5.00\%
               & 5.50\%   \\
        \hline
            Swap value &   639.98 & 509.63 & 379.28 & 248.93 & 118.58 & -11.77 \\
            European (BS) & 645.22 & 526.08 & 418.22 & 324.74 &
            246.96 & 184.52 \\
            Bermudan MR=0\% (BS) & 652.52 & 541.00 & 442.23 & 357.49 & 286.60 & 228.45 \\
            Bermudan MR=10\% (BS) & 656.70 & 547.48 & 450.62 & 367.07 &
            296.63 & 238.30 \\
            Bermudan MR=0\% (UVDD) & 675.22 & 560.57 & 455.04 & 362.39 &
            285.15 & 223.97 \\
            Bermudan MR=10\% (UVDD) & 679.24 & 565.75 & 461.58 & 370.09 &
            293.51 & 232.50 \\
            European (UVDD) & 663.95 & 546.10 & 435.07 & 335.24 &
            250.87 & 184.29 \\
        \hline
        \hline
            Strike & 6.00\%   &6.50\%   &7.00\%   &7.50\%   &8.00\%   &8.50\% \\
        \hline
            Swap value &  -142.12 & -272.47 & -402.82 & -533.17 & -663.52 & -793.87\\
            European (BS) & 135.85 & 98.84 & 71.23 & 50.95 &  36.24 &
            25.67 \\
            Bermudan MR=0\% (BS) & 181.41 & 143.75 &113.81 &90.11 &  71.40 &
            56.65 \\
            Bermudan MR=10\% (BS) & 190.64 & 152.11 & 121.18 & 96.49 & 76.83
            & 61.23 \\
            Bermudan MR=0\% (UVDD) & 177.84 & 143.48 & 118.04 & 99.17 &
            84.85 & 73.67 \\
            Bermudan MR=10\% (UVDD) & 186.11 & 151.38 & 125.55 & 106.23 &
            91.43 & 79.80 \\
            European (UVDD) & 135.00 & 100.33 & 76.63 & 60.49 &
            49.23 & 41.02 \\
        \hline
        \hline
        \multicolumn{7}{l }
        {\scriptsize MR denotes the mean-reversion level.}
    \end{tabular}
\caption{Bermudan prices for a wider range of strikes (case 1 and
8).} \label{table:Bermudan2}
\end{center}
\end{table}

\chapter{Future Smile and Smile Dynamics} \label{chapter:dynamics}

\section{Future Volatility Smile Implied by MF Models}\label{sec:future_smile}

It is well-known that the value of a path dependent option is highly
sensitive to the future volatility smile (see e.g. Ayache
\cite{Ayache 2004} and Rosien \cite{Rosien 2004}). The future
volatility smile is defined as follows. Today we observe market
quotes $\sigma_n(K)$ for European swaptions expiring at $T_n$. We
calibrate our model to the vanilla options such that the terminal
density of the underlying is consistent with the market. We now move
to a future date which is still before expiry, and use our
calibrated model to compute, at that time, values of the vanilla
options across strikes. Conditional on the state of the future date,
we can invert the option values to get the implied smile, {\em i.e.}
the future volatility smile. It should be noted that in the MF model
the state transition\footnote{For instance, the probability that
$L_n(x_n)$ goes to $L_{n+1}(x_{n+1})$.} is controlled by the
mean-reversion parameter.

Let's now explain in more detail how to obtain future smiles in the
MF models. For ease of calculation, we choose the future date,
denoted by $T_f$, to be one of the floating reset dates,
\emph{i.e.}, $T_f=T_m$ for $m=1,2,\ldots,N-1$. Suppose at time
$T_f$, $X_f=x_f$. Conditional on this state, the value of the
European swaption is determined as described in Section
\ref{sec:pricing}. In order to get the implied volatility from the
price, we still need to determine $P_n(x_f)$ and $S_n(x_f)$. This
can be done by using the functional form of $D_k(x_f)$ (k=n,..,N+1)
which can be obtained through numerical integration on the
calibrated lattice. For more details we refer to Section
\ref{section:mf_model}. Note that all these computations depend on
the conditional expectation $\phi(X_k|X_i)$, for $k>i$ and
$i=f,f+1,\cdots,N$, which is ultimately subject to the
mean-reversion parameter value.

\subsection{Future Volatility Smile Implied by the BS Mapping}
\begin{figure}[hp!]
\centering
\includegraphics[width=120mm,height=80mm]{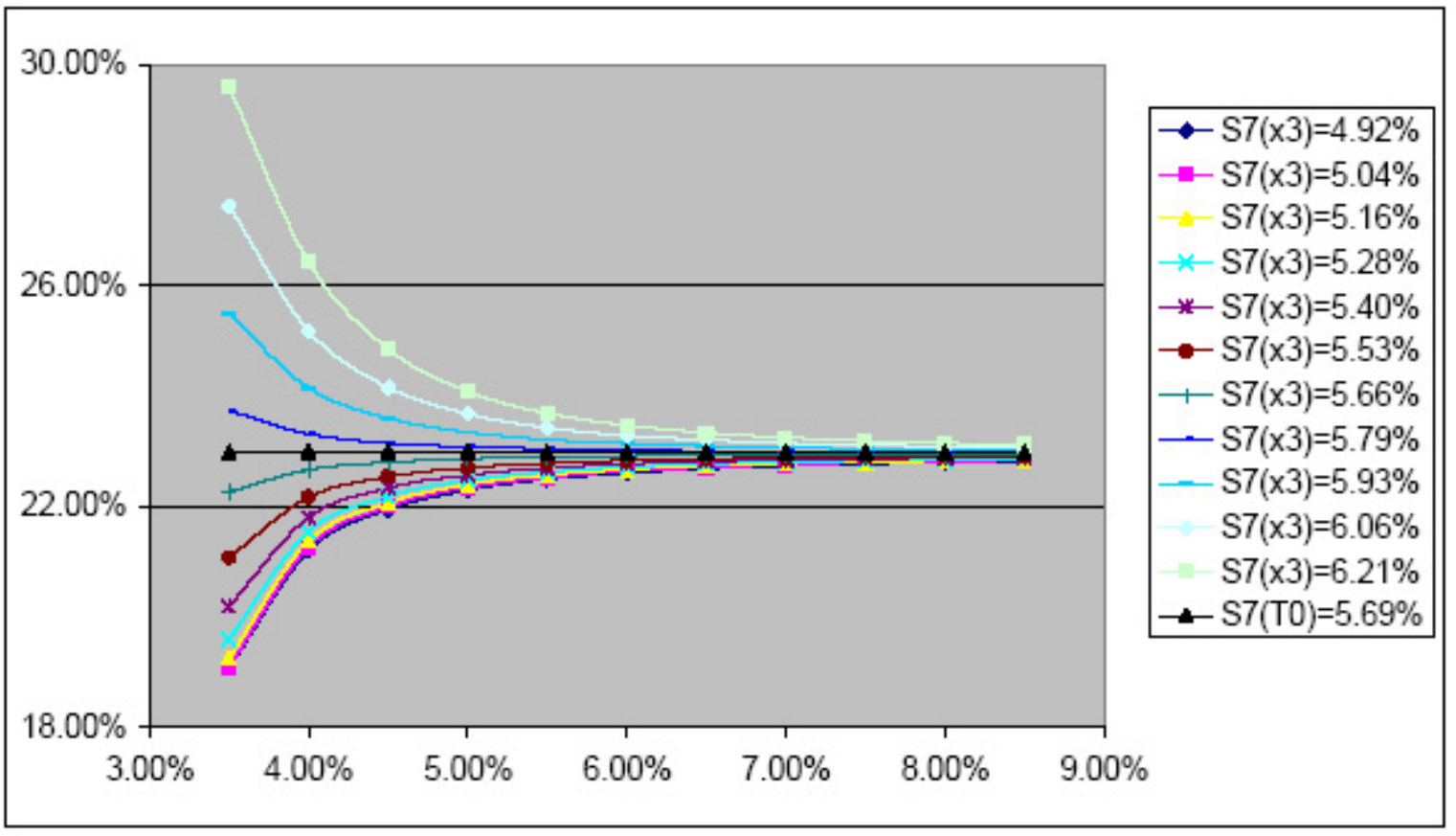}
\centering \caption{Future smiles implied by BS mapping and today's
flat volatility.} \label{fig:BS_future}
\end{figure}

We first investigate the future volatility smile implied by the
Black-Scholes digital mapping, \emph{i.e.}, the future volatility
smile implied by case 1 in Section \ref{sec:pricing_test}. The test
was run for a payer swaption expiring at the seventh floating reset
date, \emph{i.e.}, $T_7$. The at-the-money swap rate $S_7(0)$ was
around $5.69\%$, and the mean-reversion parameter was set to zero.
We calculated the future smiles standing at $T_3$ conditional on
different future states $X_3$\footnote{$X_n$ is the state variable
in MF. We chose, by trial and error, different states from the
lattice such that swap rates were in the desired range.} such that
the underlying swap rate $S_7(x_3)$ ranges from $4.92\%$ to
$6.21\%$. The corresponding future smiles are reported in Figure
\ref{fig:BS_future}. In the case of the Black-Scholes model we
expect the future smile to be flat. We see that this is not the case
in the MF model with the BS mapping. However, the future smiles
obtained for the different strikes (low and high, respectively) are
compensating each other. More precisely, the fat-tailed
distributions implied for high swap rates compensate the thin-tailed
distributions implied for low swap rates. Therefore, on an
integrated level we still may obtain a flat smile.


\subsection{Future Volatility Smile Implied by the UVDD Mapping}
\begin{figure}[h!]
\centering
\includegraphics[width=120mm,height=80mm]{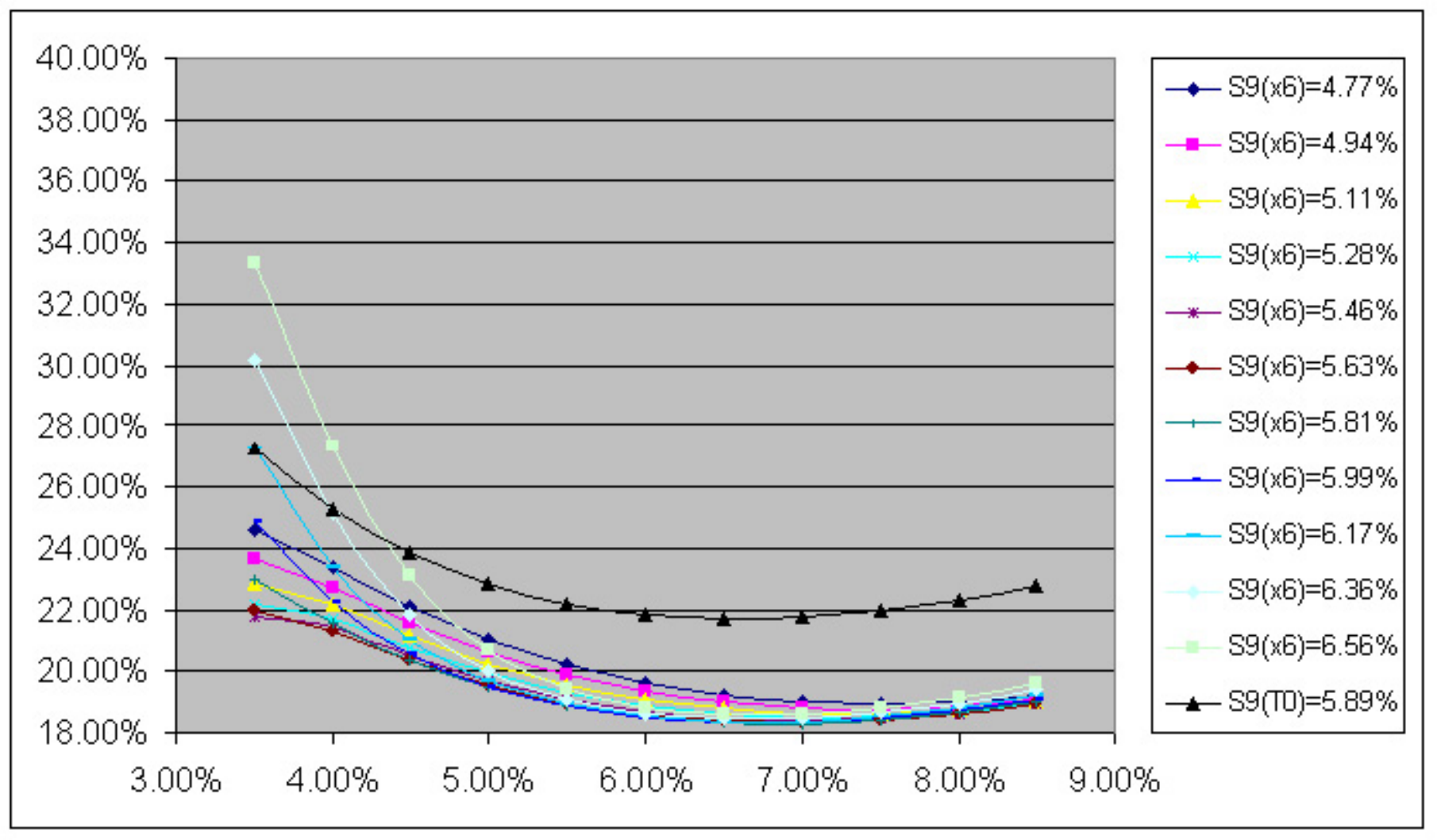}
\centering \caption{Future smiles implied by UVDD mapping for
$MR=0\%$ and today's smile.} \label{fig:UVDD_future1}
\end{figure}

\begin{figure}[h!]
\centering
\includegraphics[width=120mm,height=80mm]{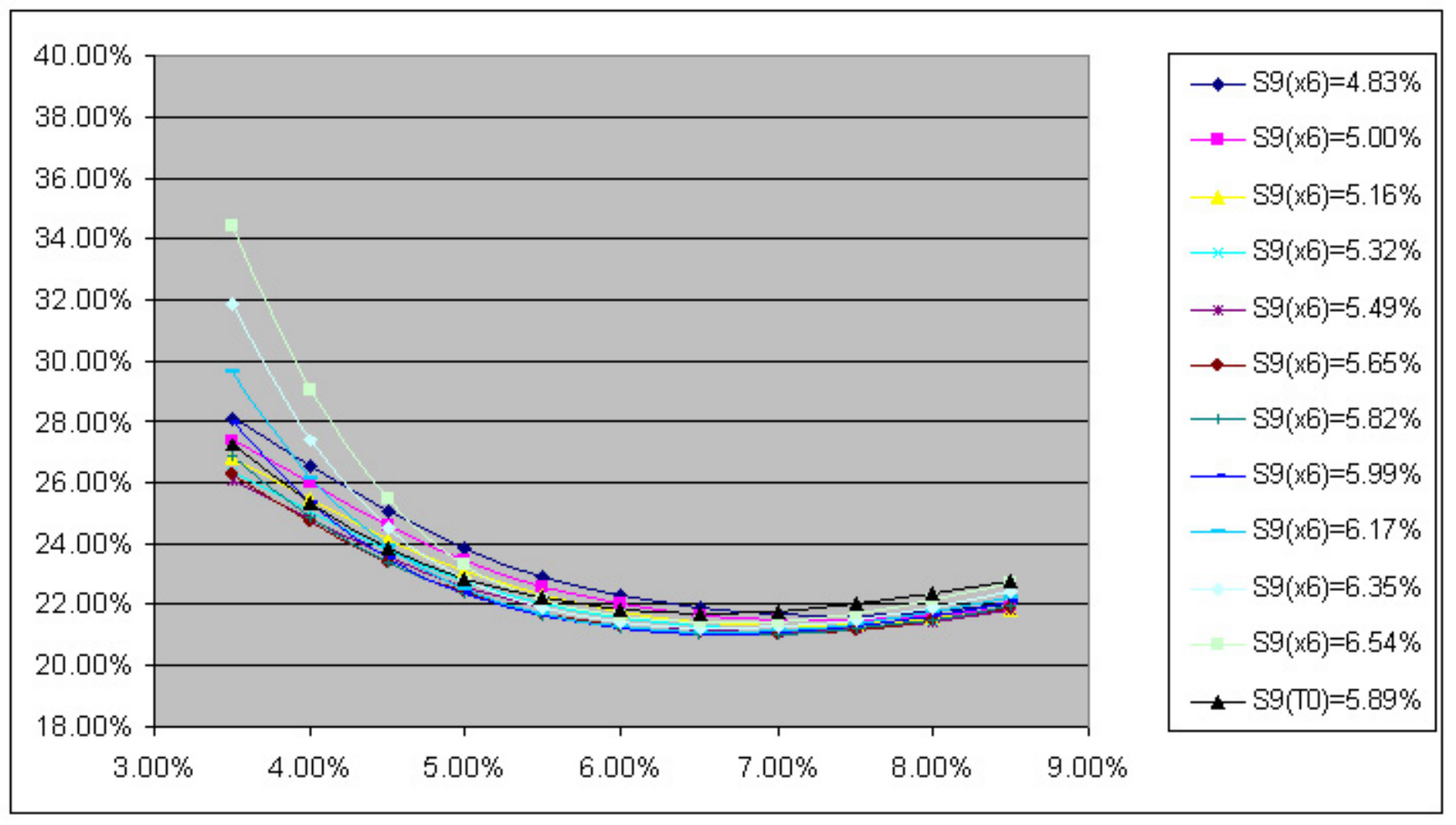}
\centering \caption{Future smiles implied by UVDD mapping for
$MR=10\%$ and today's smile.} \label{fig:UVDD_future2}
\end{figure}

\begin{figure}[h!]
\centering
\includegraphics[width=120mm,height=80mm]{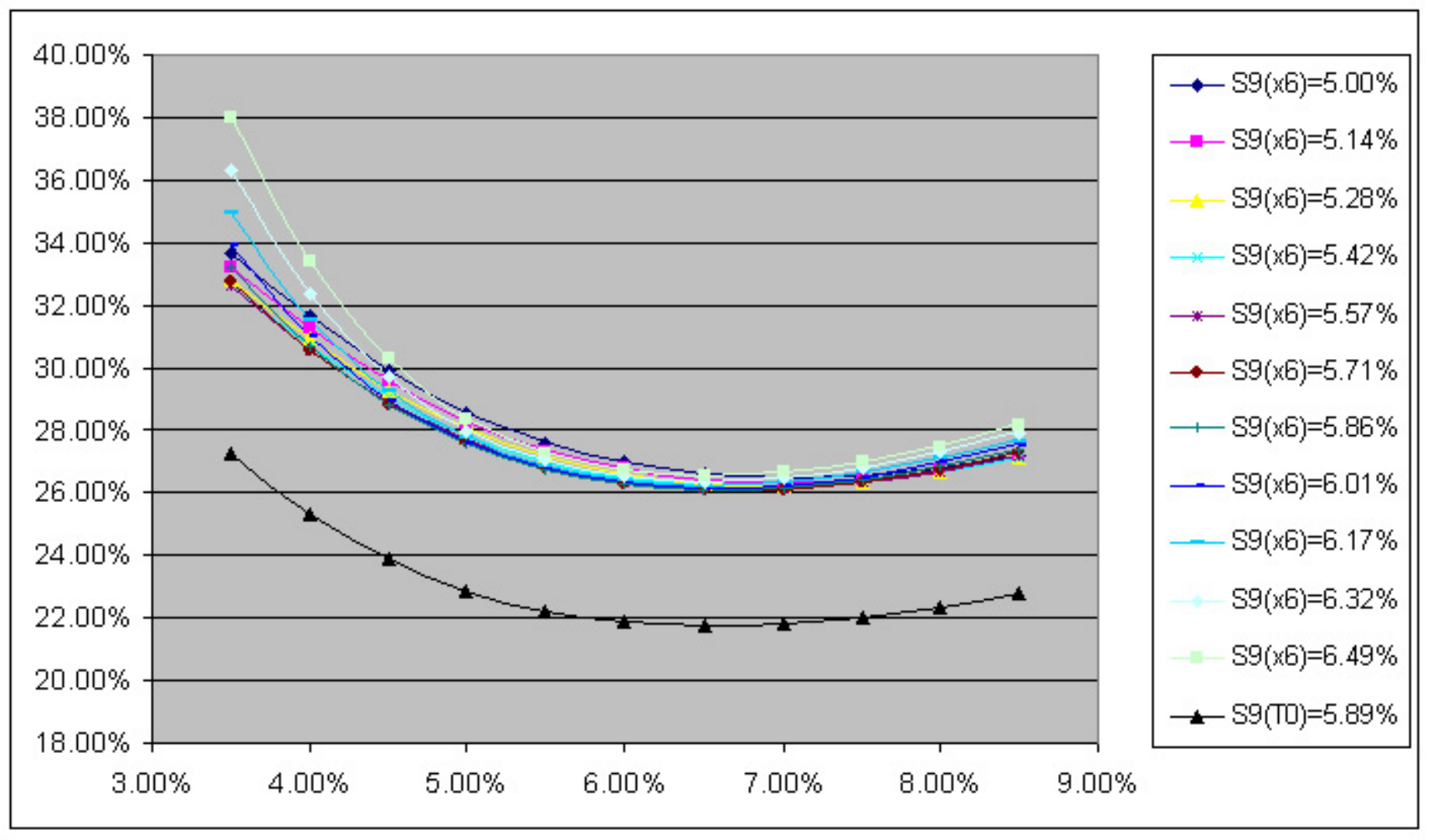}
\centering \caption{Future smiles implied by UVDD mapping for
$MR=30\%$ and today's smile.} \label{fig:UVDD_future3}
\end{figure}

We will now investigate the future volatility smile implied by the
UVDD digital mapping, more precisely, the future volatility smile
implied by case 8 in Section \ref{sec:pricing_test}. The test was
run for a payer swaption expiring at the ninth floating reset date,
\emph{i.e.}, $T_9$. The at-the-money swap rate $S_9(0)$ was
$5.89\%$. We calculated the future smiles standing at $T_6$. We
adjusted $X_6$'s value such that the underlying swap rate $S_9(x_6)$
varies approximately from $4.8\%$ to $6.5\%$. In order to study the
effect of the mean-reversion parameter, we considered three MR
levels, namely $0\%$, $10\%$, and $30\%$\footnote{A mean-reversion
speed of 30\% is very unrealistic. We set this exaggerated value
just for illustrative purposes.}, respectively. The results of these
three tests together with today's smile are reported in Figure
\ref{fig:UVDD_future1}, \ref{fig:UVDD_future2} and
\ref{fig:UVDD_future3}, respectively. \emph{We see that increasing
the mean-reversion level has the effect of increasing the overall
level of the future smiles.} This can be explained as follows. For
$X_f$ close to zero\footnote{This is also the range where the
majority of the probability mass is concentrated.}, $S_n(T_f,X_f)$,
can be approximated as follows,
\begin{eqnarray}
S_n(T_f,X_f) &\approx& S_n(T_f,x)|_{x=0} + X(T_f)\frac{\partial
S_n(T_f,x)}{\partial
x}|_{x=0} \nonumber \\
&=& \mbox{constant1} + \mbox{constant2} \times X(T_f).
\end{eqnarray}
\\
The correlation between the future underlying level $S_n(T_f,X_f)$
and the terminal underlying level $S_n(T_n,X_n)$ is given by,
\begin{equation}
Corr(S_n(T_f),S_n(T_n))\approx\rho(X_f(T_f),X_n(T_n))
=\begin{cases}\sqrt{\frac{T_f}{T_n}} & \mbox{if $a=0$}\\
\sqrt{\frac{e^{2aT_f}-1}{e^{2aT_n}-1}} & \mbox{if $a\neq
0$}\end{cases}.
\end{equation}
\\
Here we use the analytical form of $Corr(X_f(T_f),X_n(T_n))$ from
Section \ref{sec:MR}, and $a$ is the mean-reversion parameter.
Therefore, increasing the mean-reversion parameter $a$ has the
effect of reducing the correlation between the future underlying
level $S_n(T_f,X_f)$ and the terminal underlying level
$S_n(T_n,X_n)$. This implies that the average volatility within the
time period $[T_f,T_n]$ increases, which is consistent with the
phenomenon we observe in Figure \ref{fig:UVDD_future1},
\ref{fig:UVDD_future2} and \ref{fig:UVDD_future3}.
\\ \\
In conclusion, by calibrating the mean-reversion parameter to the
relevant market information\footnote{For details, please refer back
to Section \ref{sec:MR_estimation} for the relevant discussion.},
the MF model is able to control the future smiles to some extent.
\\ \\ \\

\newpage
\section{Smile Dynamics Implied by the UVDD Model}
Many previous studies \cite{Ayache 2004}\cite{Hagan
2002}\cite{Rosien 2004}\cite{Zilber 2003} point out that the smile
dynamics implied by an option pricing model indicates whether the
model is able to produce good hedge ratios. Let's take a look at an
European option and a model $M$. We define the implied volatility,
$\sigma_{imp}(S;K,T)$, as the volatility that should be used in the
Black model to match the value of this option obtained using model
$M$,
\begin{equation}
V(S;K,T) = V_{BS}(S,\sigma_{imp}(S;K,T);K,T) \label{eq:V_BS},
\end{equation}
where $V(S;K,T)$ denotes the value of the European option as
obtained by model $M$. By Equation \ref{eq:V_BS}, we can obtain the
implied Black volatility $\sigma_{imp}(S;K,T)$, which is a function
of the underlying $S$ with parameters strike $K$ and maturity $T$,
from the prices implied by the model.

The delta hedge ratio of this model, $\Delta(S;K,T)$, is given by,
\begin{eqnarray}
\Delta(S;K,T) &=& \frac{\partial V(S;K,T)}{\partial S} \;\;\;=\;\;\;
\frac{\partial V_{BS}(S,\sigma_{imp}(S;K,T);K,T)}{\partial S}
\nonumber \\
&+& \frac{\partial
V_{BS}(S,\sigma_{imp}(S;K,T);K,T)}{\partial\sigma_{imp}(S;K,T)}
\times\frac{\partial\sigma_{imp}(S;K,T)}{\partial S}
\label{eq:V_delta}.
\end{eqnarray}
From Equation \ref{eq:V_delta}, we see that the validity of the
delta ratio generated by model $M$ is subject to the sensitivity
$\frac{\partial\sigma_{imp}(S;K,T)}{\partial S}$. This is what we
mean by the \emph{smile dynamics} of a model. In Ayache \cite{Ayache
2004} and Rosien \cite{Rosien 2004}, they entitle a wider concept to
smile dynamics, which also covers the structure of
conditionals\footnote{We refer back to Section
\ref{sec:future_smile} for the relevant discussion.} apart from this
sensitivity. In our report, we stick to this definition of the smile
dynamics\footnote{This definition is equivalent to the "local smile
dynamics" in Rosien \cite{Rosien 2004}.}, \emph{i.e.},
\begin{equation} \label{eq:smile_dynamics}
\mbox{Smile Dynamics} \;\; \triangleq
\;\;\frac{\partial\sigma_{imp}(S;K,T)}{\partial S}.
\end{equation}

\begin{figure}[h!]
\centering
\includegraphics[width=110mm,height=70mm]{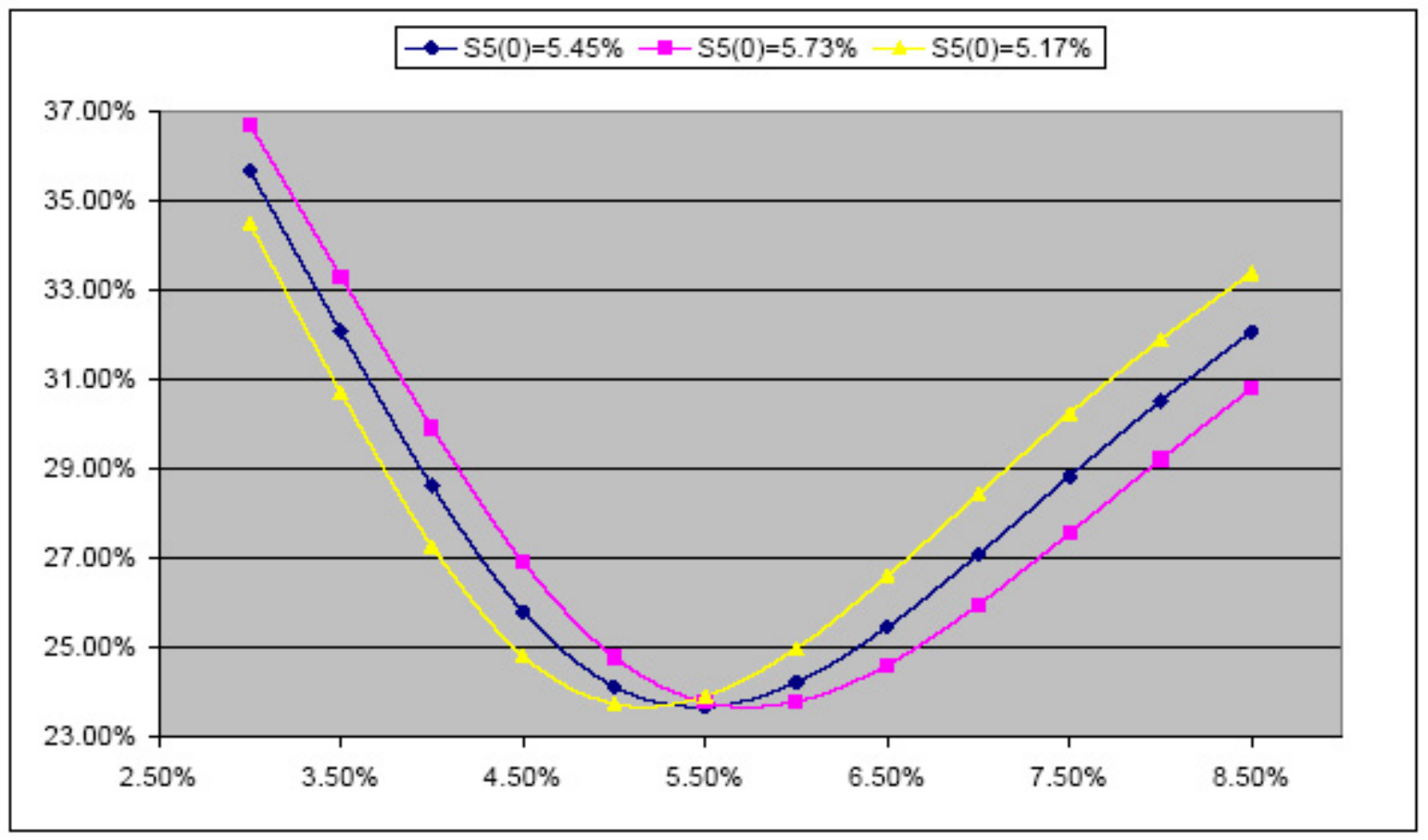}
\centering \caption{Smile dynamics implied by the UV model by
bumping the yield curve.} \label{fig:smile_dynamics3}
\end{figure}

\begin{figure}[h!]
\centering
\includegraphics[width=110mm,height=70mm]{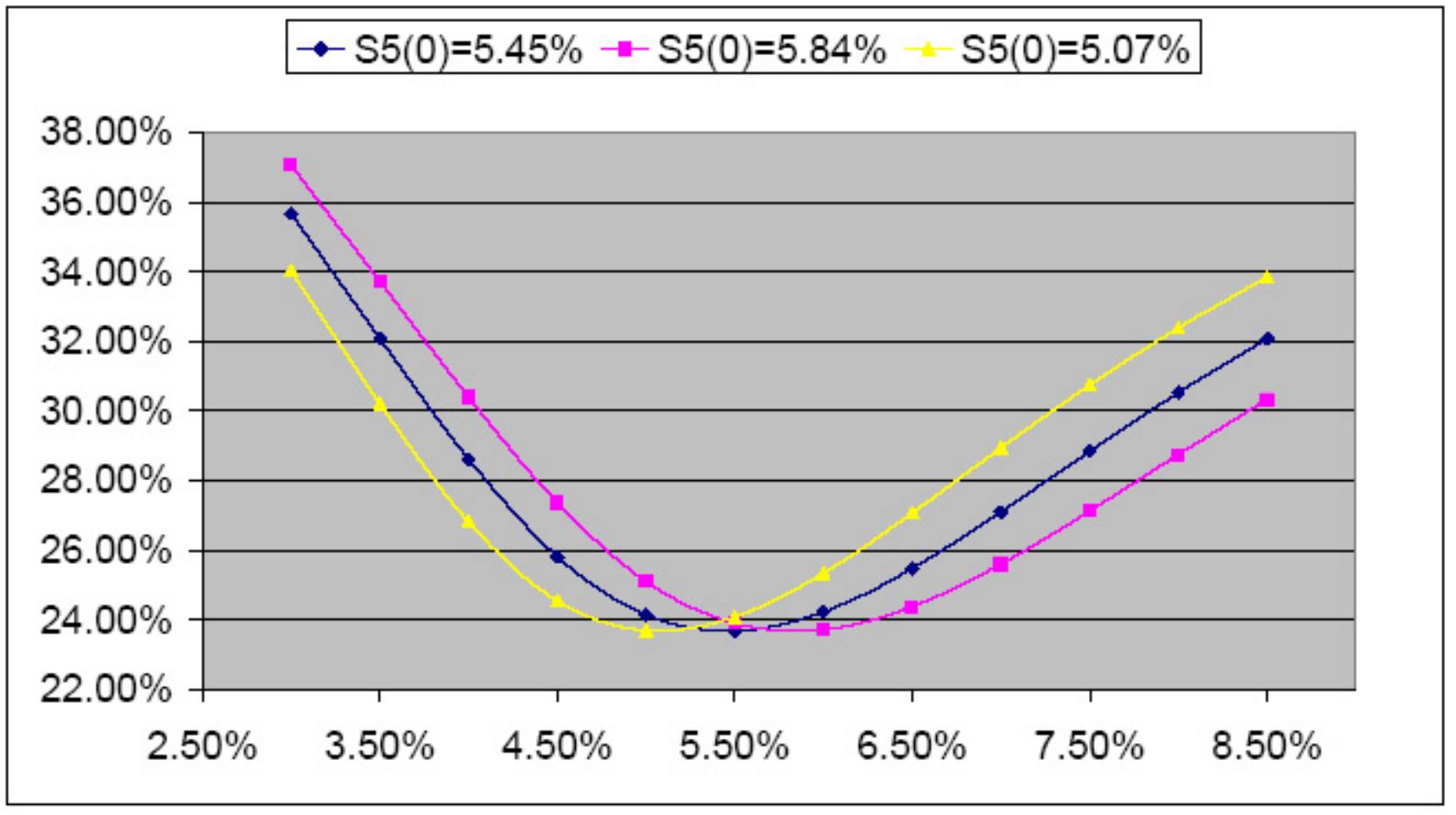}
\centering \caption{Smile dynamics implied by UV by bumping only the
discount factor $D_5(0)$.} \label{fig:smile_dynamics4}
\end{figure}

The smile dynamics has been a point of discussion for many models.
It is well-known that the stochastic volatility models show a sticky
delta behavior\footnote{A stick delta smile dynamics means the
implied volatility stays the same for every \emph{moneyness}, which
is defined as underlying's level divided by strike. This
equivalently means the volatility smile slides along the strike
axis.}, while local volatilty models may predict the opposite smile
dynamics \cite{Hagan 2002}. To study the smile dynamics of the UVDD
model, we have calculated the smile dynamics by the following cases:
\begin{itemize}
\item We first considered a digital mapping corresponding to case 6 of
Section \ref{sec:pricing_test}, \emph{i.e.}, the model has no
displacement but is a purely Uncertain Volatility (UV) model.
Moreover, a swaption expiring at the fifth floating reset date,
\emph{i.e.}, $T_5$, with at-the-money swap rate of $5.45\%$ was
considered. The first experiment was done by bumping up/down the
yield curve by 50 bp\footnote{This is equivalent to bumping only the
relevant part of the yield curve, \emph{i.e.}, from $T_5$'s yield to
$T_{11}$'s yield.} so that the underlying swap rate $S_5(0)$
increases/decreases to $5.73\%/5.17\%$, respectively. We use the
calibrated UV model corresponding to the un-bumped case. The
volatility smiles are shown in Figure \ref{fig:smile_dynamics3}. We
observe a \emph{sticky delta} smile dynamics. This is not surprising
as the UV model falls into the category of stochastic volatility
models, whose smile dynamics has typically a sticky delta effect
\cite{Rosien 2004};
\item  Because bumping the yield curve also changes the PVBP,
$P_5(0)$, (see Equation \ref{eq:European}), in the second
experiment, we have bumped up/down only the discount factor
$D_5(0)$\footnote{This is unrealistic in hedging simulations as in
practice the interest rates risk is delta-hedged using swaps.} by an
amount of 100 bp so that only $S_5(0)$ increases/dereases to
$5.84\%/5.07\%$ while $P_5(0)$ retains at the original level. The
implied volatility smiles are reported in Figure
\ref{fig:smile_dynamics4}. We observe the same sticky delta
phenomenon as was obtained in the first experiment.
\end{itemize}

\begin{figure}[h!]
\centering
\includegraphics[width=110mm,height=70mm]{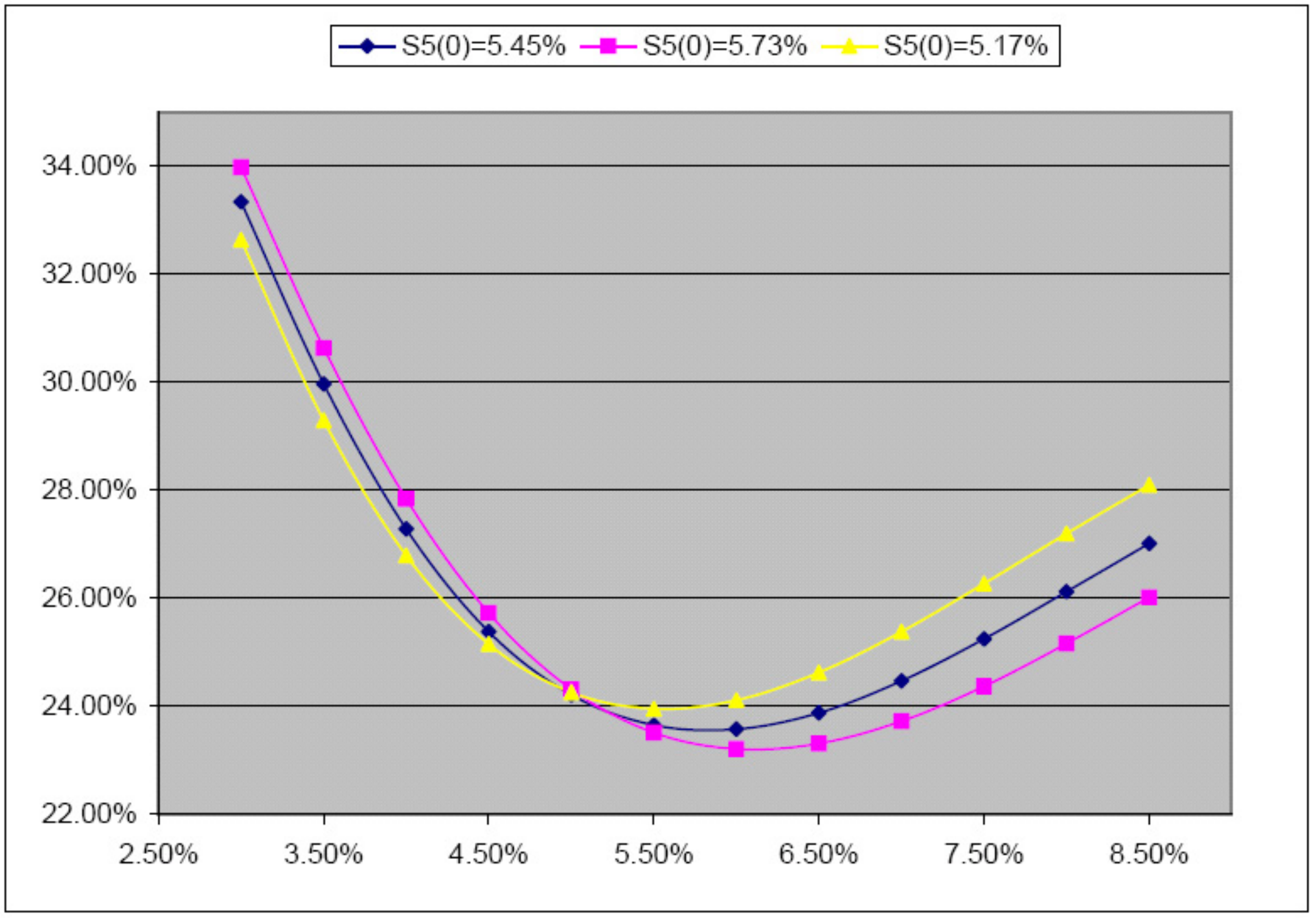}
\centering \caption{Smile dynamics implied by the UVDD model by
bumping the yield curve.} \label{fig:smile_dynamics1}
\end{figure}

\begin{figure}[h!]
\centering
\includegraphics[width=110mm,height=70mm]{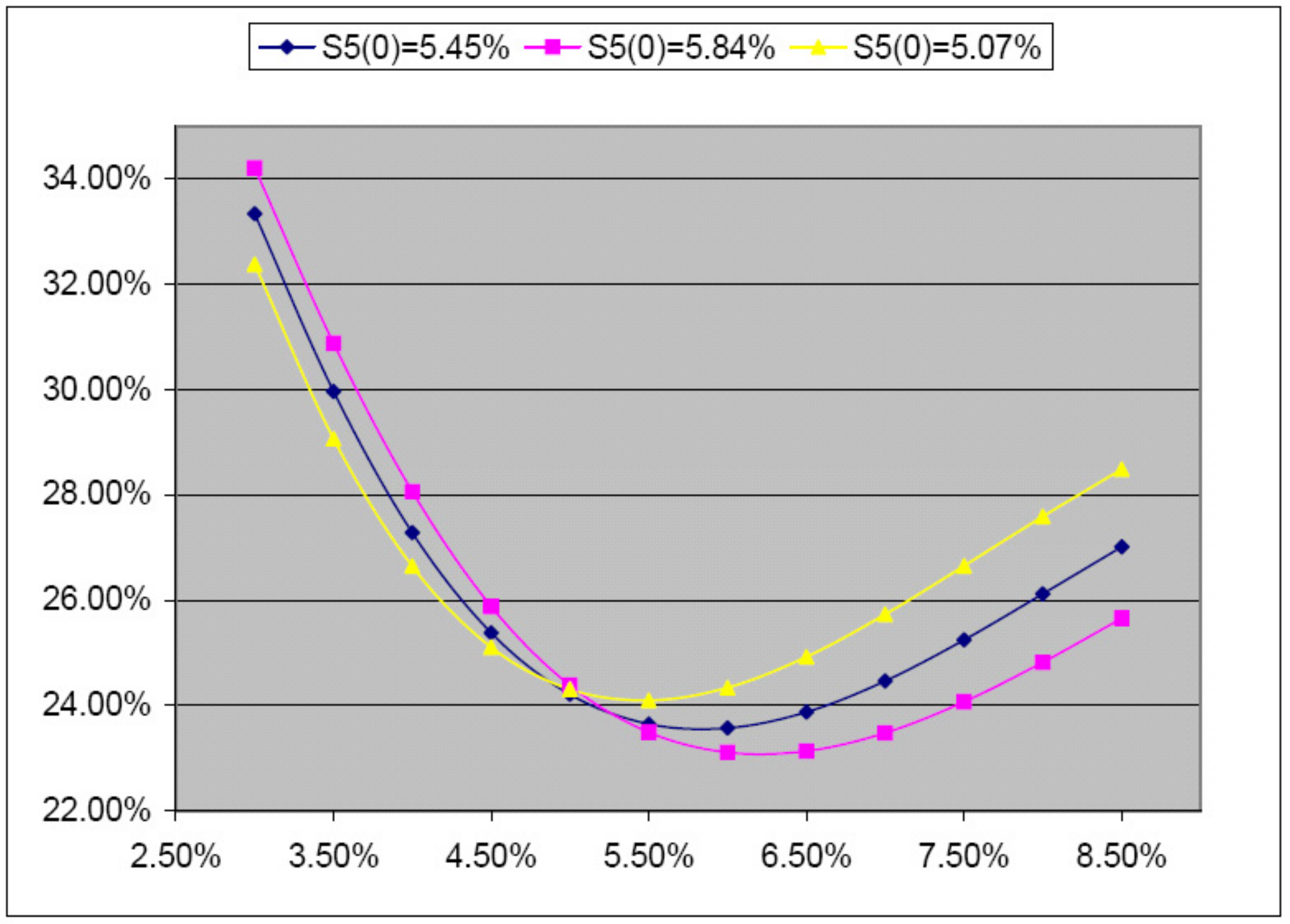}
\centering \caption{Smile dynamics implied by UVDD by bumping only
the discount factor $D_5(0)$.}
\label{fig:smile_dynamics2}
\end{figure}

We repeated these two experiments for case 8 in Section
\ref{sec:pricing_test}, \emph{i.e.}, the UVDD model. The
corresponding results are shown in Figure \ref{fig:smile_dynamics1}
and \ref{fig:smile_dynamics2}, respectively. In both figures, we see
the ATM implied volatility is sloping down and the smile moves in
the same direction as the underlying's movement. This is
qualitatively consistent with the market observation according to
Hagan \cite{Hagan 2002}.

\chapter{Calibration of UVDD Model} \label{chapter:calibration}
In this section, we will first discuss the different choices that
can be made in the calibration procedure, namely: minimization of
the error in terms of option prices or in terms of implied
volatilities. Next, we will discuss the results that have been
obtained using different settings in the calibration procedure.

\section{Calibration Methods}
We may calibrate the UVDD model by minimizing the error in terms of
option prices ({\bf OP}) across $L$ strikes, that is,
\begin{equation}
\begin{cases}
\mbox{minimizing} \;\;
F(\overrightarrow{y})=\frac{1}{L}\Sigma_{i=1}^{L}(ERR_{OP}(K_i;\overrightarrow{y}))^2
\\
\mbox{subject to} \;
\begin{cases}
\lambda^j\in [0,1] \;\mbox{for}\;j=1,\cdots,M\;\mbox{and}\;
\Sigma_{j=1}^{M}\lambda^j=1 \\
\sigma^j>0 \;\mbox{for}\;j=1,\cdots,M
\end{cases},
\end{cases}
\end{equation}
where
\begin{equation}
\begin{cases} ERR_{OP}(K_i;\overrightarrow{y})\triangleq
\frac{OP^{model}(K_i;\overrightarrow{y})-OP^{market}(K_i)}
{OP^{market}(K_i)} \\
\overrightarrow{y}\triangleq
[m,\lambda^1,\sigma^1,\lambda^2,\sigma^2,\cdots,\lambda^M,\sigma^M]
\end{cases}.
\end{equation}
\\
Alternatively, we may instead minimize the error in terms of implied
volatilities ({\bf IV}), that is,
\begin{equation}
\begin{cases}
\mbox{minimizing} \;\;
F(\overrightarrow{y})=\frac{1}{L}\Sigma_{i=1}^{L}(ERR_{IV}(K_i;\overrightarrow{y}))^2
\\
\mbox{subject to} \;
\begin{cases}
\lambda^j\in [0,1] \;\mbox{for}\;j=1,\cdots,M\;\mbox{and}\;
\Sigma_{j=1}^{M}\lambda^j=1 \\
\sigma^j>0 \;\mbox{for}\;j=1,\cdots,M
\end{cases}
\end{cases},
\end{equation}
where
\begin{equation}
ERR_{IV}(K_i;\overrightarrow{y})\triangleq
\frac{IV^{model}(K_i;\overrightarrow{y})-IV^{market}(K_i)}
{IV^{market}(K_i)}.
\end{equation}
\\
However, in practice, solving the first optimization problem does
not mean that we have an optimal solution for the other, and vice
versa. We will explain this below. Due to the relationship,
\begin{equation}
\frac{\Delta OP(K_i;\overrightarrow{y})} {\Delta
IV(K_i;\overrightarrow{y})} \approx Vega(K_i),
\end{equation}
we can derive the following relation after some algebraic
manipulations,
\begin{equation}
ERR_{IV}(K_i;\overrightarrow{y}) \approx
\frac{OP^{market}(K_i)\times IV^{market}(K_i)}{Vega(K_i)}\times
ERR_{OP}(K_i;\overrightarrow{y}).
\end{equation}
\\
First note that $IV^{market}(K_i)$ stays in a relatively narrow
range across strikes, while $OP^{market}(K_i)$ and $Vega(K_i)$ may
vary widely subject to the strike $K_i$. If $OP^{market}(K_i)$ is
very big or $Vega(K_i)$ is very small,
$ERR_{IV}(K_i;\overrightarrow{y})$ can still be large even for very
small values of $ERR_{OP}(K_i;\overrightarrow{y})$. It will be the
other way around if $OP^{market}(K_i)$ is very small or $Vega(K_i)$
is very big. This means that in practice, it is very difficult to
satisfy both criteria simultaneously.
\\ \\
We choose to minimize the error in terms of option prices instead of
implied volatilities because of the following reasons:
\begin{itemize}
\item We want to get a consistent terminal density of the underlying
by the calibration procedure. The implied density is directly
sensitive to the accuracy in terms of option prices (see Equation
\ref{eq:second_dev}). On the other hand, by Equation
\ref{eq:second_dev}, we may have
\begin{equation}
\phi(K)=\frac{\partial \widetilde
{DSN_n(t;\sigma_{imp}(K))}}{\partial \sigma_{imp}(K)} \frac{\partial
\sigma_{imp}(K)}{\partial K}
\end{equation}
Please note that $\frac{\partial \sigma_{imp}(K)}{\partial K}$ is
directly sensitive to the accuracy in terms of implied volatilities.
This means that a small minimization error in terms of implied
volatilities may lead to a large error in the value of $\phi(K)$
because of a high vega level.

\item Computing $OP^{market}(K_i)$ is much faster compared to computing
$IV^{market}(K_i)$. This is because for the latter we have to
include an extra step in the minimization procedure, in which the implied volatility is obtained from the option price. \\
\end{itemize}

In the calibration we use only two components for the UVDD model,
\emph{i.e.}, $M=2$, and thus we can alternatively use the parametric
scheme described in Equation \ref{eq:2components}. We do this
because of the following reasons:
\begin{itemize}
\item If $M=2$, every parameter plays a clear role: $\sigma$
controls the level of the smile/skew; $m$ controls the implied
volatilities' skewness; $\omega$ and $\lambda$ are responsible for
the convexity, \emph{i.e.}, the shape of the smile. But if $M>2$,
the interpretation for each parameter is not that clear any more;
\item As we will see in the next section, a mixture of two
lognormal distributions is rich enough for fitting the market
prices.
\end{itemize}

Furthermore, in the calibration, we fix $\lambda$ and thus have only
three free parameters ($\sigma^1$, $\sigma_2=\omega\sigma^1$ and
$m$). We may do this because $\omega$ and $\lambda$ control similar
features of the implied volatilities.
\\ \\
Now the calibration problem reduces to
\begin{equation}
\begin{cases}
\mbox{minimizing} \;\;
F(\overrightarrow{y})=\frac{1}{L}\Sigma_{i=1}^{L}(ERR_{OP}(K_i;\overrightarrow{y}))^2
\\
\mbox{subject to} \;\;\; \sigma^1>0, \sigma^2>0 \;\mbox{and}\; m\geq
0
\end{cases},
\end{equation}
where
\begin{equation}
\overrightarrow{y}\triangleq [\sigma^1,\sigma^2,m].
\end{equation}
We set $m\geq 0$ because the case of $m<0$ would generate an
unrealistic shape of implied volatilities\footnote{For details,
please refer to Section \ref{sec:UVDD}}.
\\ \\
We are using the NL2SOL minimizer in the calibration\footnote{For
details of the NL2SOL algorithm, please refer to Dennis-Gay-Walsh
\cite{DGW 1981}.}. NL2SOL is an unconstrained minimizer. Thus we
need to transform our bounded model parameters to unbounded ones:

\begin{eqnarray}
x&=&\log(\sigma^1) \\
y&=&\log(\sigma^2) \\
z&=&\log{(\frac{h_m}{m}-1)} \label{eq:tranform_m}.
\end{eqnarray}
\\
Equation \ref{eq:tranform_m} is just for restricting $m$ within
$(0,h_m)$. This will be used in Section
\ref{section:calibration_results} (case 6).

\section{Calibration Results} \label{section:calibration_results}

We run our tests based on the market data corresponding to Data Set
II in Appendix \ref{sec:DataSet2}, which is for the EURO market, and
used the setting of Trade II in Appendix \ref{sec:TestTrades}. We
calibrate the following models to all co-terminal swaptions, $ESN_n$
for n=1,..,20:

\begin{itemize}
\item case 1: a Black model, that is, we take the ATM quote as the flat volatility.
This case serves as a benchmark for the relative error when smile is
not taken into account;
\item case 2: a lognormal model, that is, a reduced Displaced Diffusion model with
$m_n=0$. In this case, we are not necessary fitting the ATM
volatility;
\item case 3: a Displaced Diffusion model with $m_n\geq 0$;
\item case 4: a UVDD model for $m_n\geq 0$ and $\lambda_n=0.75$;
\item case 5: the same model as in case 4, but the calibration is done in terms of
implied volatilities instead of options prices;
\item case 6: a UVDD model for $0\leq m_n\leq 0.10$ and
$\lambda_n=0.75$.
\end{itemize}

\begin{figure}[hp!]
\centering
\includegraphics[width=145mm,height=80mm]{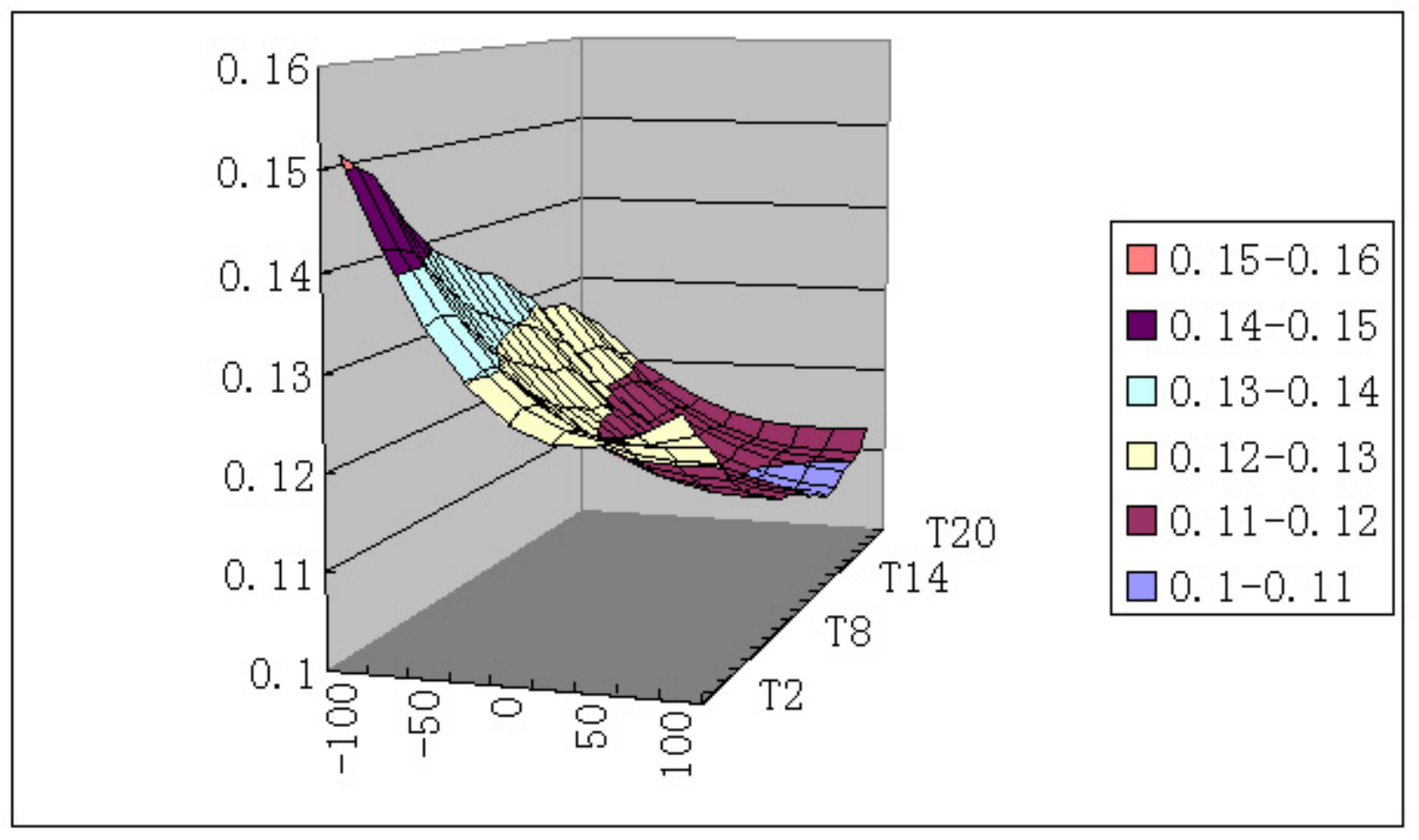}
\centering \caption{Input volatility surface.}
\label{fig:calibration_input}
\end{figure}

\begin{figure}[hp!]
\centering
\includegraphics[width=120mm,height=80mm]{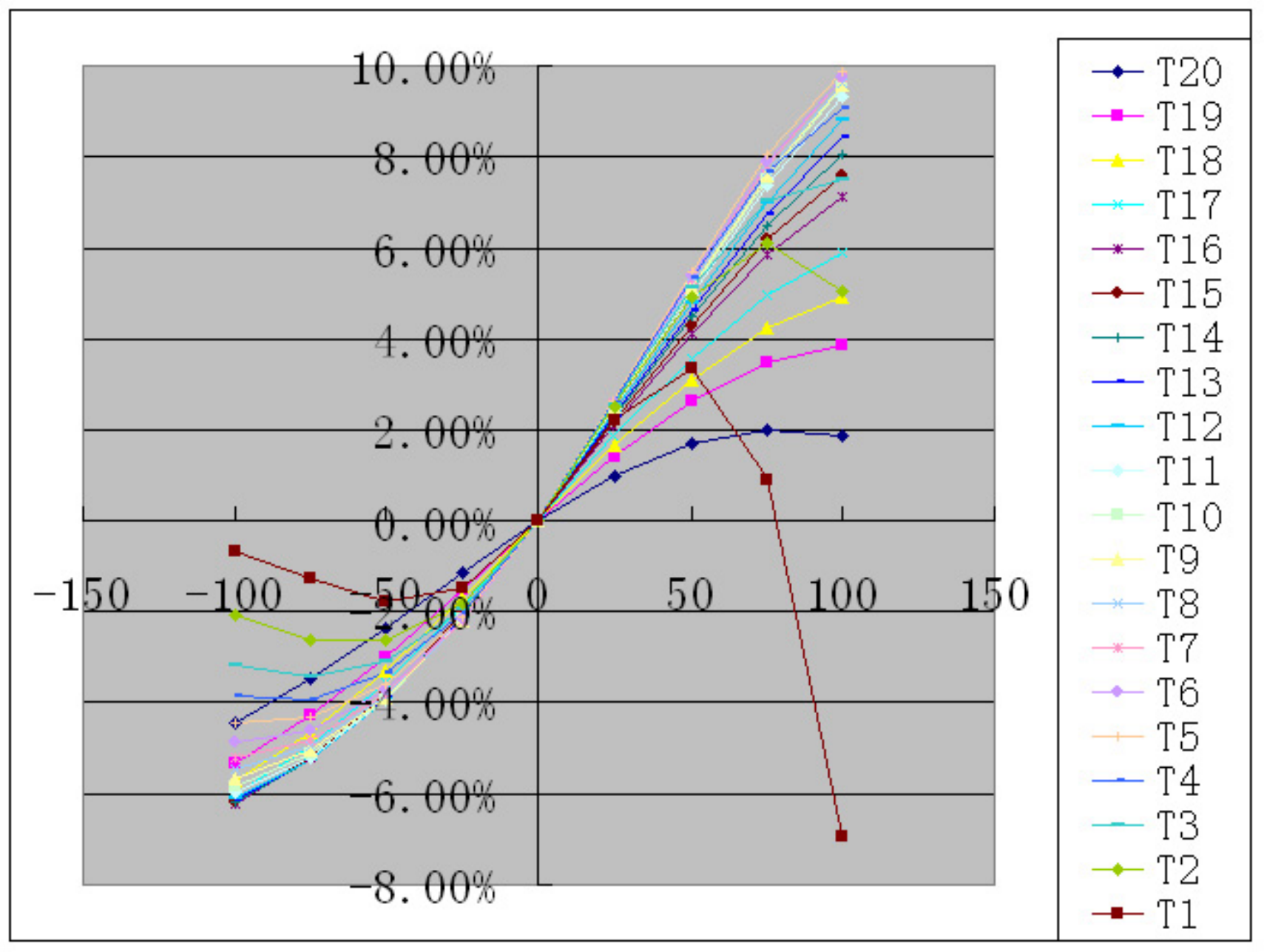}
\centering \caption{Relative error across strike obtained for the
Black model with ATM volality in terms of option prices; the average
absolute error is 3.93\%.} \label{fig:calibration_BSATM}
\end{figure}

\begin{figure}[hp!]
\centering
\includegraphics[width=120mm,height=80mm]{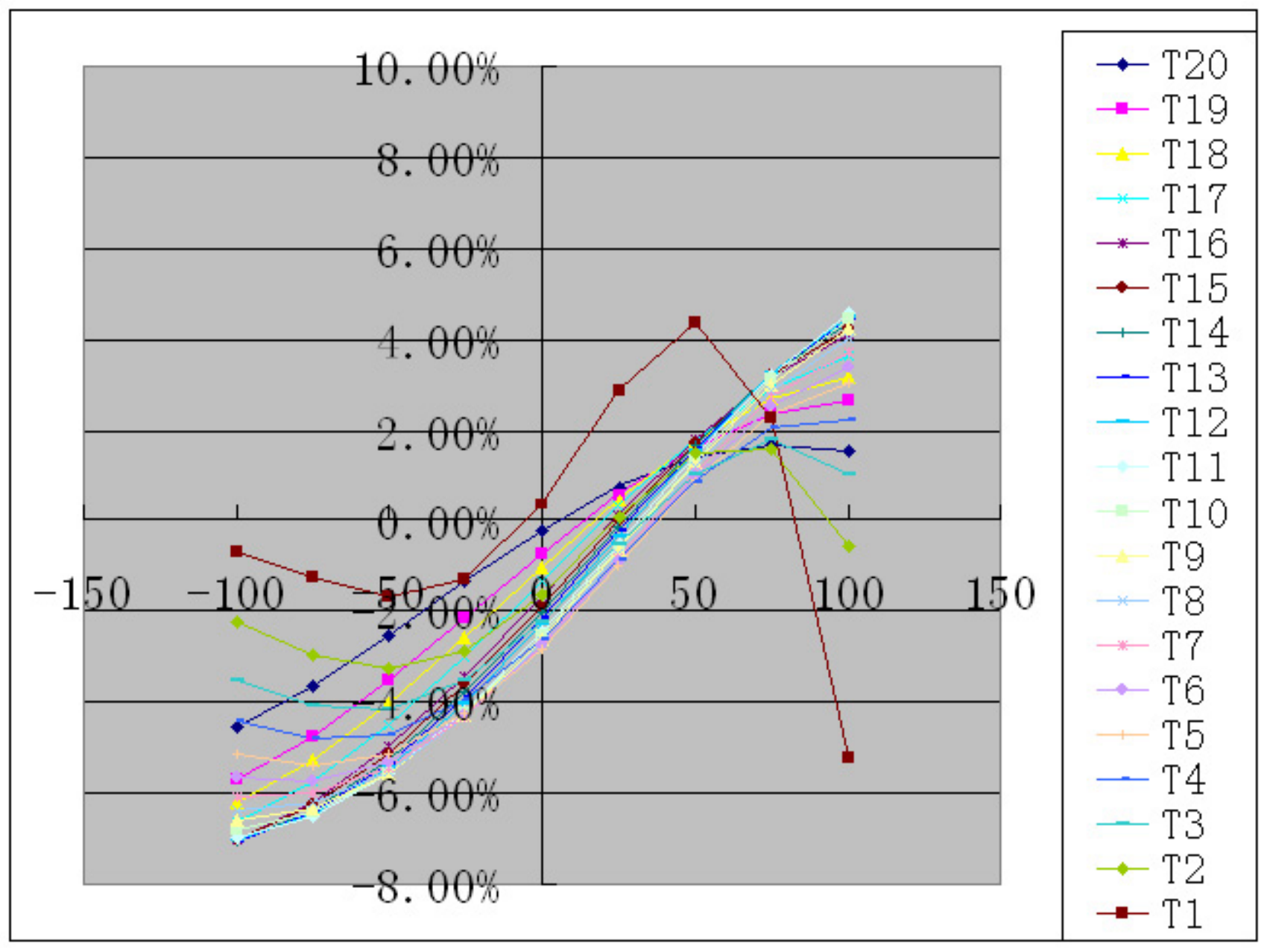}
\centering \caption{Relative error across strike obtained for the
calibration of the lognormal model in terms of option prices; the
average absolute error is 3.26\%.} \label{fig:calibration_BS}
\end{figure}

\begin{figure}[hp!]
\centering
\includegraphics[width=120mm,height=80mm]{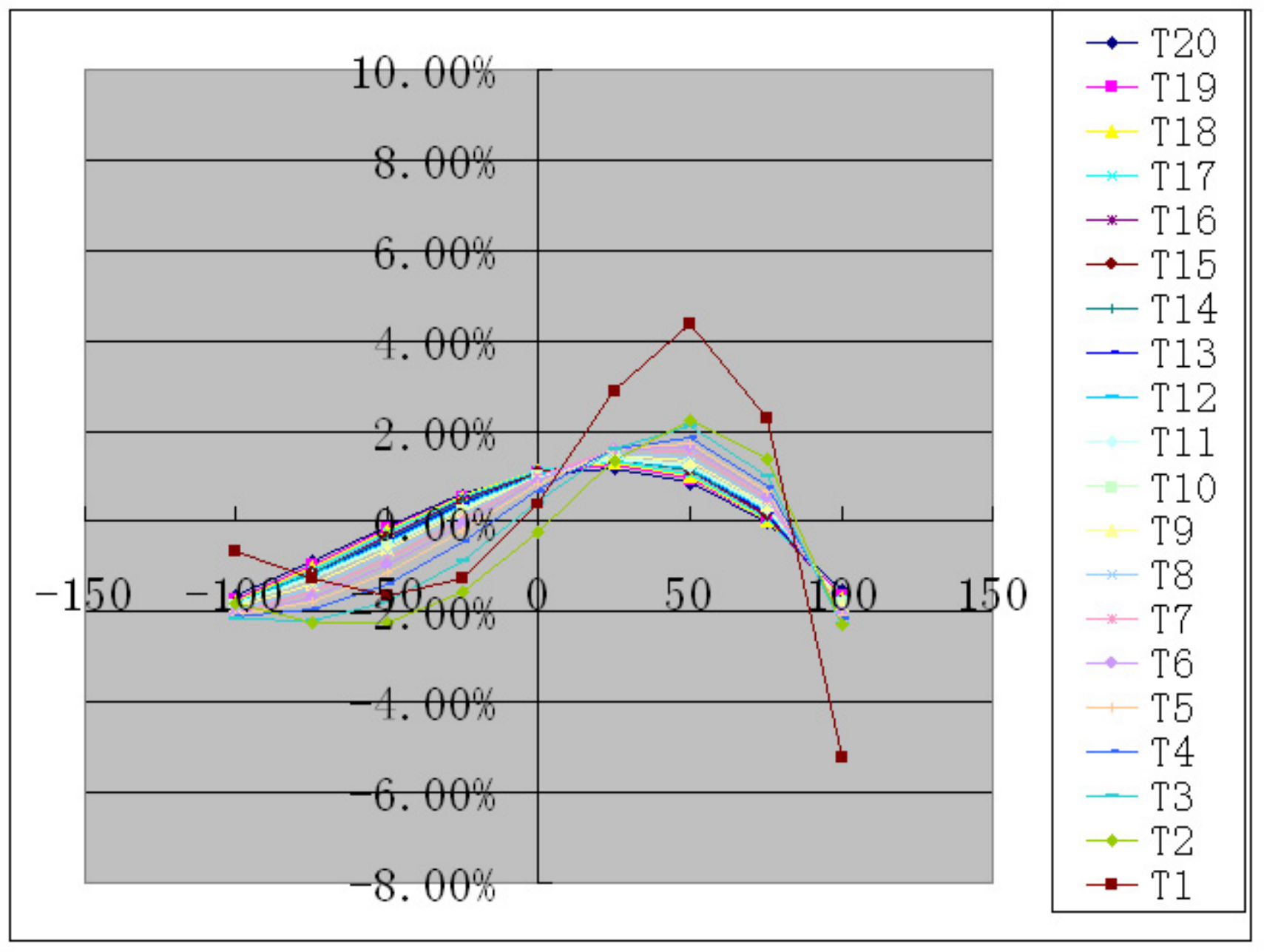}
\centering \caption{Relative error across strike obtained for the
calibration of the Displaced Diffusion model in terms of option
prices; the average absolute error is 1.22\%.}
\label{fig:calibration_DD}
\end{figure}

\begin{figure}[hp!]
\centering
\includegraphics[width=120mm,height=80mm]{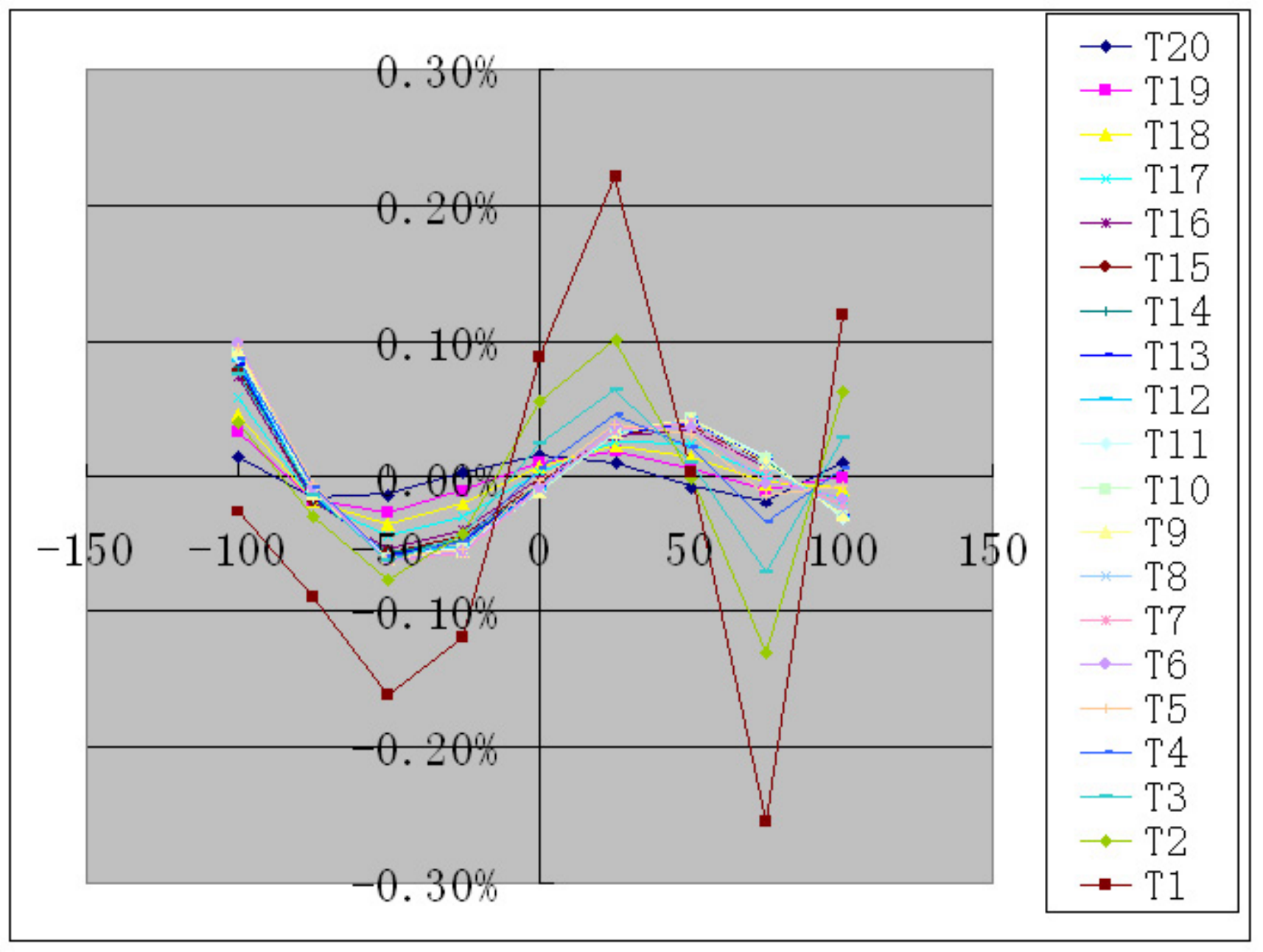}
\centering \caption{Relative error across strike obtained for the
calibration of the UVDD model in terms of option prices; the
absolute absolute error is 0.04\%.} \label{fig:calibration_UVDD}
\end{figure}

\begin{figure}[hp!]
\centering
\includegraphics[width=120mm,height=80mm]{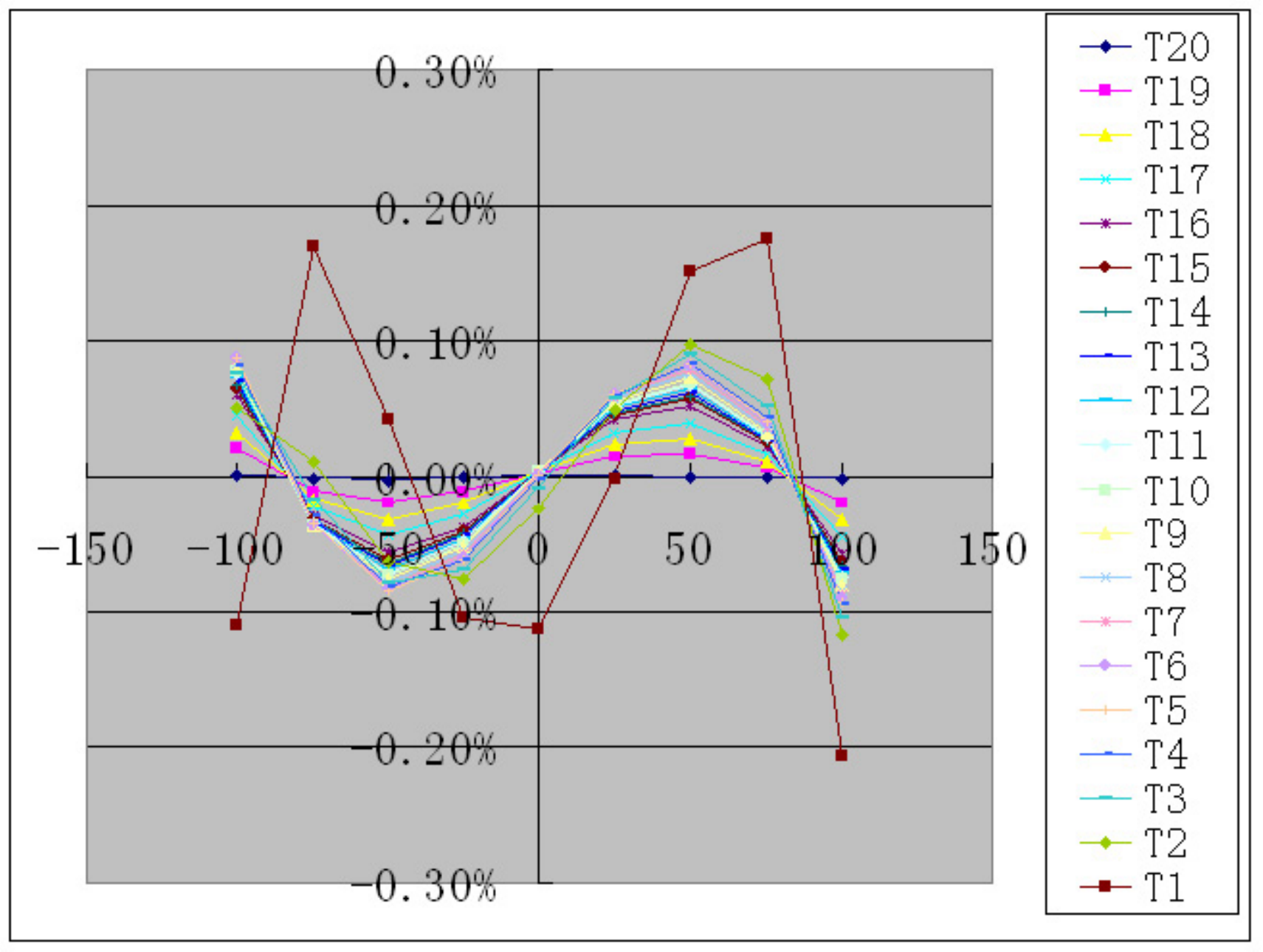}
\centering \caption{Relative error across strike obtained for the
calibration of the UVDD model in terms of implied volatilities; the
absolute absolute error is 0.05\%.} \label{fig:calibration_UVDD_IV}
\end{figure}

\begin{figure}[hp!]
\centering
\includegraphics[width=120mm,height=80mm]{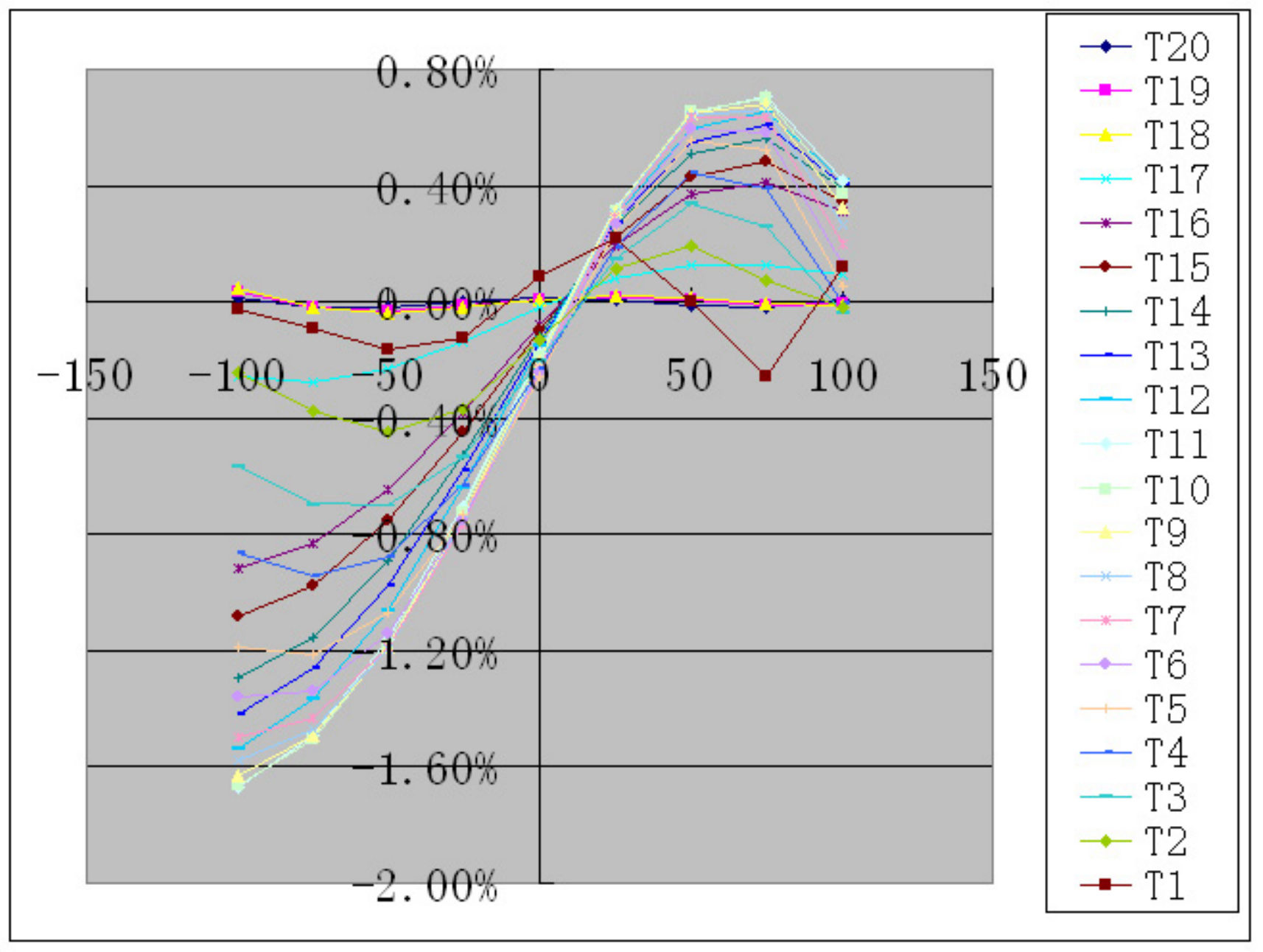}
\centering \caption{Relative error across strike obtained for the
calibration of the UVDD model with $m_n$ within $(0,0.10)$ in terms
of option prices; the average absolute error is 0.49\%.}
\label{fig:calibration_UVDD_mp1}
\end{figure}

\begin{figure}[h!]
\centering
\includegraphics[width=120mm,height=80mm]{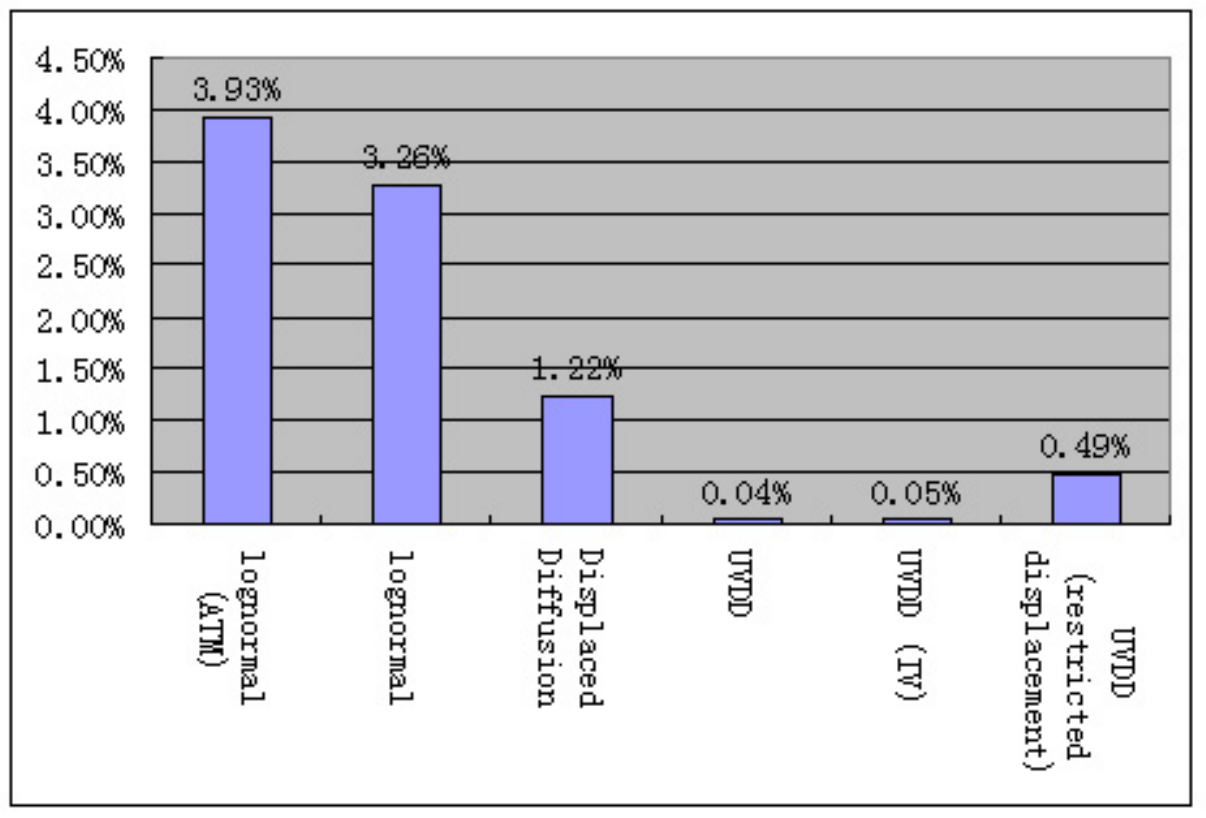}
\centering \caption{Comparison of average absolute error for all the
cases.} \label{fig:calibration_compare}
\end{figure}

For each option maturity $T_i$ $(i=1,2,\cdots,20)$, we calibrate the
model to market prices\footnote{Except for case 5, in which the
calibration is to implied volatilities.} for strikes where the
offset relative to the ATM point varies from -100bp to 100bp, in
total 9 quotes\footnote{More precisely, there are quotes with offset
of -100bp, -75bp, -50bp, -25bp, 0, 25bp, 50bp, 75bp and 100bp. See
Data Set II in Appendix \ref{sec:DataSet2}.}. These 20 input
volatility skews/smiles are shown in Figure
\ref{fig:calibration_input}. The relative error for the calibration
of the above models in terms of option prices\footnote{Except for
case 5, in which the relative error is in terms of implied
volatilities.} is shown from Figure \ref{fig:calibration_BSATM} to
\ref{fig:calibration_UVDD_mp1}, respectively. The average absolute
error is also listed in the legend of each figure.
\\ \\
From Figure \ref{fig:calibration_BSATM} to \ref{fig:calibration_DD},
we see that the Displaced Diffusion model improves the fitting to
market prices significantly than the Black model. From Figure
\ref{fig:calibration_DD} and \ref{fig:calibration_UVDD}, we see that
the UVDD model further improves the fitting to market prices
significantly than the Displaced Diffusion model. In this test, we
don't see from Figure \ref{fig:calibration_UVDD} and
\ref{fig:calibration_UVDD_IV} any difference between the
calibrations in terms of option prices and implied volatilities. A
comparison for the relative error in all the cases is shown in
Figure \ref{fig:calibration_compare}.
\\ \\
Sometimes, if we want to get a very good fit to the market, the
calibrated $m_n$ parameter can be extremely large\footnote{This
doesn't happen in the test here.}, for instance, more than 100, and
in the same time the calibrated $\sigma_n^1$ and $\sigma_n^2$ can be
extremely small, for instance, less than 0.01bp. $m_n$ being 100
means that a negative swap rate very close to -10000\% is allowed in
the model, which is obviously unrealistic. Therefore, we prefer a
local solution which leads to realistic parameters over a global
solution which may be unrealistic. This is why we have restricted
the displacement parameter in case 6. We see in Figure
\ref{fig:calibration_UVDD_mp1} that restricting $m_n$ within
$(0,0.10)$ still gives us a reasonably good fit.

\begin{figure}[hp!]
\centering
\includegraphics[width=\textwidth,height=10mm]{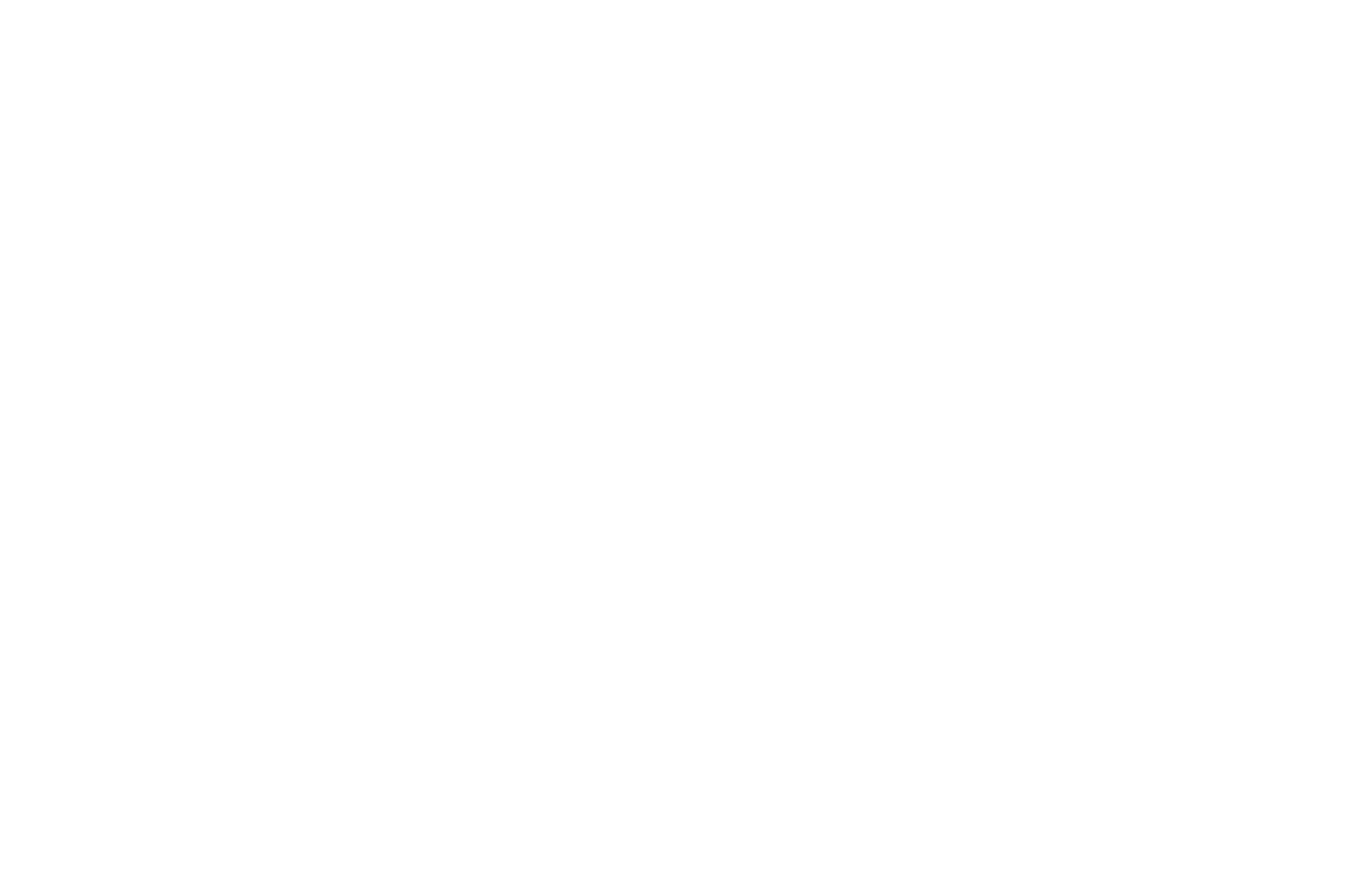}
\end{figure}

\newpage

\section{Terminal Density Implied by the UVDD Model}

\begin{figure}[hp!]
\centering
\includegraphics[width=100mm,height=63mm]{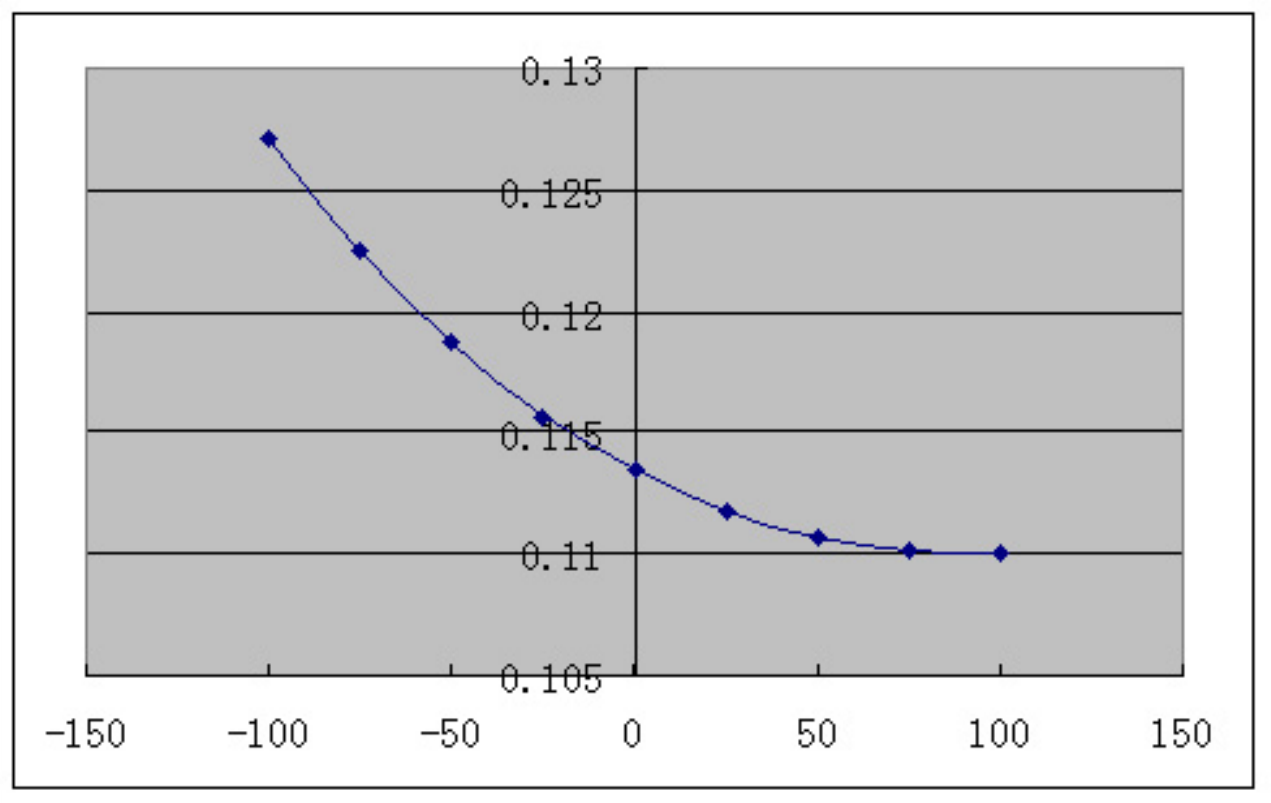}
\centering \caption{Implied volatility smile for $T_{18}$.}
\label{fig:pdf3}
\end{figure}

\begin{figure}[hp!]
\centering
\includegraphics[width=100mm,height=63mm]{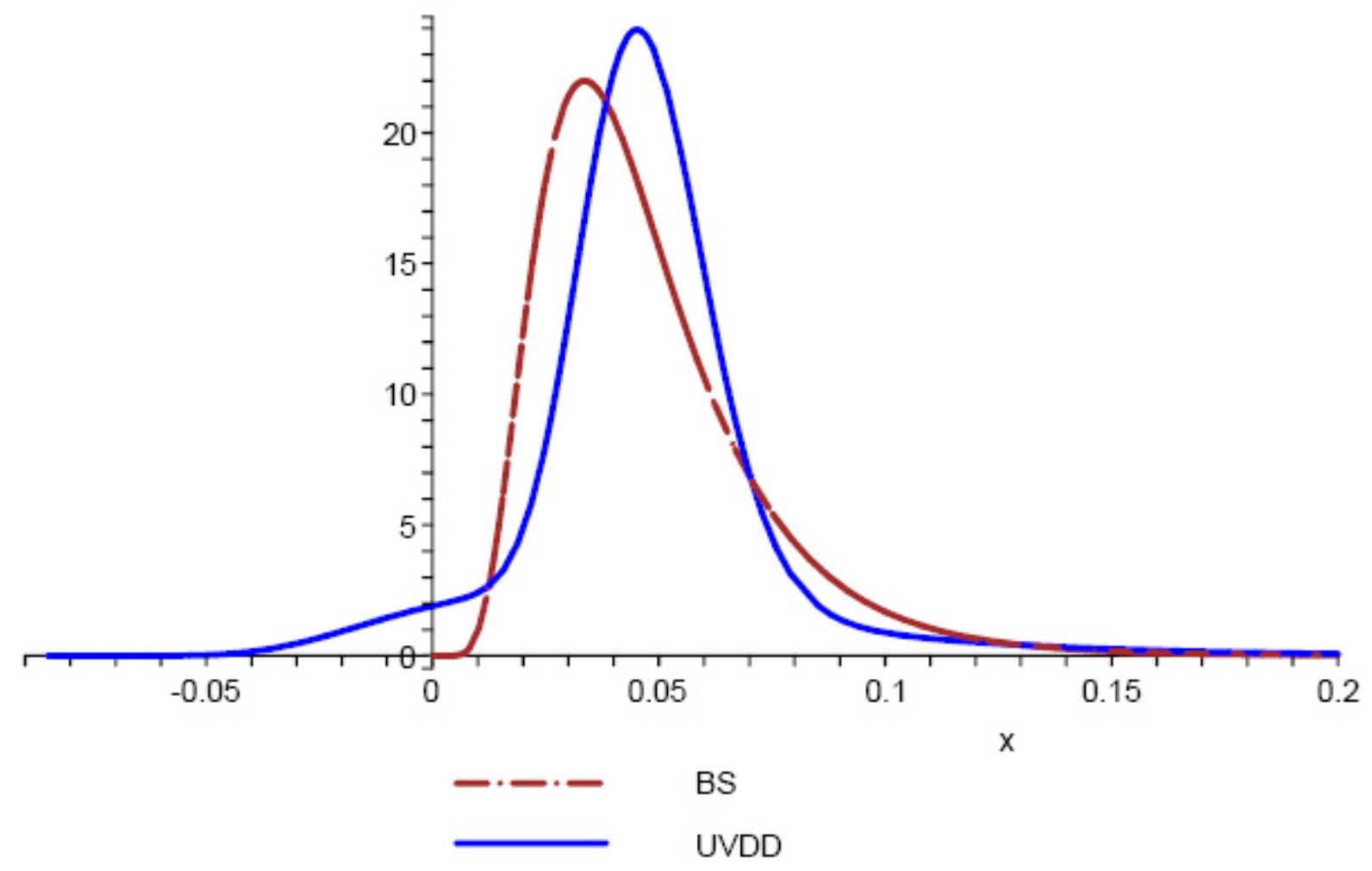}
\centering \caption{Probability densities of the swap rate
$S_{18}(T_{18})$.} \label{fig:pdf1}
\end{figure}

\begin{figure}[hp!]
\centering
\includegraphics[width=100mm,height=63mm]{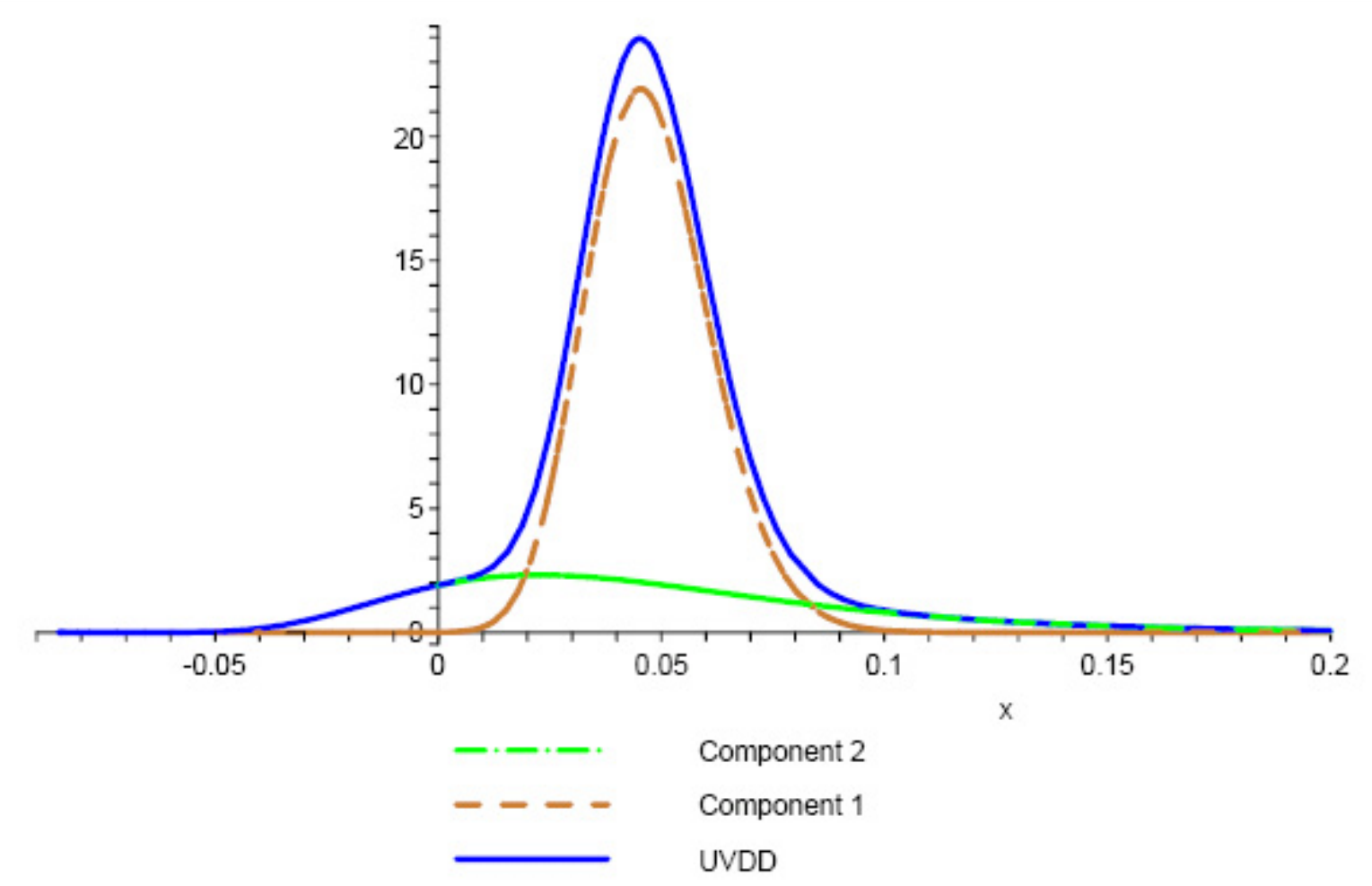}
\centering \caption{Decomposition of the probability density
function for $S_{18}(T_{18})$.} \label{fig:pdf2}
\end{figure}

Another interesting quantity is the terminal density implied by the
model for the underlying swap rate. We study this for the underlying
rate with maturity at $T_{18}$ corresponding to case 4 of Section
\ref{section:calibration_results}. The input implied volatility
quotes are shown in Figure \ref{fig:pdf3}. The ATM strike level is
$4.75\%$. The probability density function is plotted in Figure
\ref{fig:pdf1}, and the calibrated model parameters are listed
below.
\begin{equation}
\nonumber
\begin{cases}
\sigma_{18}^1=2.45\% \\
\sigma_{18}^2=8.79\% \\
m_{18}=8.52\%
\end{cases}.
\end{equation}
\\
In Figure \ref{fig:pdf1}, the implied distribution by taking only
the ATM volatility, \emph{i.e.}, the counterpart lognormal
distribution, is plotted as well. We see that the UVDD model allows
for negative swap rates (down to almost $-8.79\%$) which is
unrealistic. However the negative swap rates have small densities
($Prob(S_{18}(T_{18})\leq 0)=3.95\%$). Such a small fraction of
negative swap rates leads to a left-side fat tail, and thus helps
the model to generate the skew effect. Figure \ref{fig:pdf2} shows
the two lognormal components of the implied distribution, each of
them scaled by their weight factor.

\chapter{Hedging Simulations}\label{chapter:Hedging}
\section{Overview of the Hedging Simulations}
We have performed our hedging simulations on a 10 year Bermudan
swaption with the right to exercise at floating reset dates $T_n$,
for $n=1,2,\cdots,10$. The trade is running from May 28th 2004 to
July 29th 2005 (in total 14 months). The trade specification is
shown in Table \ref{table:hedge_trade} of Appendix
\ref{sec:TestTrades}. The market data for the hedge tests, part of
which was created synthetically based on the available
data\footnote{All the available market data is for the EURO
market.}, is presented in Section \ref{sec:hedgedata}. We have
performed both delta and delta+vega hedgings to the smile and
non-smile Bermudans\footnote{The smile Bermudan is the abbreviation
for a Bermudan swaption which is valued by taking volatility smiles
into account. The non-smile Bermudan is the abbreviation for a
Bermudan swaption which is valued without taking volatility smiles
into account.}. The details of the hedging strategies will be
explained in Section \ref{sec:hedge_setup}. The calculation of delta
and vega ratios will be elaborated in Section \ref{sec:sensitivity}.
Below we summarize the most important conclusions from the conducted
hedging simulations:

\begin{itemize}
\item The
smile model\footnote{The "smile model" is an abbreviation for the MF
model with UVDD digital mapping.} outperforms the non-smile
model\footnote{The "non-smile model" is an abbreviation for the MF
model with Black-Scholes digital mapping.} in both delta hedging and
delta+vega hedging simulations. In both the smile and non-smile
cases, a delta hedging reduces the profit$\&$loss effect of the
unhedged Bermudan trade significantly, and a delta+vega hedging
further improves the delta hedging performance significantly.
Besides, a delta+vega hedging reduces significantly the oscillation
of the hedged NPV compared to the corresponding delta hedging, but
doesn't affect the drift level of the hedged NPV. The above
conclusions will be drawn gradually from Section
\ref{sec:hedgeresult1} to \ref{sec:hedgeresult3};
\item The change from rolling the vega positions from daily to monthly
has little impact on the delta+vega hedged NPV. The relevant details
can be found in Section \ref{sec:hedgeresult1};
\item Increasing the mean-reversion parameter reduces the drift of the
hedged NPV, but doesn't affect its oscillation. The relevant details
are in Section \ref{sec:hedgeresult4}.
\end{itemize}

\newpage
\section{Market and Synthetic Data} \label{sec:hedgedata}

\subsection{Available Market Data}
\label{sec:marketdata}

\subsubsection{Yield Curve Data}

\begin{figure}[h!]
\centering
\includegraphics[width=100mm,height=55mm]{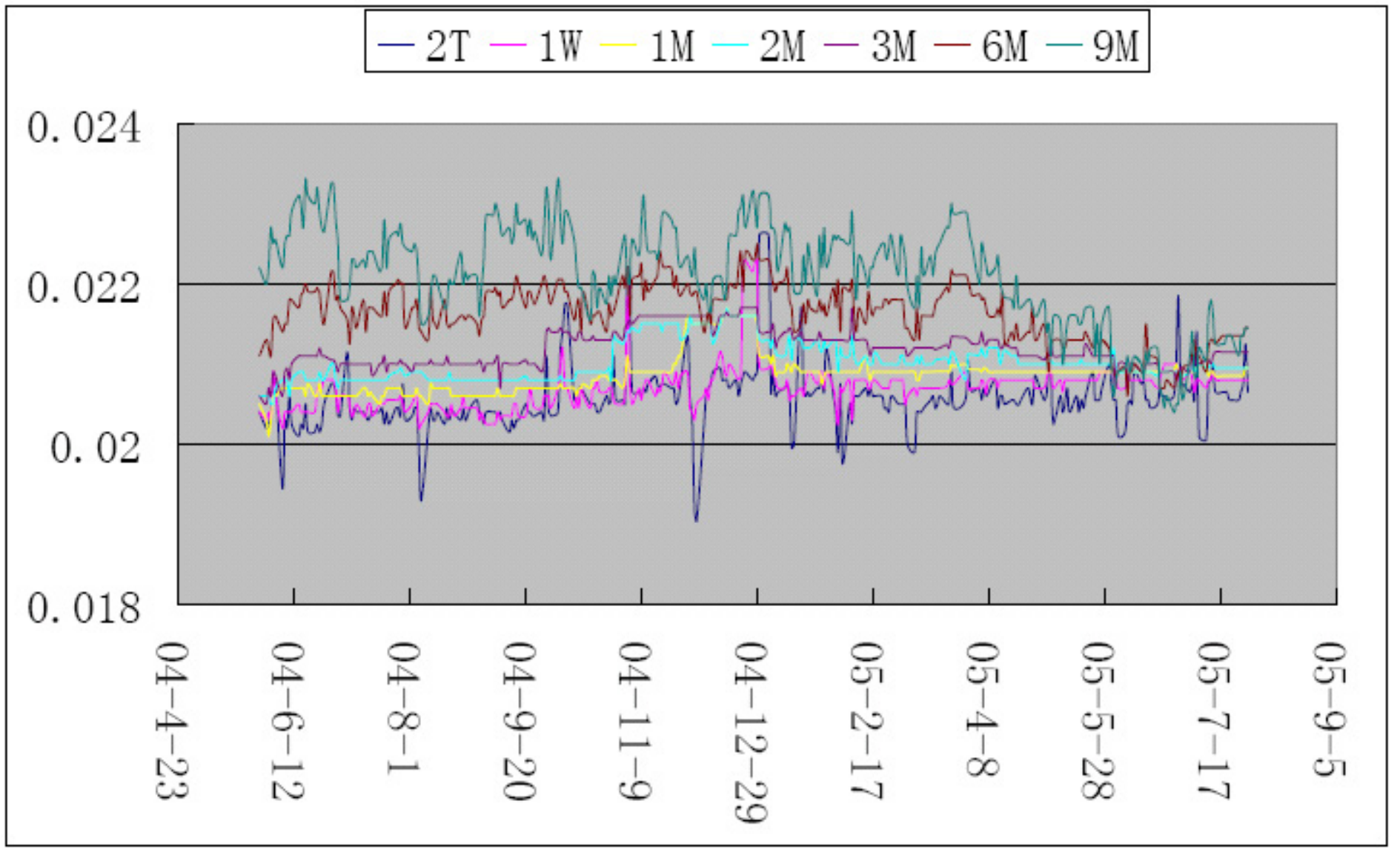}
\centering \caption{Historical deposit rates.}
\label{fig:Deposits}
\end{figure}

\begin{figure}[h!]
\centering
\includegraphics[width=100mm,height=55mm]{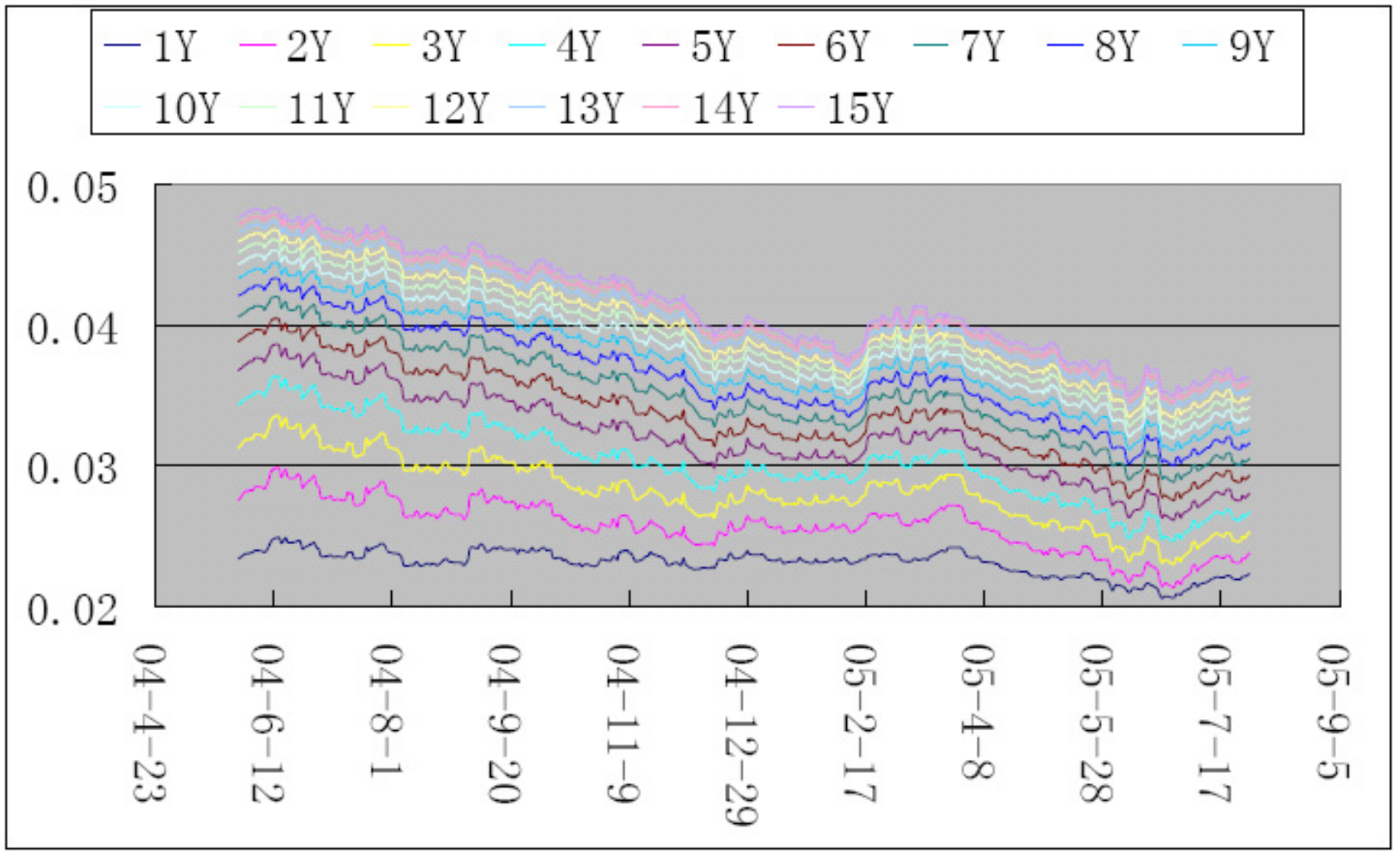}
\centering \caption{Historical spot-starting swap rates.}
\label{fig:SpotSwaps}
\end{figure}

\begin{figure}[h!]
\centering
\includegraphics[width=100mm,height=55mm]{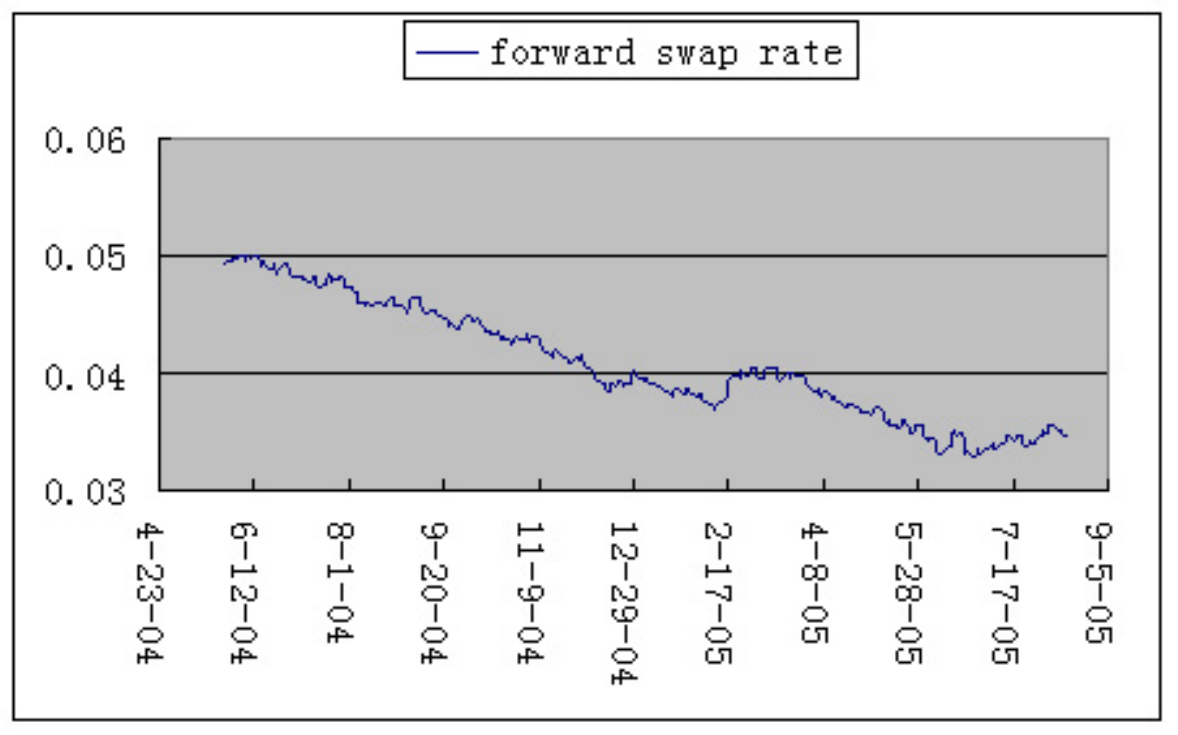}
\centering \caption{Forward swap rates for $T_1$.}
\label{fig:forward_rate_T1}
\end{figure}

There are two kinds of yield curve related data. The first one is
the deposit rates of 2 days, 1 week, 1 month, 2 months, 3 months, 6
months and 9 months, all shown in Figure \ref{fig:Deposits}. The
second one is the spot-starting swap rates with tenors from 1 year
to 15 years, shown in Figure \ref{fig:SpotSwaps}. The deposit rates
are used to construct the short-end of the yield curve. The
spot-starting swap rates are used to construct the long-end of the
yield curve. For every day in the 14 month hedge period, the yield
curve is bootstrapped from these deposit rates and spot-starting
swap rates.\footnote{The bootstrapping is done by using ABN AMRO's
Common Analytics Library. We use this library as a black box.} From
the constructed yield curves, we calculate the forward swap rates.
For example, Figure \ref{fig:forward_rate_T1} shows the time series
of the forward swap rate corresponding to the underlying co-terminal
swap which starts at $T_1$.

\newpage
\subsubsection{Swaption Volatility Data}
\begin{figure}[h!]
\centering
\includegraphics[width=100mm,height=55mm]{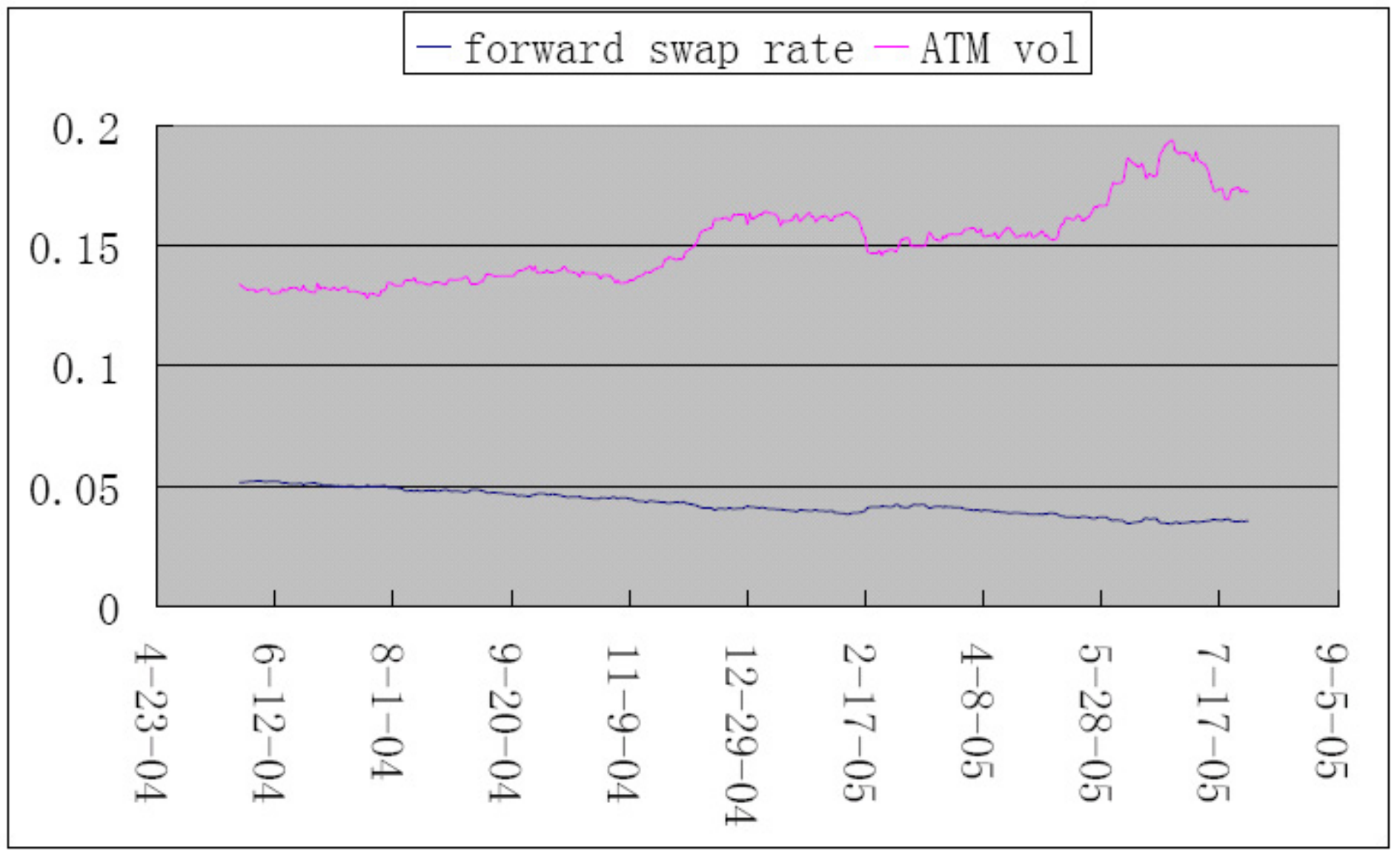}
\centering \caption{ATM volatilities and forward swap rates for $T_2$.}
\label{fig:ATM vol_T2}
\end{figure}

We have access to daily ATM implied volatilities of European
swaptions. Each European swaption has two attributes, the expiry and
tenor. The expiry can typically take the following 11 values: 1
month, 3 months, 6 months, 1 year, 2 years, 3 years, 4 years, 5
years, 7 years, 10 years and 15 years. The tenor takes one of the 10
values from 1 year to 10 years. Thus every day there are in total
110 ATM quotes available for the European swaptions. From these
quotes, we get the ATM volatilities by applying a two-dimensional
linear interpolation. For example, Figure \ref{fig:ATM vol_T2} shows
the time series of the ATM volatilities of the European swaption
corresponding to underlying co-terminal swap which starts at $T_2$.
The time series of the corresponding forward swap rates is plotted
in the figure as well.

\begin{figure}[h!]
\centering
\includegraphics[width=120mm,height=75mm]{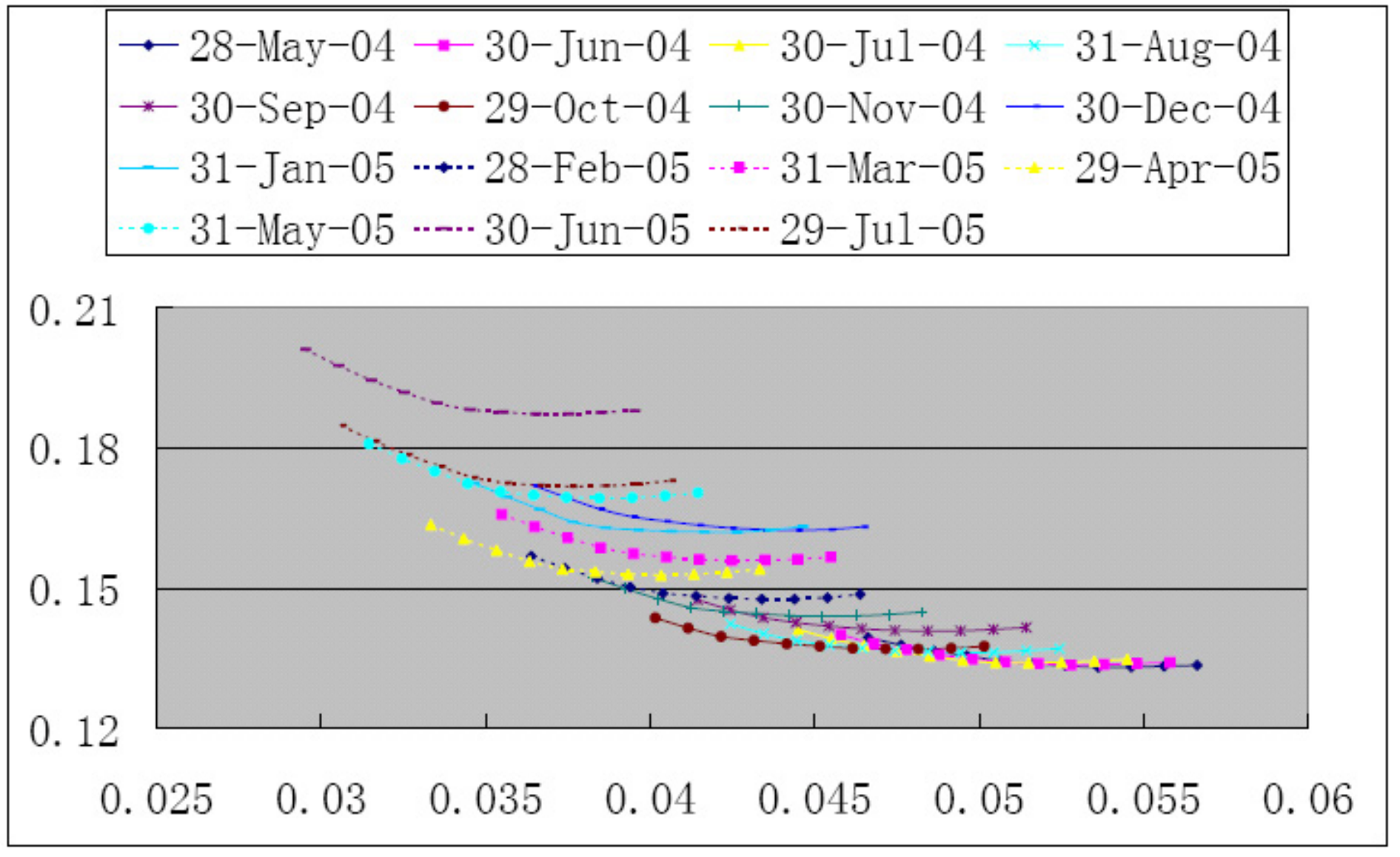}
\centering \caption{End of month smiles for $T_2$.}
\label{fig:EoM_smiles_T2}
\end{figure}

Besides we have only access to end-of-month smile quotes for each
month of the period. Each European swaption has three attributes,
namely, expiry, tenor and strike. Each attribute can take a value in
a certain range. Similar to above, we get the required European
swaptions' smile quotes by applying a three-dimensional linear
interpolation. Figure \ref{fig:EoM_smiles_T2} shows the end-of-month
smiles of the European swaption corresponding to the underlying
co-terminal swap which starts at $T_2$. Each of these smiles are
composed of 11 quotes for strikes with the offset to the ATM point
varying from -50bp to 50bp. More precisely, there are quotes for
strikes with an offset to the ATM level of -50bp, -40bp, -30bp,
-20bp, -10bp, 0, 10bp, 20bp, 30bp, 40bp and 50bp.

\subsection{Creating Synthetic Smiles}
\label{sec:syntheticdata}

For a hedging simulation based on a one-day time step, we need daily
smile quotes even just for book-keeping of the value of the
Bermudan. Thus we need to create smile data for non-end-of-month
dates. This is achieved as follows:
\begin{itemize}
\item For each end-of-month date, we
calibrate the swaptions' prices to the UVDD model to get the model
parameters $\sigma_n^1$, $\sigma_n^2$, $\omega_n$ and $m_n$. In the
calibration, we use exactly the same setting corresponding to case 6
in Section \ref{section:calibration_results}. That is, we have two
log-normal components with $\lambda_n=0.75$, and $m_n$ is restricted
within the range $[0,0.10]$;\footnote{For more details of the
calibration, please refer back to Chapter
\ref{chapter:calibration}.}
\item For each of the other days,
we get the values of $\omega_n$ and $m_n$ by linear interpolation of
the parameters corresponding to the previous and next end-of-month
dates. We do the interpolation in terms of these two parameters,
because the former is an indicator of the smile shape and the latter
of the skew effect.\footnote{For details, please refer back to
Section \ref{sec:UVDD}.} The parameter $\sigma_n^1$ has been
adjusted such that the implied (Black) ATM volatility equals the
market quote.

\end{itemize}

\begin{figure}[h!]
\centering
\includegraphics[width=120mm,height=75mm]{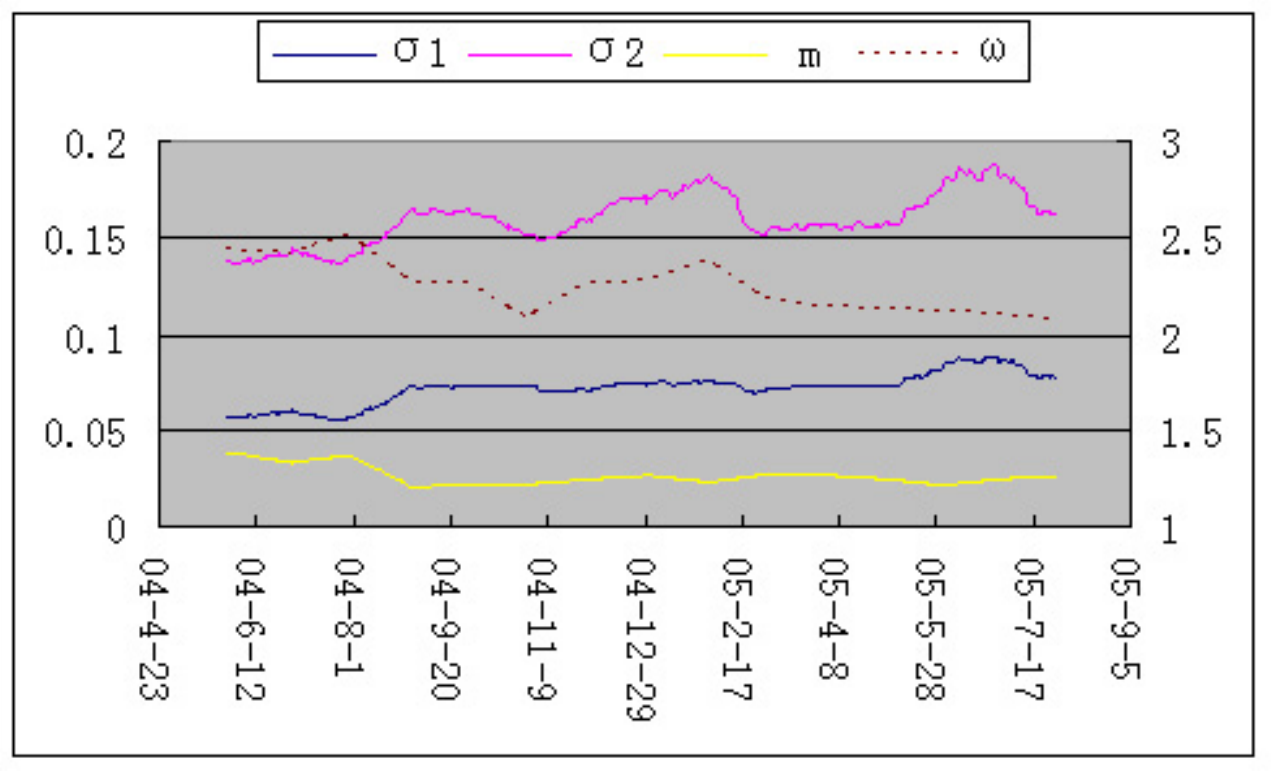}
\centering \caption{UVDD model parameters for $T_2$.}
\label{fig:UVDD_parameters_T2}
\end{figure}

Now we have created the required smile data in terms of the UVDD
model parameters for the complete 14 months period on a daily level.
Figure \ref{fig:UVDD_parameters_T2} shows the time series of the
UVDD model parameters. In Figure \ref{fig:UVDD_parameters_T2}, the
range in which $\omega_n$ varies is shown in the y-axis on the
right.

\section{Hedge Test Setup} \label{sec:hedge_setup}

We set up our hedge test as follows.
\\ \\
{\bf 1.} At the very beginning of the first day, we long a Bermudan
swaption by shorting money from our bank account. We keep the
Bermudan till the last day of the hedge period. We construct a hedge
portfolio containing all the hedging instruments. At the very
beginning of the first day, the hedge portfolio contains nothing.
\\ \\
{\bf 2.} Everyday we value the Bermudan and the hedge portfolio. The
hedged NPV\footnote{NPV denotes net present value.} corresponding to
that day is given by
\begin{equation} \label{eq:hedged NPV}
\mbox{"hedged NPV"} = \mbox{"Bermudan value"} + \mbox{"hedge
portfolio value" + "bank account value"}.
\end{equation}
\\
{\bf 3.} After that, we add the hedge portfolio's value to our bank
account and liquidate the instruments in the hedge portfolio
(constructed in the previous day).
\\ \\
{\bf 4.} Next, we calculate the vega sensitivities of the Bermudan
and take positions of European swaptions to neutralize these vega
sensitivities. We deduct money from our bank account for setting up
these positions ({\bf vega hedging}).
\\ \\
{\bf 5.} Then, we calculate the delta sensitivities of the Bermudan
and the hedge portfolio as a whole. We take positions in
spot-starting swaps and deposits to neutralize these delta
sensitivities. We again deduct money from our bank account for
setting up these positions ({\bf delta hedging}).
\\ \\
{\bf 6.} Finally, at the end of the day, we add the accrued interest
for that day to our bank account.
\\ \\
{\bf 7.} We repeat steps {\bf 2.} to {\bf 6.} on a daily basis until
the last day of the hedge period.
\\ \\
The above procedure is for a delta+vega\footnote{How to calculate
the sensitivities will be elaborated in the next section.} hedging
simulation. For only a delta hedging simulation, we have to skip the
4th step.

\section{Sensitivity Calculation} \label{sec:sensitivity}

In Pelsser \cite{Pelsser 1998-c}, hedging simulations for non-smile
Bermudans were conducted. When a vega hedging was set up, ATM
European swaptions were used to neutralize the vega sensitivity of
the Bermudan. The vega of an ATM European swaption is given by the
following formula,

\begin{equation} \label{eq:BS_vega}
\frac{\partial ESN_n(t;K)}{\partial \bar{\sigma}}|_{K=S_n(t)} =
P_n(t)S_n(t)\phi(d_+) \sqrt{T_n - t},
\end{equation}
where
\begin{equation}
\begin{cases} \nonumber
\phi(x) = \frac{1}{\sqrt{2\pi}}e^{-\frac{1}{2}x^2}
\\
d_+ = \frac{1}{2}\bar{\sigma}\sqrt{T_n - t}
\\ \mbox{$\bar{\sigma}$ is the Black volatility}
\end{cases}.
\end{equation}
\\
Of course, only ATM European swaptions are used to which the
Bermudan shows vega sensitivities. What's important to mention is,
that even for European swaptions, it is assumed that the volatility
is flat across all strikes. This means that when we liquidate the
European positions on the next day, these are not
marked-to-market\footnote{This is because today's ATM option will
very probably become an ITM/OTM option tomorrow.}, but
marked-to-model. This was a fairly good approximation when those
hedging simulations were performed. Because at that time, the smile
effect of the swaption market was much less pronounced than
nowadays.
\\ \\
We conduct vega hedging against a smile Bermudan in terms of the
UVDD volatilities $\sigma_n^1$. For example, for our Bermudan trade,
we can calculate the sensitivity with respect to
$\sigma_n^1,n=1,2,\cdots,10$, \emph{i.e.}, in total 10 vega
sensitivities. We can do this by using the following analytical
formula\footnote{Equation \ref{eq:UVDD_vega} is achieved by
differentiating Equation \ref{eq:UVDD_European} with respect to
$\sigma_n^1$.},
\begin{equation} \label{eq:UVDD_vega}
\frac{\partial ESN_n(t;K)}{\partial \sigma_n^1}|_{K=S_n(t)} =
 P_n(t)(S_n(t)+m_n) [\lambda_n\phi(d_+^1) +
(1-\lambda_n)\phi(d_+^2)\omega_n ]\sqrt{T_n - t},
\end{equation}
where
\begin{equation}
\begin{cases} \nonumber
\phi(x) = \frac{1}{\sqrt{2\pi}}e^{-\frac{1}{2}x^2}
\\
d_+^1 = \frac{1}{2}\sigma_n^1\sqrt{T_n - t}
\\
d_+^2 = \frac{1}{2}\omega_n\sigma_n^1\sqrt{T_n - t}
\end{cases}.
\end{equation}
\\
Then we need to take a certain amount of position for each European
to vega hedge against the smile Bermudan. The next day when we
liquidate the European positions, we mark the positions to
market.\footnote{The mark-to-market in the smile case is a bit
tricky, since most of the days' smile data are created by using the
UVDD model. For details, we refer back to Section
\ref{sec:syntheticdata}.}
\\ \\
The deposits and spot-starting swaps are the inputs for constructing
the yield curve. Thus the changes of the option value with respect
to the change of these instruments' {\emph rates} are defined as its
delta ratios. The way to calculate the delta ratios is the same for
both the smile and non-smile cases.
\\ \\
A relevant (payer) spot-starting swap, with the fixed coupon rate at
the par level, is used to neutralize the sensitivity to the
corresponding spot-starting swap rate. The ratio of the change of
the swap value to the change of the par swap rate is
\begin{eqnarray}
\frac{\partial SV(t;K)}{\partial S}|_{K=S} &=& \frac{\partial
[PVBP(S-K)]}{\partial S}|_{K=S}
\;\;\;\;\;\;\;\;\;\;\;\;\;\;\;\;\;\;\;\;
\mbox{// by Eq. \ref{eq:swap_value}}\nonumber\\
&=& \{PVBP + (S-K)\frac{\partial PVBP}{\partial S}\}|_{K=S} \nonumber\\
&=& PVBP,             \label{eq:swap_delta}
\end{eqnarray}
where $S$ denotes the par swap rate.
\\ \\
A relevant deposit is used to neutralize the sensitivity to the
corresponding deposit rate. The sensitivity of the value of the
deposit to the deposit rate is
\begin{eqnarray}
\frac{\partial ("deposit\;value")}{\partial ("deposit\;rate")}&=&
\frac{\partial (\frac{1} {1 + "deposit\;rate"\times
\delta t}) }{\partial ("deposit\;rate")} \nonumber \\
&=& - \frac{\delta t}{(1+"deposit\;rate"\times \delta t)^2},
\label{eq:deposit_delta}
\end{eqnarray}
where $\delta t$ denotes the accrued time associated with the
deposit.
\\ \\
Except for the sensitivities described in Equation \ref{eq:BS_vega},
\ref{eq:UVDD_vega}, \ref{eq:swap_delta} and \ref{eq:deposit_delta},
all the other vega and delta sensitivities are computed numerically
by the "bump and revalue" method. The "bump and revalue" method is
just a simple finite-difference approach to approximate the first
derivative,
\begin{equation}
V'(x) \approx \frac{V(x+b)-V(x)}{b},
\end{equation}
where $V(x)$ is the value of an instrument which depends on the
underlying factor $x$, and $b$ is the bump size. In our test, the
bump size is always set to 1bp, except for the vega of the non-smile
Bermudan, which is calculated with a bump size of 10bp. This has
been determined by experimenting with different settings. Note that
increasing or decreasing the bump size by a factor 10 has little
impact for the hedge results.

\section{Results of Hedge Tests}

\subsection{Comparison between Hedging against Smile Bermudan and the Original Hedging against Non-smile Bermudan}
\label{sec:hedgeresult1}
\begin{figure}[h!]
\centering
\includegraphics[width=100mm,height=65mm]{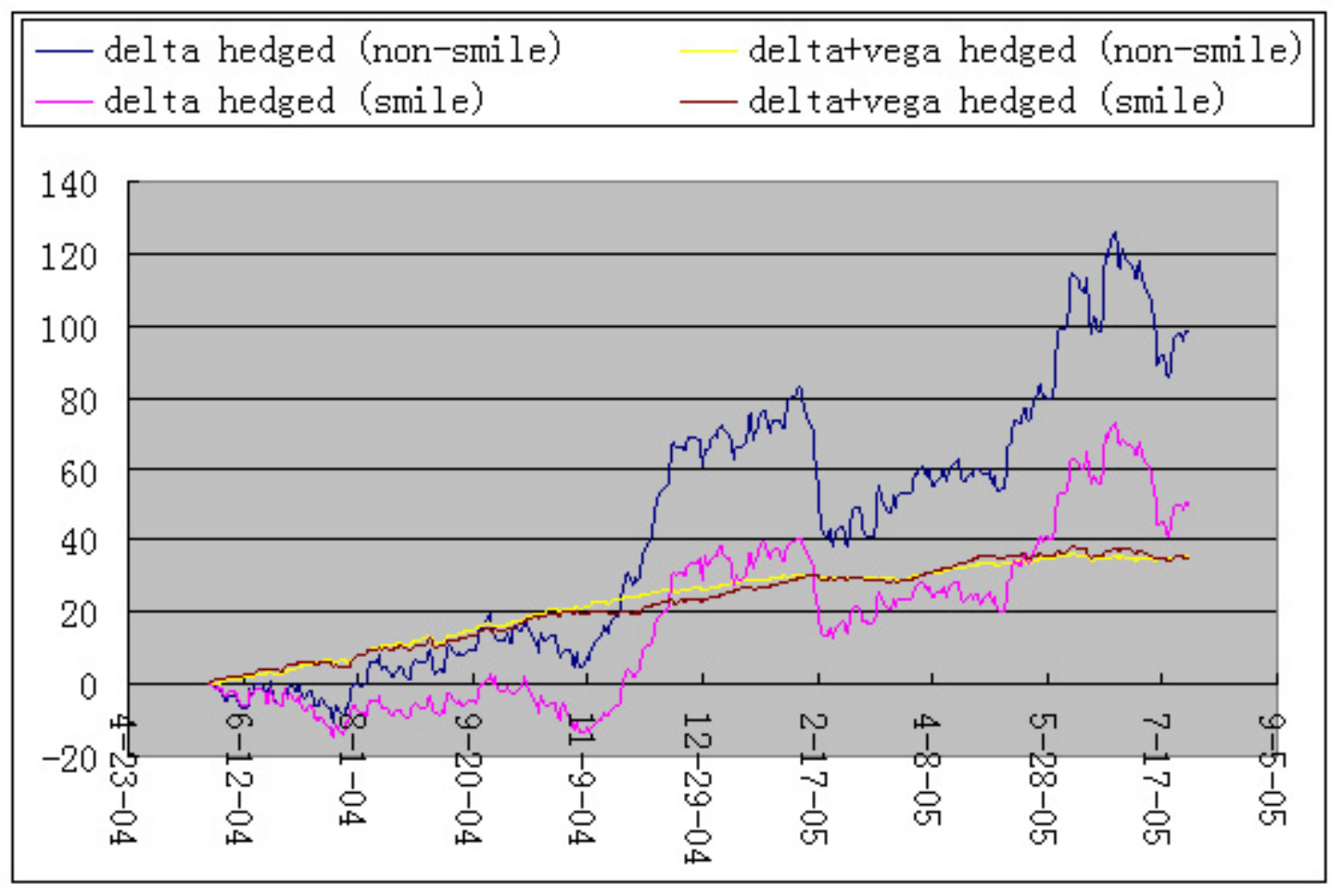}
\centering \caption{Delta and delta+vega hedged NPV (Europeans
marked to model in the non-smile case).} \label{fig:NTM-Kp040}
\end{figure}

\begin{figure}[h!]
\centering
\includegraphics[width=100mm,height=70mm]{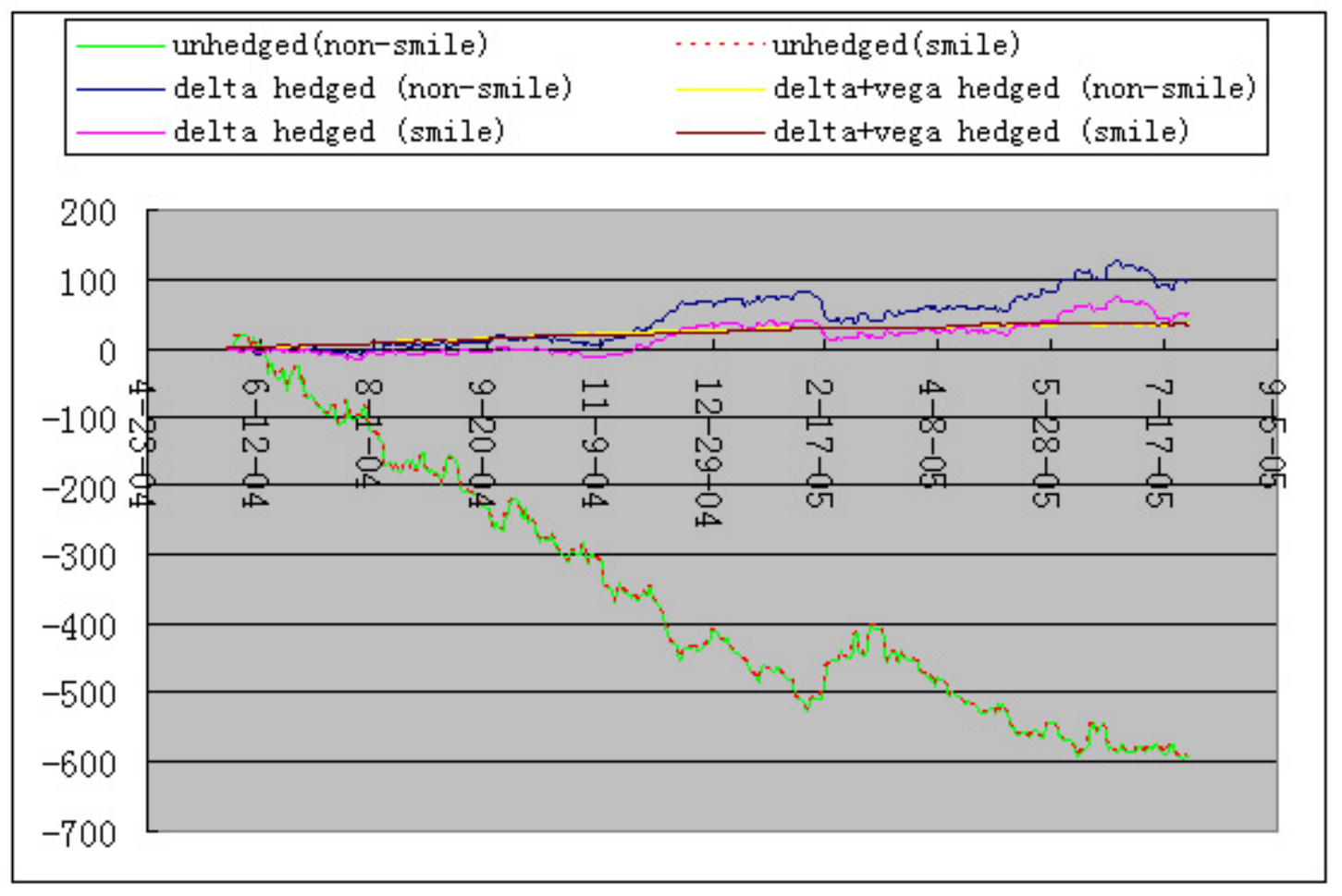}
\centering \caption{Adding the unhedged NPV to Figure
\ref{fig:NTM-Kp040}.} \label{fig:NTM-Kp040-unhedged}
\end{figure}

For hedging the non-smile Bermudans, we first stick to the approach
used in Pelsser \cite{Pelsser 1998-c} as a benchmark. In Figure
\ref{fig:NTM-Kp040}, we show the delta and delta+vega hedged NPV for
hedging smile and non-smile Bermudans. The mean-reversion parameter
is set to zero.\footnote{Again we don't quantify the mean-reversion
parameter. $0\%$ can be seen as a benchmark mean-reversion level. In
Section \ref{sec:hedgeresult4}, we will discuss the impact of the
mean-reversion level on the hedge performance.} The strike of the
Bermudan is set to $4.0\%$. This is a near-the-money level, which
has the maximal exposure to vega risk. If we check Figure
\ref{fig:forward_rate_T1}, the Bermudan is running from a little
in-the-money to a little out-of-the-money during the whole hedge
period. We see from Figure \ref{fig:NTM-Kp040} that the smile model
has a better delta hedging performance than the non-smile model. But
they have similar delta+vega hedging performances.\footnote{Please
note that this is not a fair comparison for delta+vega hedging
because the Europeans in the non-smile case should be marked to
market instead.} If we put the unhedged NPV along with the hedged
NPV, which is shown in Figure \ref{fig:NTM-Kp040-unhedged}, any of
the hedged NPVs has a much smaller order of magnitude.

\subsubsection{Monthly Vega-hedging vs Daily Vega-hedging}
\begin{figure}[h!]
\centering
\includegraphics[width=100mm,height=63mm]{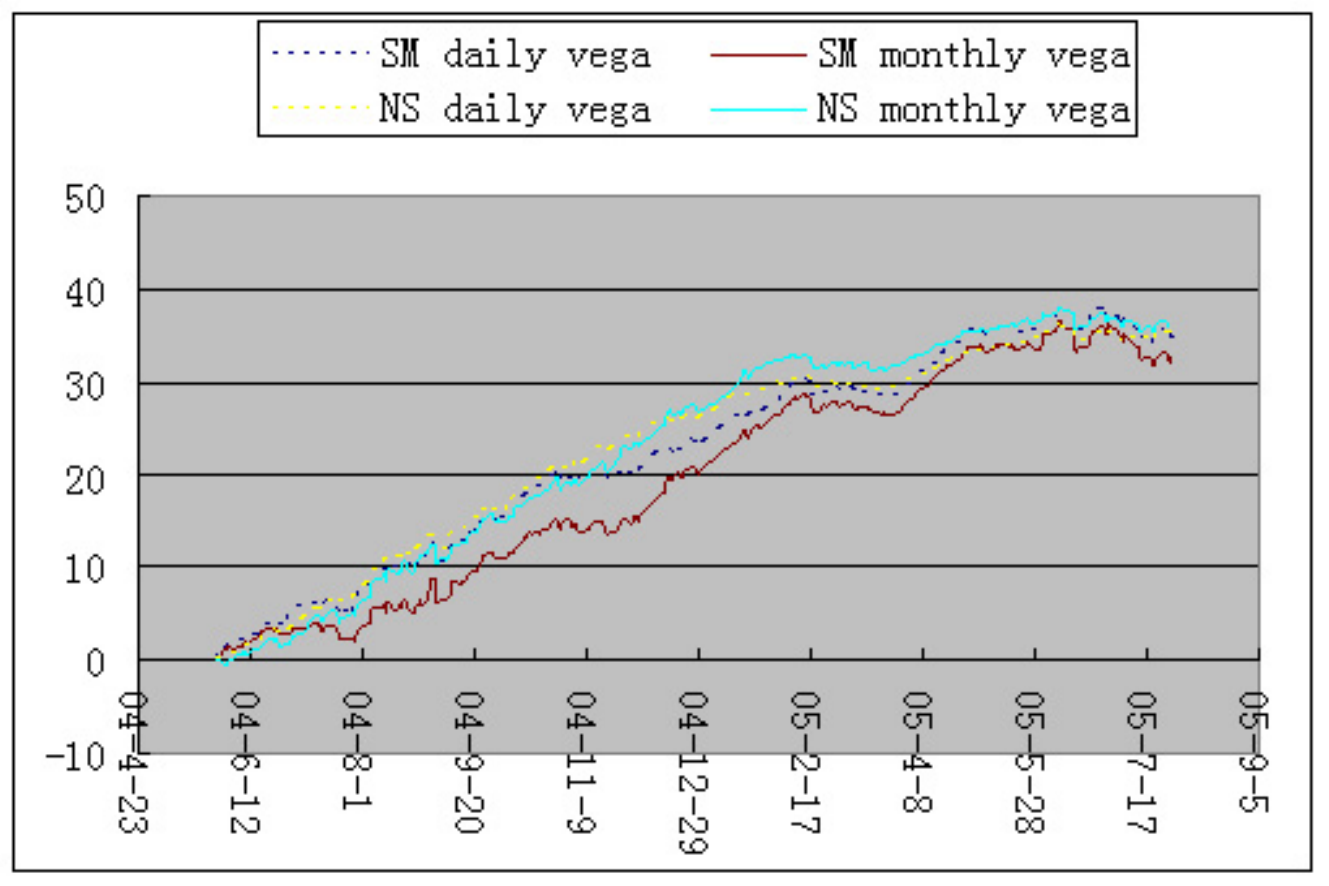}
\centering \caption{Monthly vega-hedging vs daily vega-hedging.}
\label{fig:vega-monthly-vs-daily}
\end{figure}

The transaction cost for trading European swaptions is fairly
expensive. In practice, we can not do the vega hedging on a daily
basis, but on a monthly basis. Figure
\ref{fig:vega-monthly-vs-daily} shows the delta+vega hedged NPV when
the European positions are only rolled at the end of each month,
together with the original daily delta+vega hedged NPV. We see that
in both the smile and non-smile cases, the change from a daily
rolling to monthly rolling for vega positions has little impact to
the hedged NPV. The vega hedging in all the forth-coming tests is
done on a daily basis.

\subsubsection{Experiments for Different Settings of the UVDD Model}
We have also tested some other settings for the smile vega hedging.
These tests are all related to adjustments of the calibration to the
end-of-month smile data\footnote{For relevant details, we refer back
to Section \ref{sec:syntheticdata}.}:
\begin{itemize}
\item case 1: When we calibrate the UVDD model, we set the first
components' weight $\lambda_n$ to $0.5$ instead of the original
$0.75$;
\item case 2: When we calibrate the UVDD model, we restrict the
displacement coefficient within $(0,0.05)$ instead of the original
$(0,0.10)$;
\item case 3: We calibrate the UVDD model to only 3 quotes instead
of the original 11 quotes. More precisely, there are quotes with
relative strikes to the ATM level of -50bp, 0 and 50bp.
\end{itemize}

\begin{figure}[h!]
\centering
\includegraphics[width=100mm,height=59mm]{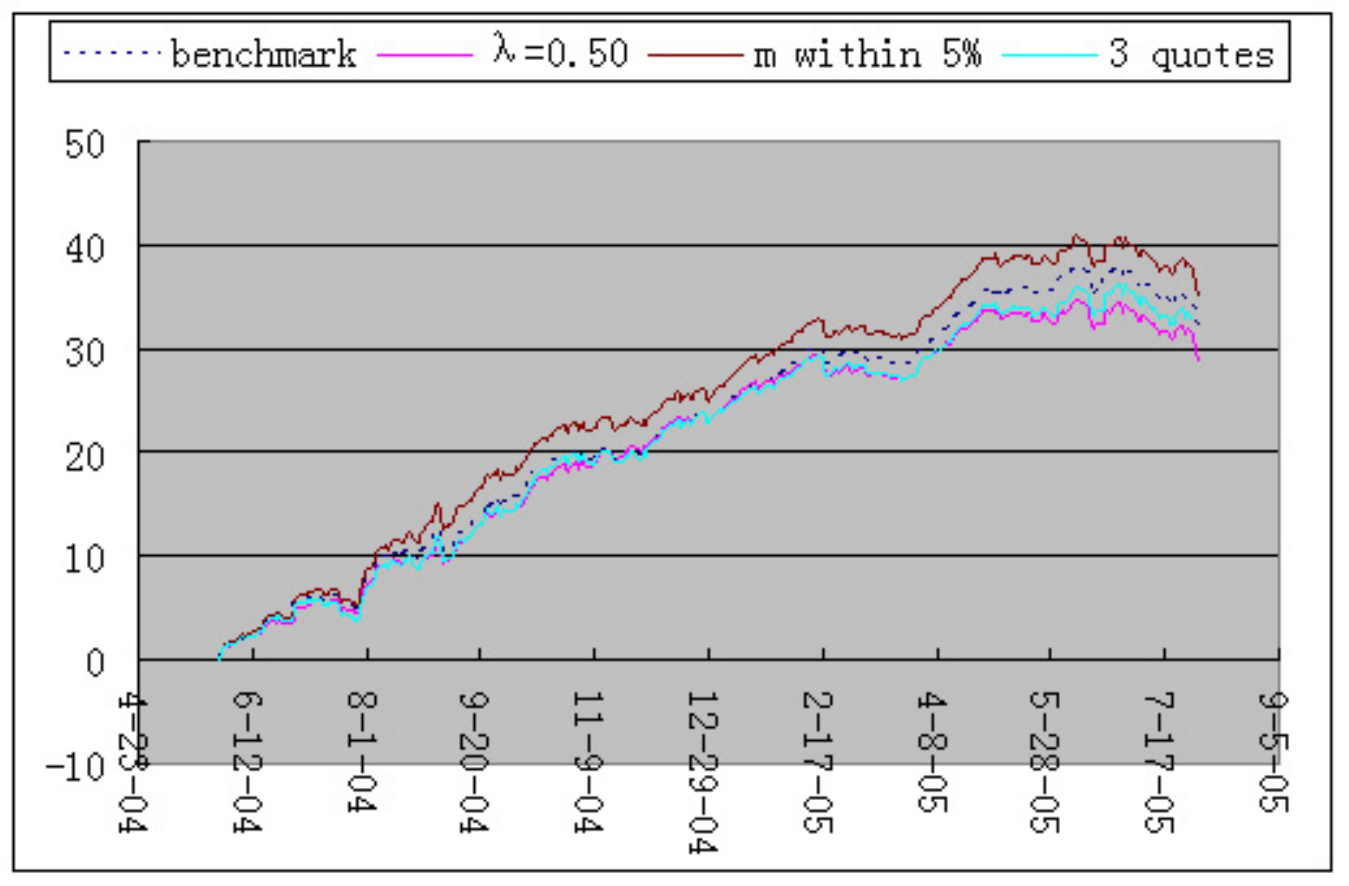}
\centering \caption{Some tunings for the smile vega-hedging.}
\label{fig:smile-vega-tuning}
\end{figure}

We show in Figure \ref{fig:smile-vega-tuning} the delta+vega hedged
NPV for each of the three cases described above. The benchmark
series is the original delta+vega hedged NPV in the smile case. We
see that all these different settings have little impact to the
hedged NPV.

\subsection{Marking the Vega-hedging to Market in case of Non-smile Bermudan}
\label{sec:hedgeresult2}

\begin{figure}[h!]
\centering
\includegraphics[width=100mm,height=75mm]{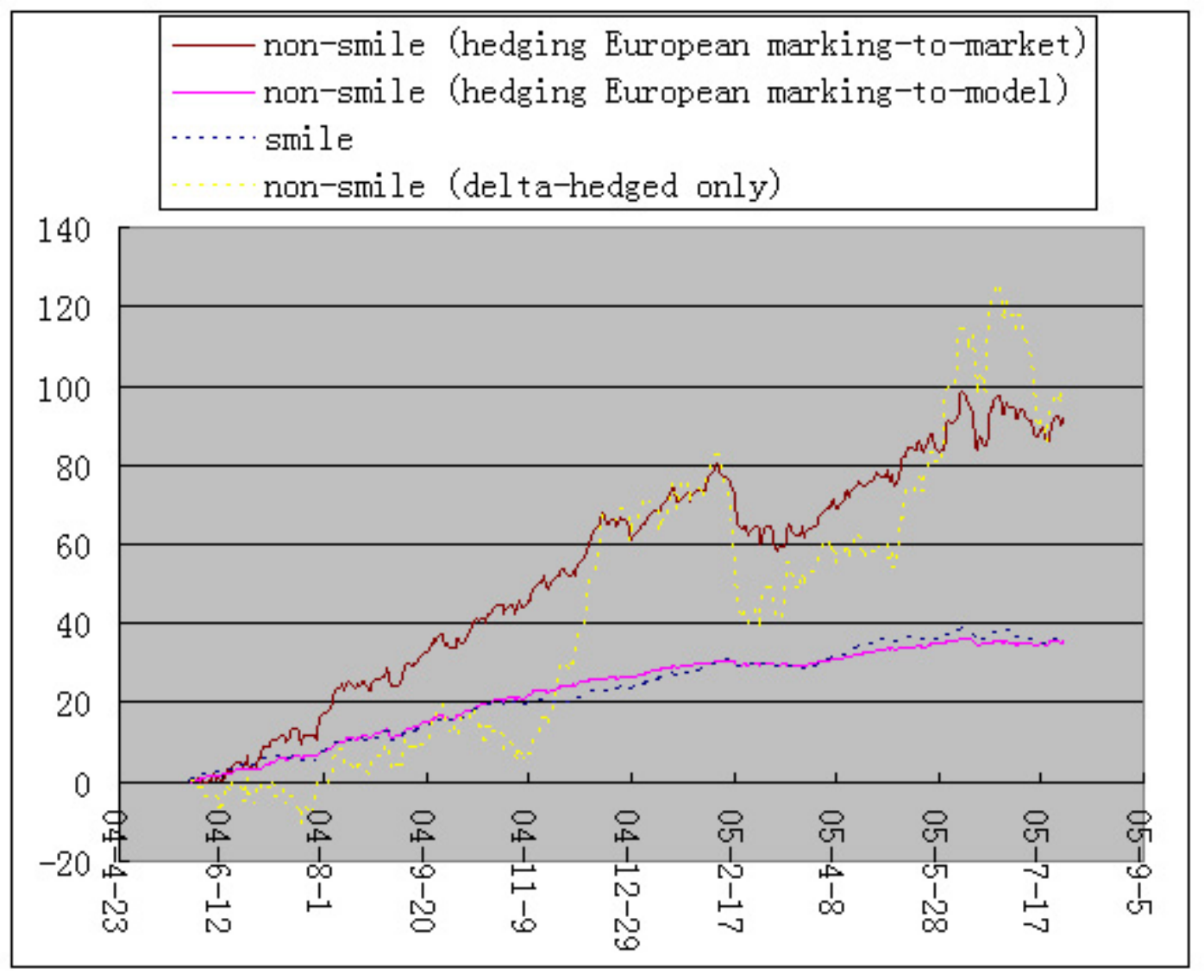}
\centering \caption{Effect of Europeans being marked to model in the
non-smile case for its delta+vega hedged NPV.}
\label{fig:MtM-NTM-Kp040}
\end{figure}

For the tests described in the previous section, for the non-smile
delta+vega hedging simulations, the hedging European swaptions are
not marked to market when they are being liquidated. However, for a
fair comparison of the hedge performance between the smile and
non-smile models, it is consistent to mark European swaptions in the
non-smile case to market instead of to the (Black) model. Figure
\ref{fig:MtM-NTM-Kp040} shows the new delta+vega hedged NPV in the
non-smile case according to the adjustment. For completeness, we
have also included in the figure the following data: the original
non-smile delta+vega hedged NPV, where the Europeans are marked to
model; the smile delta+vega hedged NPV; the non-smile delta hedged
NPV.
\\ \\
We see from Figure \ref{fig:MtM-NTM-Kp040} that the
marking-to-market version of the non-smile delta+vega hedged NPV has
a larger (positive) drift than the marking-to-model one. This can be
explained as follows. From Figure \ref{fig:SpotSwaps},
\ref{fig:forward_rate_T1} and \ref{fig:ATM vol_T2}, we see that the
underlying co-terminal swap rates have an overall decreasing trend.
In our hedge test, we long a Bermudan and short\footnote{In most
cases, a Bermudan has positive vegas.} ATM (payer) Europeans to kill
the vega sensitivities. The next day, the Europeans are most likely
to be out of the money because of the decreasing trend of the
underlying. We see from Figure \ref{fig:EoM_smiles_T2} that the
out-of-the-money, but close to ATM, volatility quote is always lower
than the ATM quote. This means that on the next day the Europeans
marked to market have lower values than those when marked to (Black)
model. Because of the \emph{short} positions, the marking-to-market
version's hedge portfolio has a higher value than the
marking-to-model one. This directly leads to a larger drift of the
hedged NPV for the former by Equation \ref{eq:hedged NPV}. This
phenomenon can also be explained in another way. In the original
non-smile case, although we take the correct prices for the (ATM)
European positions, we are generating the wrong hedge ratios. The
next day, if we still mark the Europeans to the wrong model, which
generates the wrong ratios, we would get a quite satisfactory hedge
result. But if we mark them to market, the hedge performance becomes
much worse.
\\
\begin{figure}[h!]
\centering
\includegraphics[width=100mm,height=55mm]{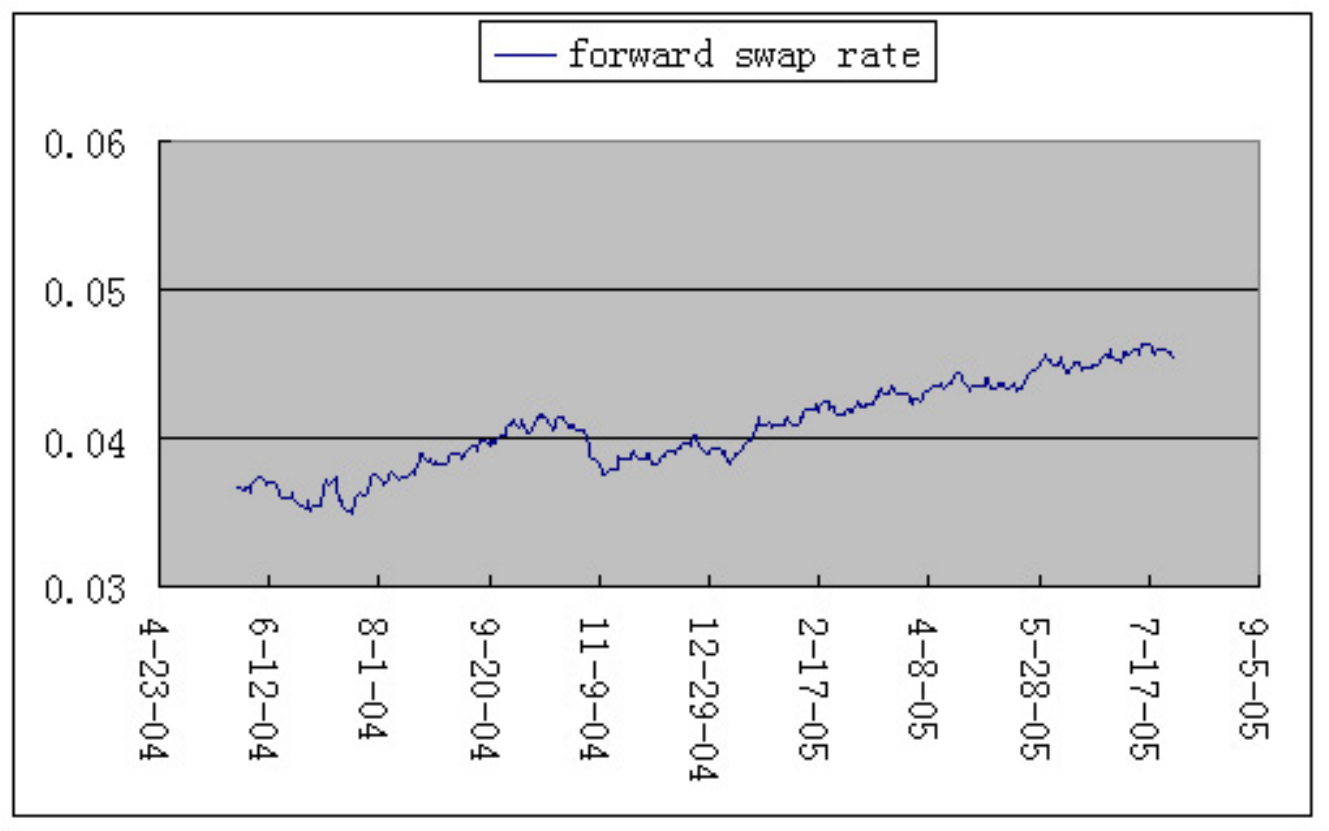}
\centering \caption{Forward swap rates for $T_1$ after reversing the
historical data.} \label{fig:forwSwapRate[T1]-reverse}
\end{figure}

\begin{figure}[h!]
\centering
\includegraphics[width=100mm,height=65mm]{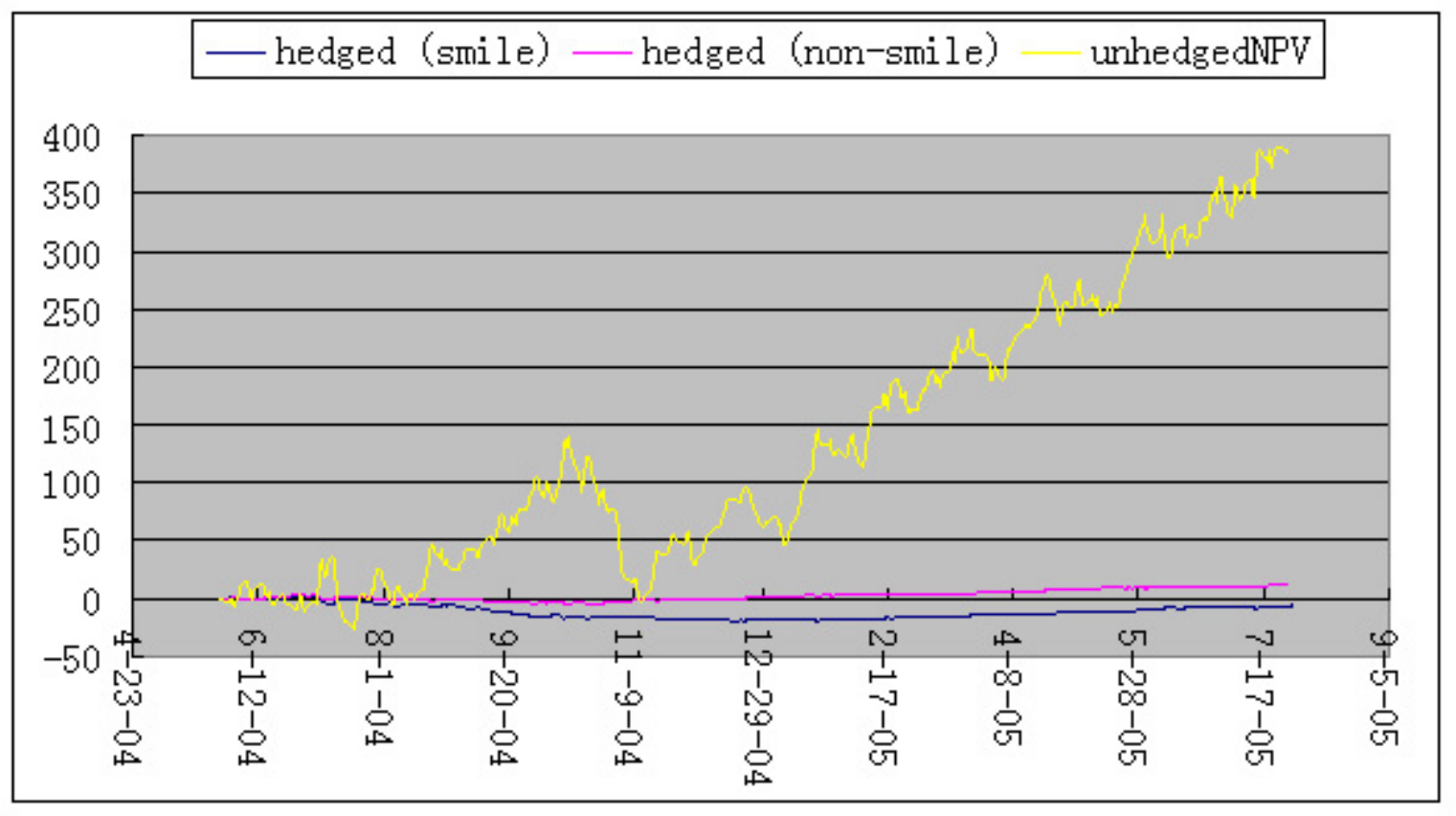}
\centering \caption{Counterpart figure to Figure
\ref{fig:NTM-Kp040-unhedged} after reversing the historical data.}
\label{fig:NTM-Kp040-reverse}
\end{figure}

\begin{figure}[h!]
\centering
\includegraphics[width=100mm,height=70mm]{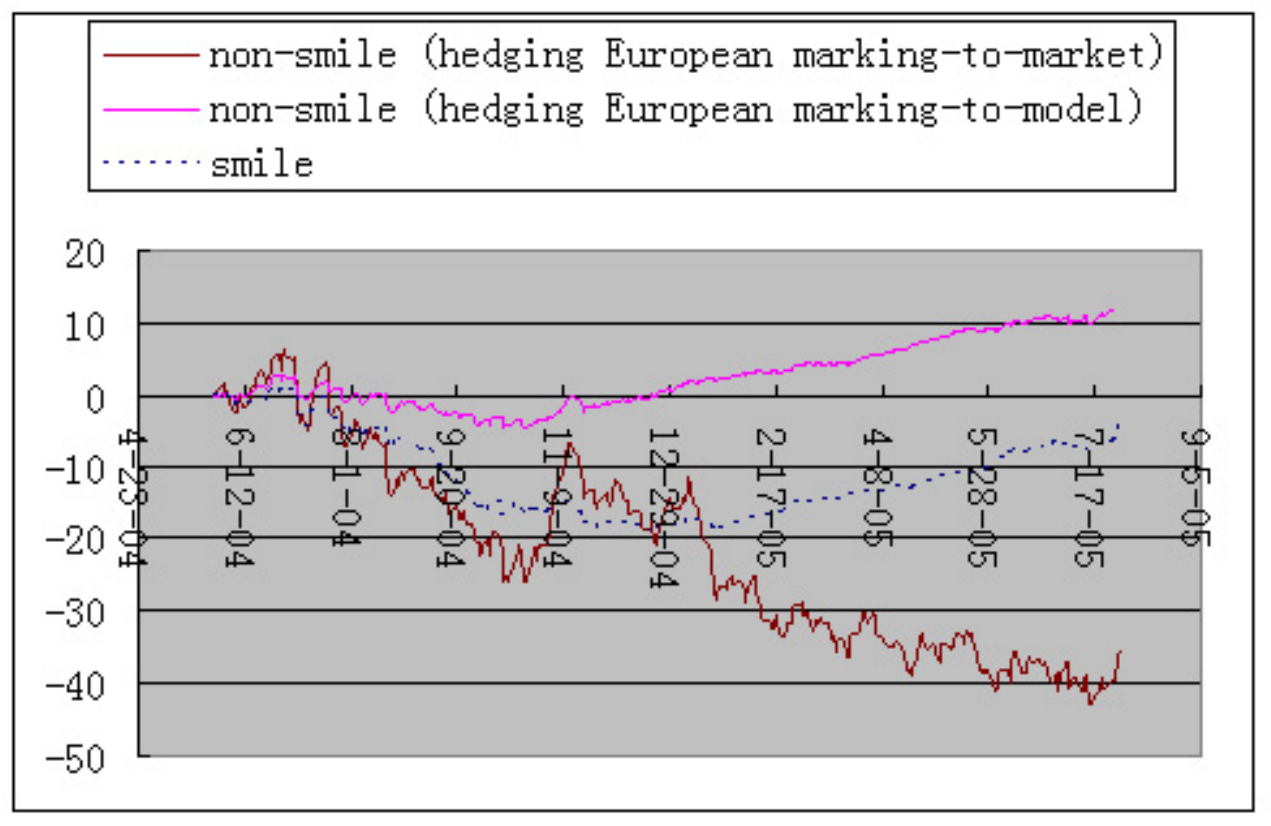}
\centering \caption{Counterpart figure to Figure
\ref{fig:MtM-NTM-Kp040} after reversing the historical data.}
\label{fig:MtM-NTM-Kp040-reverse}
\end{figure}

If we reverse the historical data, we should expect the opposite
drift effect in the non-smile case between the marking-to-market and
marking-to-model versions. Figure \ref{fig:forwSwapRate[T1]-reverse}
shows that the time series of the forward swap rate corresponding to
the underlying co-terminal swap which starts at $T_1$. Now the
underlying has an increasing trend, starting a little
out-of-the-money and ending a little in-the-money (strike $4.0\%$).
Figure \ref{fig:NTM-Kp040-reverse} is the counterpart figure to
Figure \ref{fig:NTM-Kp040-unhedged} after reversing the historical
data, but without the delta hedged NPV. As expected, we see that the
unhedged NPV now has a large positive drift instead of the original
very negative one. Figure \ref{fig:EoM_smiles_T2} shows that the
in-the-money volatility quote is always higher than the ATM quote.
This means that on the next day the Europeans marked-to-market have
lower values than those when marked to (Black) model. This leads to
the opposite effect for the drift. Figure
\ref{fig:MtM-NTM-Kp040-reverse} is the counterpart figure to Figure
\ref{fig:MtM-NTM-Kp040} after reversing the historical data, but
without the delta hedged NPV. We see from Figure
\ref{fig:MtM-NTM-Kp040-reverse} that the marking-to-market version
of the non-smile delta+vega hedged NPV has a larger ({\emph
negative}) drift than the marking-to-model one. This is exactly the
drift effect we expect for reversing the historical data.

\subsection{Hedge Tests for ITM/OTM Trades}
\label{sec:hedgeresult3}

\begin{figure}[h!]
\centering
\includegraphics[width=100mm,height=67mm]{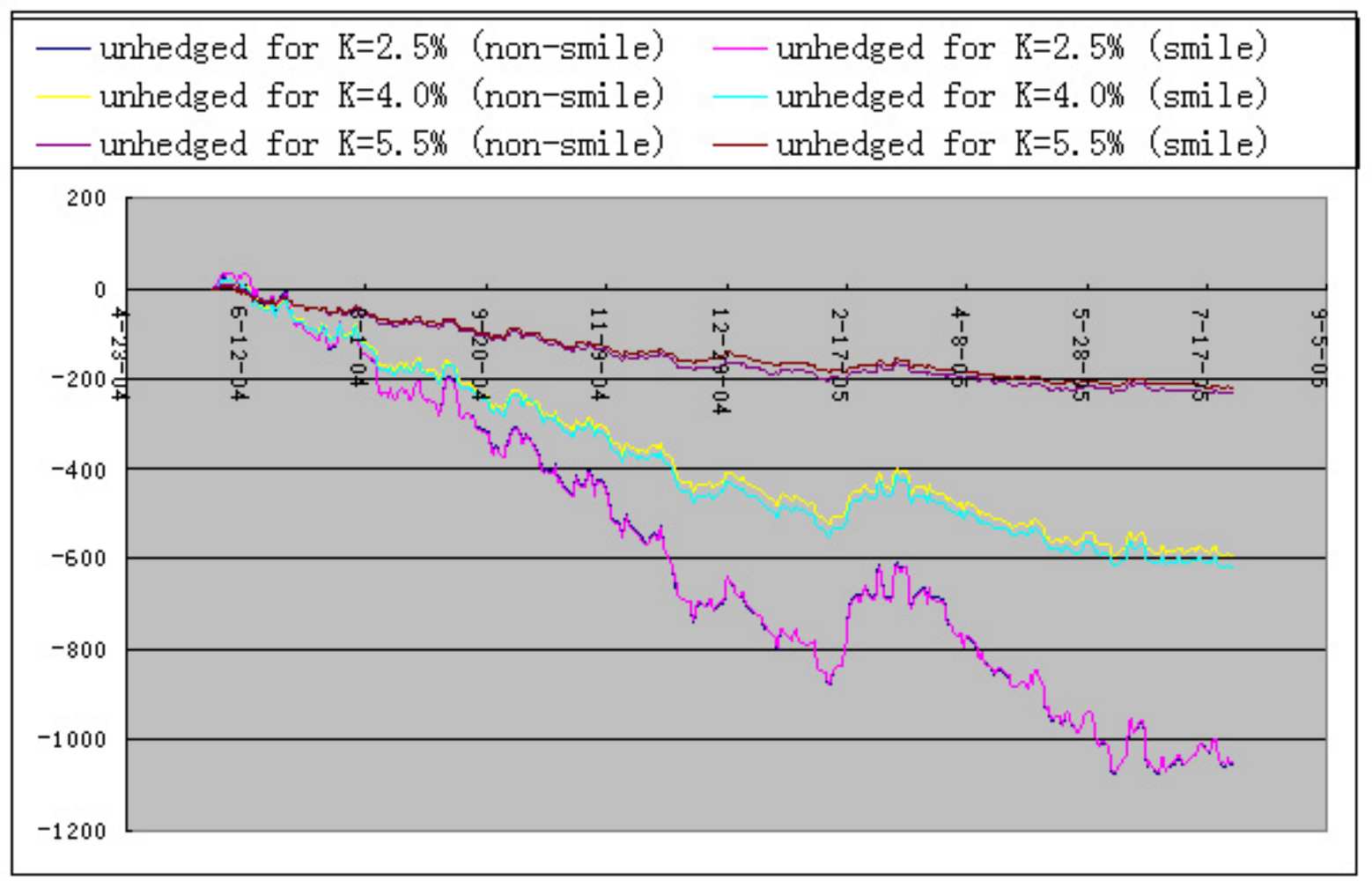}
\centering \caption{Unhedged NPV for the ITM/near-the-money/OTM
trades.} \label{fig:unhedged-strikes}
\end{figure}

In Section \ref{sec:hedgeresult1} and \ref{sec:hedgeresult2}, we
only conducted hedge tests for the near-the-money strike ($4.0\%$),
which is most sensitive to vega risk. In this section, we perform
the counterpart hedging simulations for the in-the-money ({\bf ITM})
strike ($2.5\%$) and the out-of-the-money ({\bf OTM}) strike
($5.5\%$).
\\ \\
Let's first get an impression of the magnitudes of the unhedged NPV.
Figure \ref{fig:unhedged-strikes} shows the unhedged NPV for the
ITM/OTM trades, together with the previous near-the-money ones. We
see that the ITM unhedged NPV has the largest drift and OTM a fairly
small one. The near-the-money unhedged NPV fall in between.

\begin{figure}[h!]
\centering
\includegraphics[width=100mm,height=65mm]{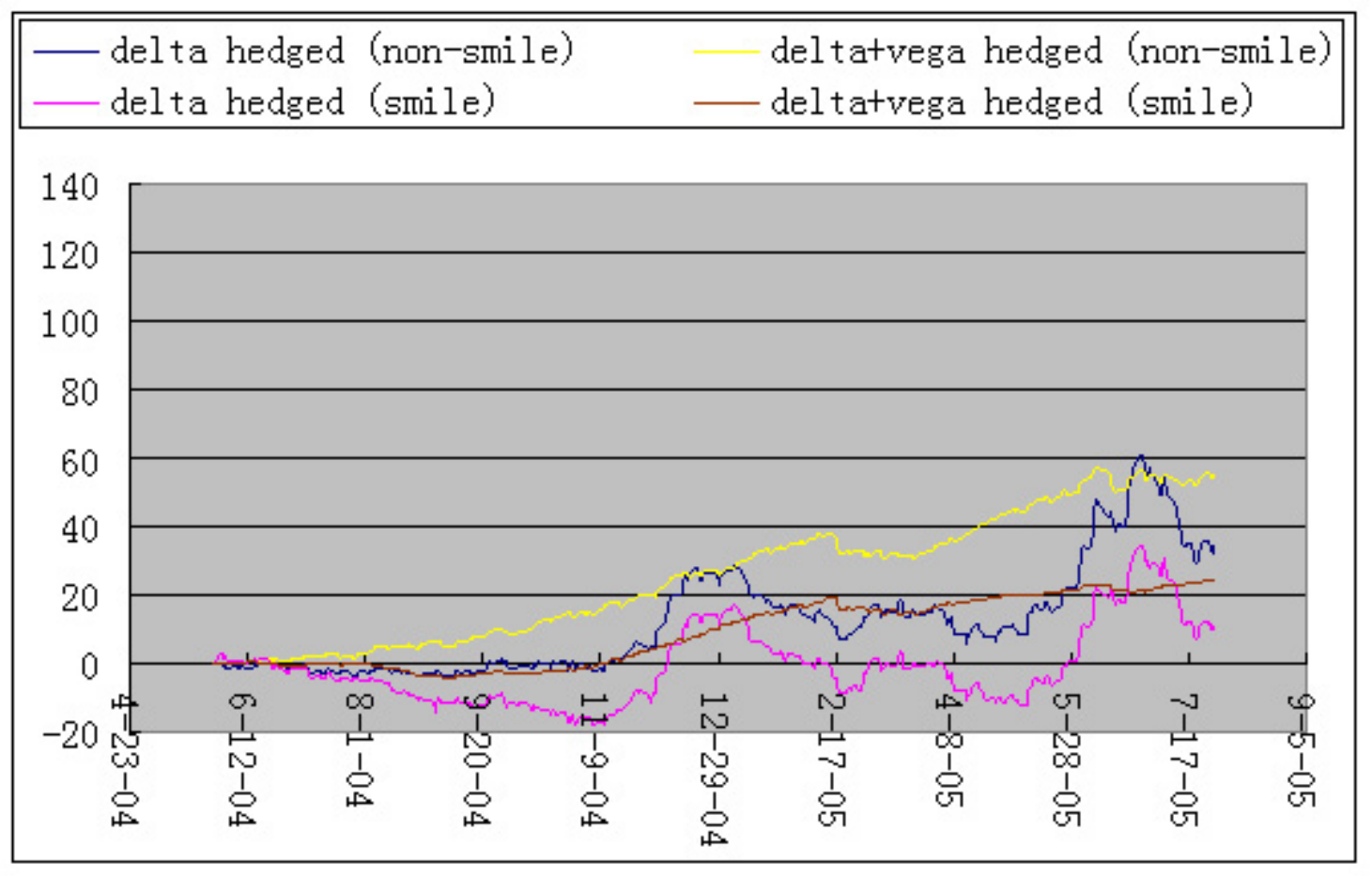}
\centering \caption{Delta and delta+vega hedged NPV for the ITM
trade.}
\label{fig:MtM-ITM-Kp025}
\end{figure}

\begin{figure}[h!]
\centering
\includegraphics[width=100mm,height=65mm]{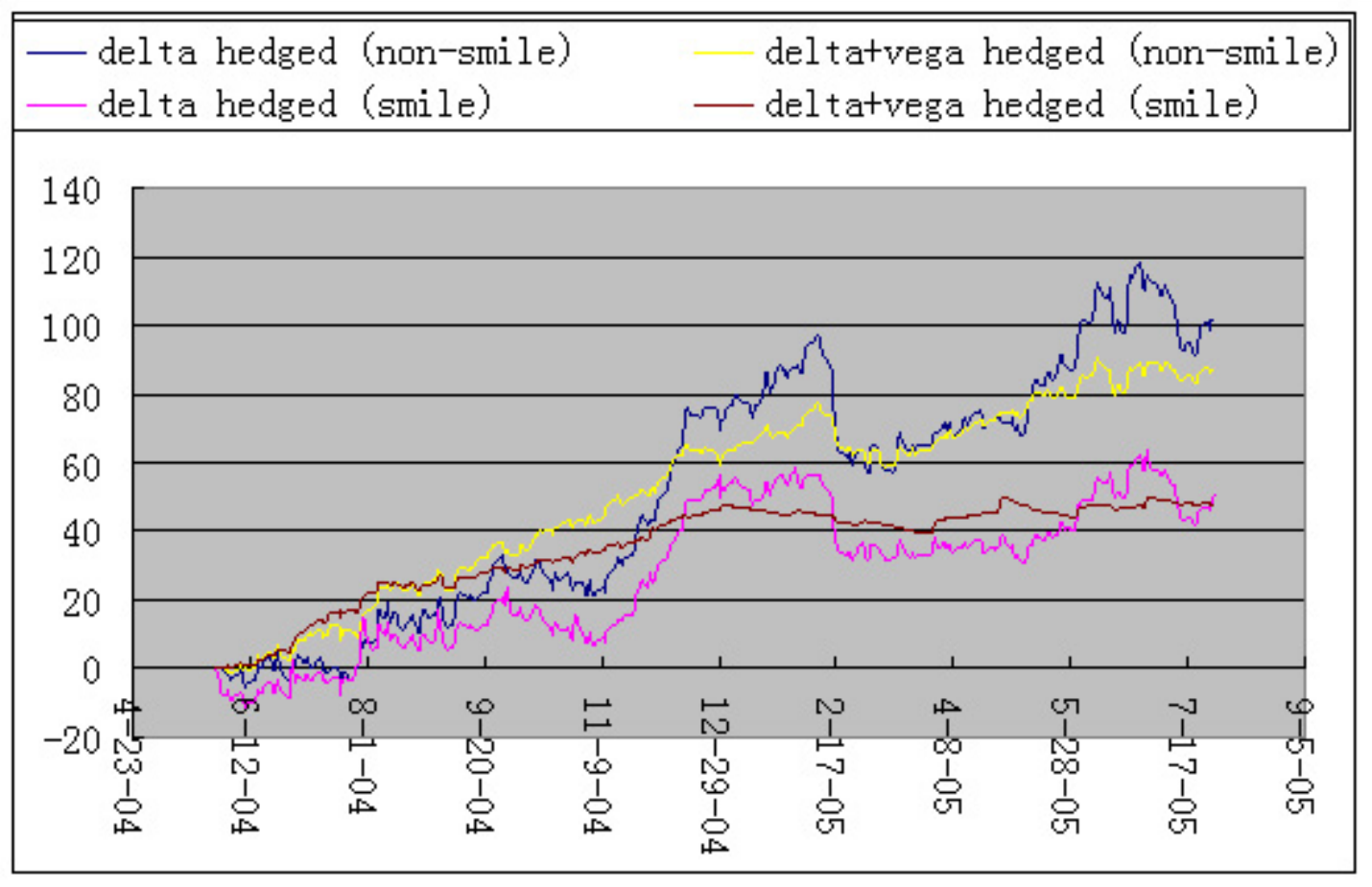}
\centering \caption{Delta and delta+vega hedged NPV for the OTM
trade.}
\label{fig:MtM-OTM-Kp055}
\end{figure}

\begin{figure}[h!]
\centering
\includegraphics[width=120mm,height=60mm]{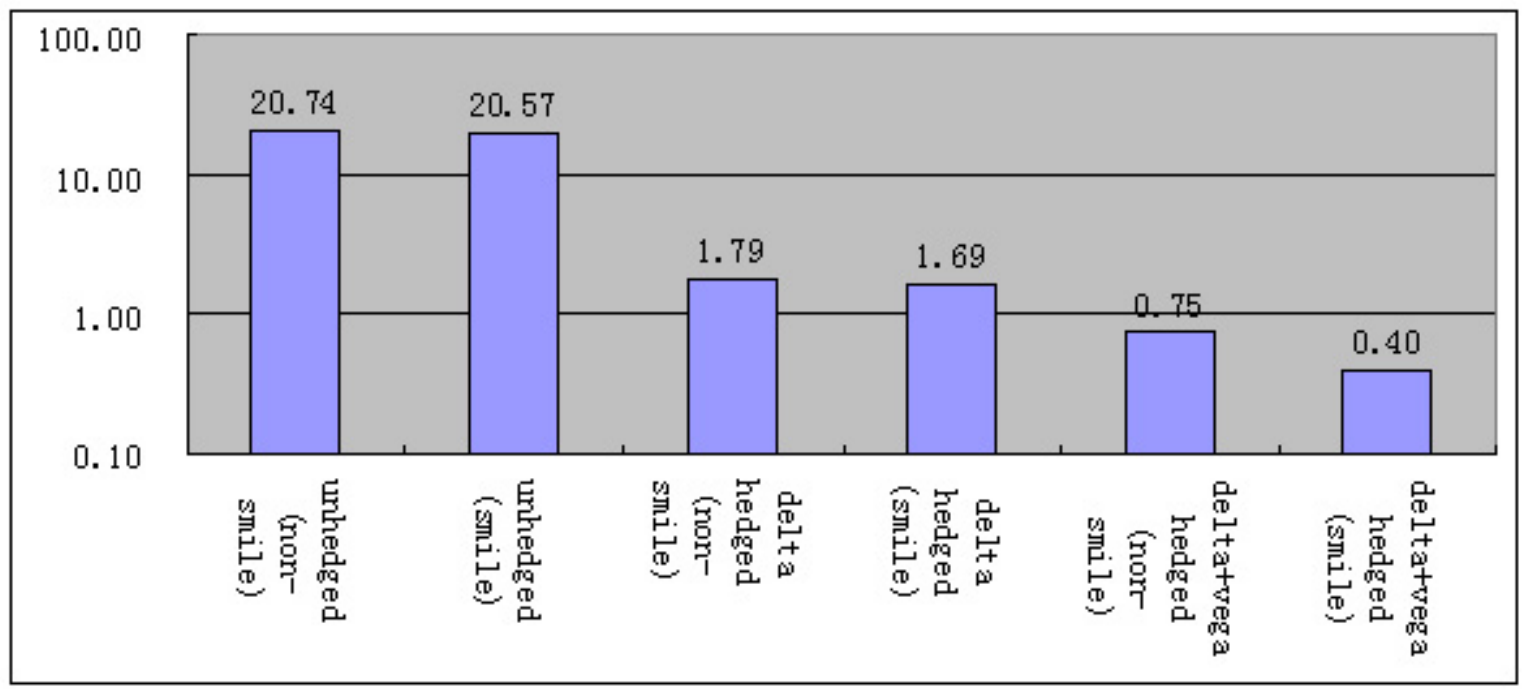}
\centering \caption{Standard deviation of the unhedged and hedged
daily P$\&$L ($K=2.5\%$).} \label{fig:PnL-vol-ITM-Kp025}
\end{figure}

\begin{figure}[h!]
\centering
\includegraphics[width=120mm,height=60mm]{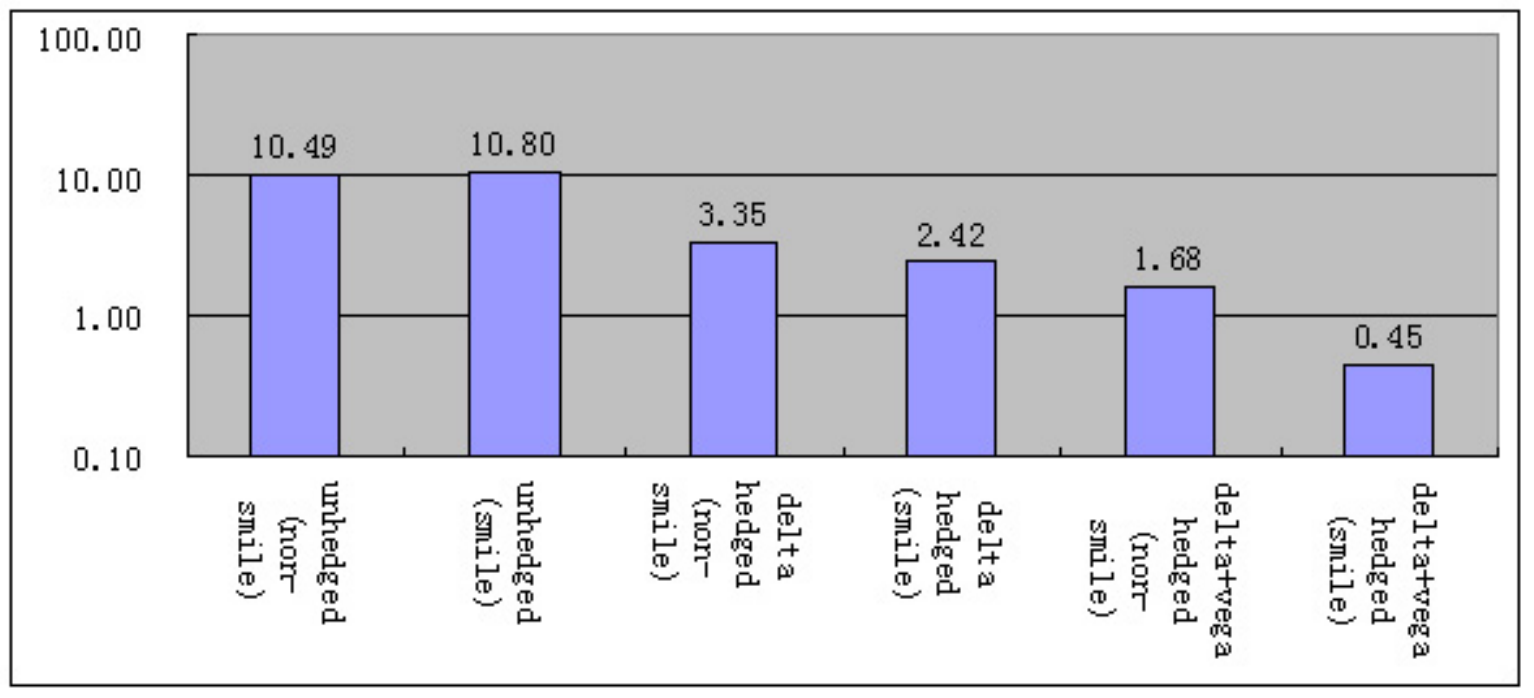}
\centering \caption{Standard deviations of the unhedged and hedged
daily P$\&$L ($K=4.0\%$).}
\label{fig:PnL-vol-NTM-Kp040}
\end{figure}

\begin{figure}[h!]
\centering
\includegraphics[width=120mm,height=60mm]{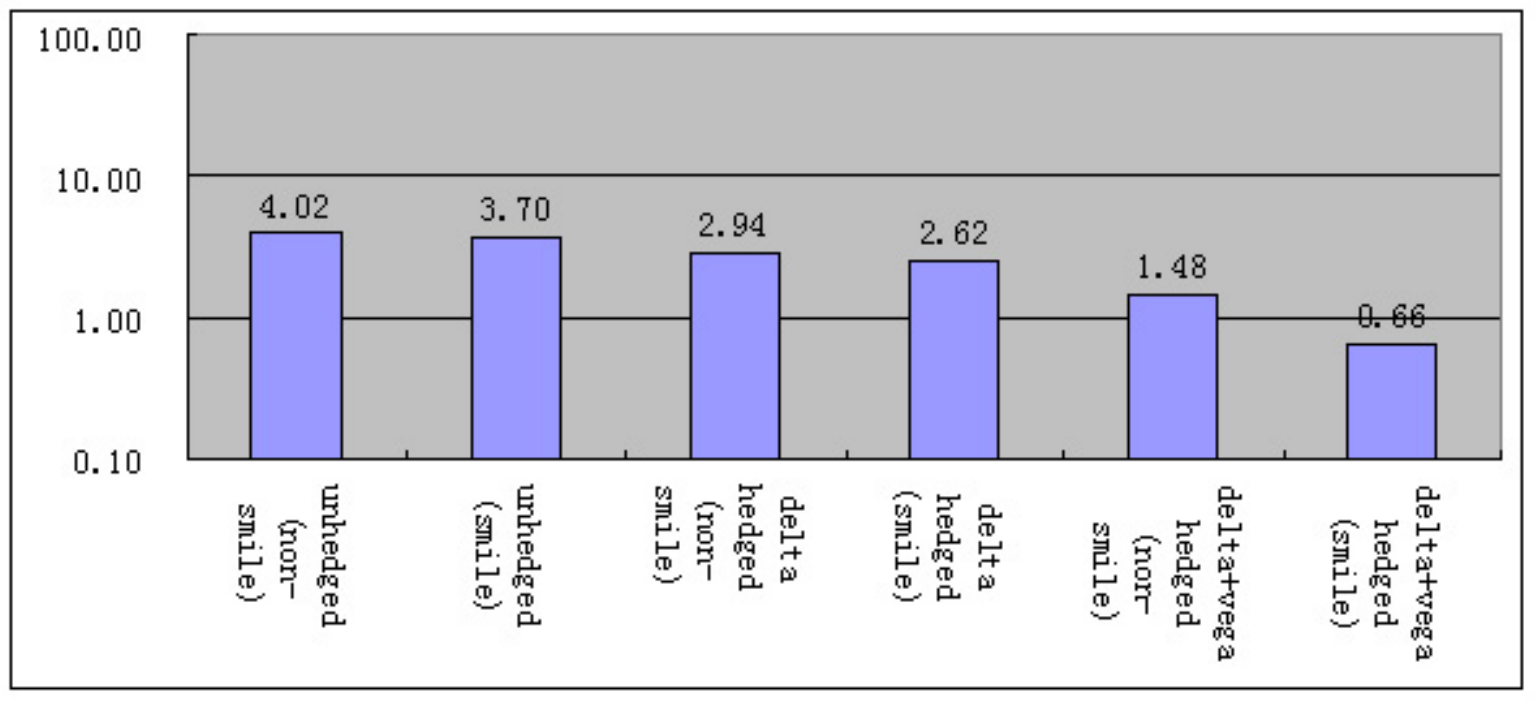}
\centering \caption{Standard deviations of the unhedged and hedged
daily P$\&$L ($K=5.5\%$).}
\label{fig:PnL-vol-OTM-Kp055}
\end{figure}

In Figure \ref{fig:MtM-ITM-Kp025} and \ref{fig:MtM-OTM-Kp055}, we
show the delta and delta+vega hedged NPV for hedging smile and
non-smile Bermudans with strike levels of $2.5\%$ and $5.5\%$,
respectively. Note that in the non-smile cases, the hedging
Europeans are marked to market when they are being
liquidated.\footnote{This applies to this whole section.} Similar to
the results of the near-the-money trade, the smile model outperforms
the non-smile model in both delta and delta+vega hedgings.
\\ \\
We also observe that in both smile and non-smile cases, a delta+vega
hedging reduces significantly the oscillation of the hedged NPV as
compared to the delta hedging. However, it doesn't affect the drift
level of the hedged NPV. This phenomenon can be observed throughout
the trades across the three strikes (ITM/near-the-money/OTM). In
Figure \ref{fig:PnL-vol-ITM-Kp025}, \ref{fig:PnL-vol-NTM-Kp040} and
\ref{fig:PnL-vol-OTM-Kp055}, we show the standard deviations of the
unhedged and hedged daily profit and loss (P$\&$L) for each of the
three strikes, respectively. The y-axes in these figures are all in
logarithmic scale. The main conclusions are:
\begin{itemize}
\item Delta hedging reduces significantly the standard deviation of the unhedged
daily P$\&$L. A delta+vega hedging further reduces significantly the
standard deviation of the delta hedged daily P$\&$L;
\item The smile model outperforms
the non-smile model in both the standard deviations of the delta and
delta+vega hedged daily P$\&$L.
\end{itemize}

\newpage
\subsection{Impact of the Mean-reversion Parameter on the Hedge Performance}
\label{sec:hedgeresult4}

In all the tests discussed before, the mean-reversion parameter has
been set to zero. In this section, we test the impact of the
mean-reversion level on the hedge performance. We stick to the
delta+vega hedging and a Bermudan with the strike level of $4.0\%$
is considered.

\begin{figure}[h!]
\centering
\includegraphics[width=\textwidth,height=98mm]{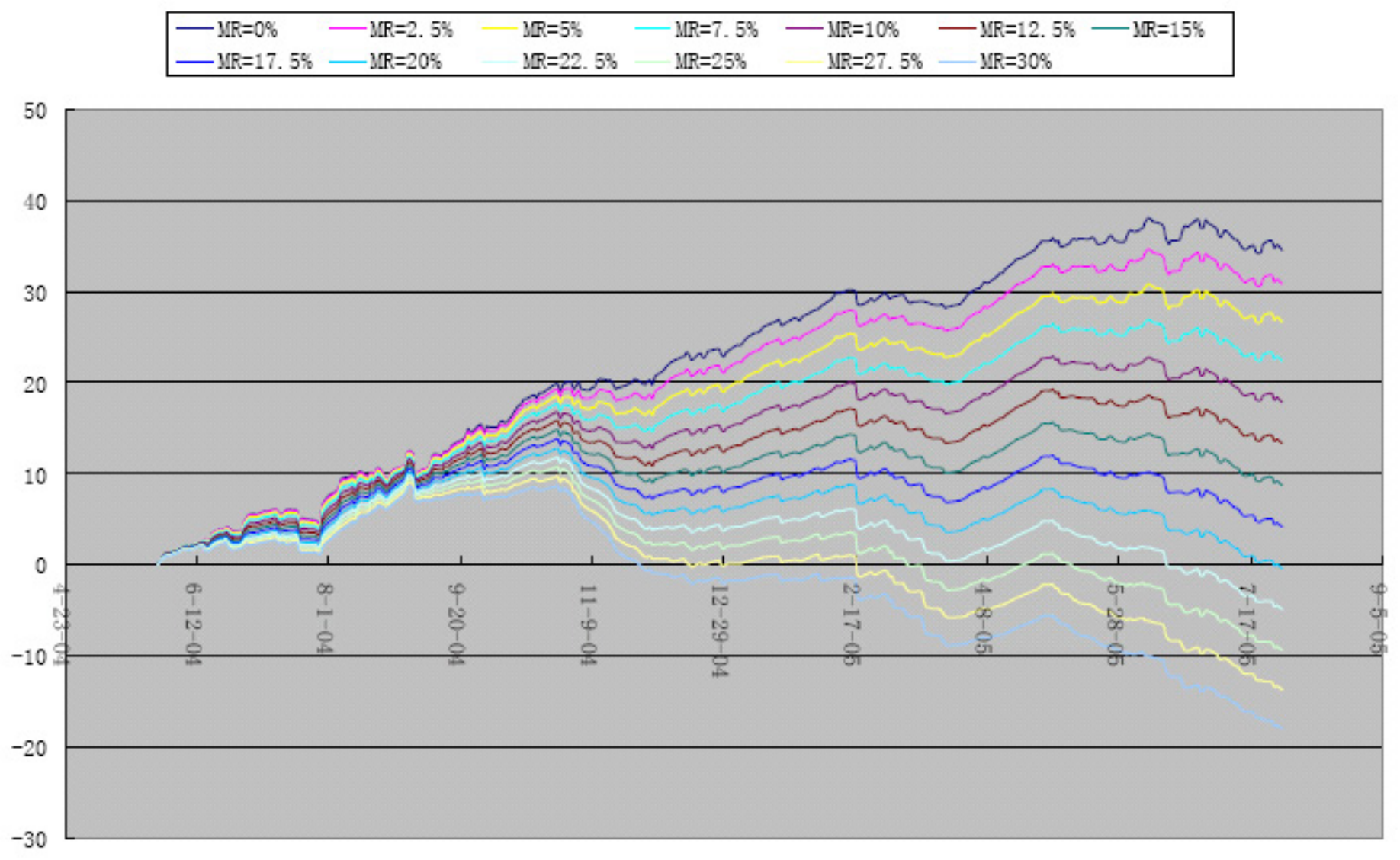}
\centering \caption{Delta+vega hedged NPV in the smile case by
setting different values of the mean-reversion (MR) parameter.}
\label{fig:smile-MR}
\end{figure}

\begin{figure}[h!]
\centering
\includegraphics[width=\textwidth,height=68mm]{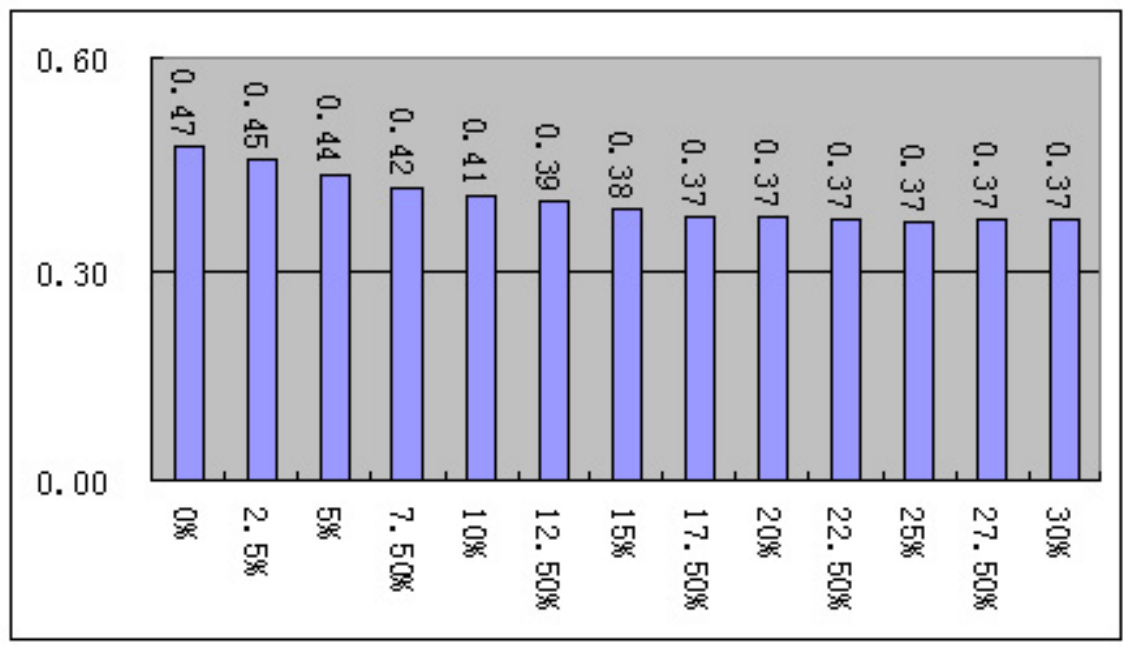}
\centering \caption{Standard deviations of hedged daily P$\&$L
corresponding to those tests in Figure \ref{fig:smile-MR}.}
\label{fig:smile-MR-PnL-vol}
\end{figure}

In Figure \ref{fig:smile-MR}, we show the hedge results which have
been obtained by using different values of the mean-reversion
parameter. In Figure \ref{fig:smile-MR-PnL-vol}, we show the
standard deviations of hedged daily P$\&$L corresponding to those
tests in Figure \ref{fig:smile-MR}. From these two figures, we
clearly see that increasing the mean-reversion level decreases the
level of the hedged NPV's drift, but has little impact of its
oscillation.
\\ \\
By following the approach described in Section
\ref{sec:MR_estimation}, our estimate for the mean-reversion
parameter is around $3\%$. Please note this is just a rough
estimate. The estimated mean-reversion parameter by that approach is
not accurate due to the following two reasons:

\begin{itemize}
\item When we choose historical data of different periods, we get
different estimated levels;
\item Even if we fix a period of historical data, when the
Bermudan trade in our hedge test is running towards its first
exercise date, the $\rho$ and $\sqrt{\frac{T_n}{T_k}}$ in Equation
\ref{eq:MR_corr_factor} are both changing over time. This leads to a
time-varying mean-reversion level.\footnote{For details of terminal
correlations and the mean-reversion parameter, please refer back to
Section \ref{sec:MR_all}.}
\end{itemize}

We are not in favor of quantifying the mean-reversion level by that
approach because of the following reasons:

\begin{itemize}
\item The lognormal assumption in Equation
\ref{eq:MR_esti_assumption} conflicts the UVDD's assumption that the
terminal density is a mixture of lognormal distributions;
\item For determining the marginal density of each individual terminal swap
rate, we make use of the current market information (European
swaptions). But for determining the joint distribution of the
underlying swap rates, we make use of the past information
(historical data of swap rates). It is not clear whether these two
sources of information are consistent to each other.
\item Whether the mean-reversion level is a constant over time is questionable.
But even if we used Section \ref{sec:MR_estimation}'s approach to
implement a dynamic mean-reversion estimator, we would introduce
another uncertainty, a time-varying or even stochastic MR, which can
not be killed by European swaptions in a hedging simulation. One
ideal way to solve this problem is to determine the mean-reversion
parameter by extracting information from another relevant
path-dependent option, which is liquid enough. In this way, we may
kill the MR uncertainty in hedging as well as extract consistent
market information for the marginal and joint densities of the
underlying swap rates.
\end{itemize}

\newpage
\subsection{Discussion of the Residual Drift of the Hedged NPV}
\label{sec:hedgeresult5}

Although the smile model achieves a better hedging performance than
the non-smile model, there is still a residual drift in the hedged
NPV. This might be due to the following reasons:

\begin{itemize}

\item A misspecified mean-reversion parameter. This has already been elaborated in
Section \ref{sec:hedgeresult4};

\item A mismatch of cash-flows. We are using spot-starting
swaps instead of forward swaps for delta-hedging.\footnote{To get
the time series of even one forward swap running towards its expiry
date is very much time-consuming for us. This is not feasible for
the project.} The payments out of the Bermudan are not exactly
offset with the cash-flows out of the spot-starting swaps, but are
only offset on an aggregate basis. Pelsser \cite{Pelsser 1998-c}
performed an "exact" hedge, which uses discount bonds to exactly
offset the cash-flows from the Bermudan. The test result shows that
the "exact" hedge reduces the drift significantly;

\item Other misspecified model parameters. Pelsser \cite{Pelsser 1998-c}
explained that even for a European option, a misspecified model
parameter may lead to a drift for the delta-hedged NPV, even if the
hedging is done continuously. The magnitude of the drift is
proportional to the level of gamma. The only way to kill this risk
is to gamma hedge the position. Unfortunately, a Bermudan swaption
on a multi-dimensional underlying can not be gamma hedged, because
we don't have enough hedging instruments to kill the cross-term
sensitivities $\frac{\partial^2 BSN}{\partial S_i
\partial S_j}$, for $i\neq j$.

\end{itemize}

\chapter{Conclusions $\&$ Suggestions for Future Research}
\label{chapter:conclusions}

A volatility smile has been successfully incorporated into the
Markov-functional model by using the UVDD digital mapping. The new
model has a significant impact for pricing Bermudan swaptions,
especially for the deep ITM/OTM strikes. The MF model based on the
UVDD mapping also improves the hedging performance significantly
compared to the MF model based on the Black-Scholes mapping. This is
consistent with the fact that the smile model has some freedom to
control the implied future smiles and generates fairly good smile
dynamics.\footnote{For details, please refer back to Chapter
\ref{chapter:dynamics}.}
\\ \\
However, the method for estimation of the mean-reversion parameter
could probably be improved. As discussed in Section
\ref{sec:hedgeresult4}, the present approach has some crucial
drawbacks. A better approach would be to use relevant path-dependent
options which directly contain the information for the joint
distribution of the underlying swap rates.

\appendix
\chapter{Notation and Preliminary Knowledge}
\section{Notation and Preliminary Knowledge} \label{sec:notation}
In this appendix, we would like to explain the notation and some
preliminary knowledge relevant to this report.

\begin{itemize}
\item We choose a tenor structure $T_{1}, T_{2}, ...,T_{N+1}$ where
$T_{n}(n=1,2,...,N)$ denotes the n-th floating reset date. In other
words, the LIBOR rate $L_{n}$ has a tenor of $[T_{n},T_{n+1}]$ with
length $\alpha_{n}=T_{n+1}-T_{n}$; the swap rate $S_{n}$ has a tenor
of $[T_n,T_{N+1}]$ with cash exchange at time $T_{n+1},
...,T_{N+1}$.
\item Let $D_{n}(t)$ denote the value at time t of a discount bond
maturing at $T_{n}$.
\item The forward LIBOR rate $L_{n}(t)$ is defined as
\begin{equation} \label{eq:libor}
L_{n}(t)= \frac{1}{\alpha_{n}}(\frac{D_{n}(t)}{D_{n+1}(t)}-1).
\end{equation}
\item Let $P_{n}(t)$ denote the PVBP, namely \textit{present value of a
basispoint}, on tenor $[T_{n},T_{N+1}]$.
\begin{equation} \label{eq:PVBP}
P_{n}(t)=\sum_{k=n+1}^{N+1}\alpha_{k-1}D_k(t).
\end{equation}
The following linear relationship is of use in MF model's digital
mapping\footnote{This will be explained in Section
\ref{sec:BS_mapping}.}:
\begin{eqnarray} \label{eq:R_PVBP}
\frac{P_n(T_{n+1})}{D_{N+1}(T_{n+1})}
&=&\frac{\sum_{k=n+1}^{N+1}\alpha_{k-1}D_k(T_{n+1})}{D_{N+1}(T_{n+1})} \nonumber\\
&=&\frac{\alpha_n+\sum_{k=n+2}^{N+1}\alpha_{k-1}D_k(T_{n+1})}{D_{N+1}(T_{n+1})}\nonumber\\
&=&\frac{\alpha_n+P_{n+1}(T_{n+1})}{D_{N+1}(T_{n+1})}
=\frac{\alpha_n}{D_{N+1}(T_{n+1})}+\frac{P_{n+1}(T_{n+1})}{D_{N+1}(T_{n+1})}.
\end{eqnarray}

\item A payer swap pays the fixed leg and receives the floating leg;
a receiver swap receives the fixed leg and pays the floating leg.

\item The forward par swap rate $S_{n}(t)$ on tenor $[T_{n},T_{N+1}]$
can be expressed as\footnote{For derivation, we refer to Chapter 25
of Bjork \cite{Bjork 2003}.}
\begin{equation} \label{eq:swap}
S_{n}(t)=\frac{D_{n}(t)-D_{N+1}(t)}{P_{n}(t)}.
\end{equation}

\item Denote the forward measure for the numeraire $P_{n}(t)$ by
$Q^{n,N+1}$. Then the forward swap rate $S_{n}(t)$ is a
$Q^{n,N+1}$-martingale.\footnote{For proof, we refer to Chapter 25
of Bjork \cite{Bjork 2003}.}

\item Let $SV_n(t;K)$ denote the value at time t of a swap with a fixed
rate K on tenor $[T_{n},T_{N+1}]$ with unit notional amount. The
value of the swap is\footnote{For proof, we refer to Chapter 25 of
Bjork \cite{Bjork 2003}.}
\begin{equation} \label{eq:swap_value}
SV_n(t;K)=P_n(t)\varphi(S_n(t)-K),
\end{equation}
where $\varphi$ is 1 for a payer swap and -1 for a receiver swap.
\item Let $SV(t;K)$ denote the value at time t of the swap with a fixed rate
K on tenor $[T_{n},T_{N+1}]$ with unit notional amount where
$T_{n-1}<t\leq T_n$. In other words, $SV(t;K)$ denotes the value of
the swap that has the closet starting date among all the co-terminal
swaps with the same fixed rate K.

\item Let $DSN_n(t;K)$ denote the value at time t of a Digital
swaption expiring at time $T_n$ with strike K on a swap on tenor
$[T_{n},T_{N+1}]$ with unit notional amount. A Digital swaption pays
the amount of $P_n(T_n)$ if it expires in the money and nothing
otherwise. It can be expressed under the forward measure $Q^{n,N+1}$
by using the martingale property,
\begin{eqnarray}
\nonumber DSN_n(t;K) & = &
P_n(t)\mathbb{E}_{t}^{n,N+1}[\frac{DSN_n(T_n;K)}{P_n(T_n)}] \\
\nonumber & = & P_n(t)\mathbb{E}_{t}^{n,N+1}[\frac{P_{n}(T_n)
I_{\{\varphi (S_n(T_n)- K)>0\}}}{P_n(T_n)}] \\
& = & P_n(0)\mathbb{E}_{t}^{n,N+1}[ I_{\{\varphi(S_n(T_n)-K)>0\}}]
\nonumber \\& = & P_n(t)\int_{-\infty}^{\infty} I_{\{ \varphi (y -
K)>0\}}\phi(y) dy \label{eq:DSN},
\end{eqnarray}
where $\phi$ denotes the probability density function under
$Q^{n,N+1}$ of the terminal swap rate $S_n(T_n)$ and $\varphi$ is 1
for a payer Digital swaption and -1 for a receiver one. Assuming a
lognormal distribution, we derive the Black formula of a Digital
swaption by some algebra
\begin{equation} \label{eq:DSN_BS}
DSN_n(t;K)=P_n(t)\Phi(\varphi\frac{\log(S_{n}(t)/K)-\frac{1}{2}\bar{\sigma}_n^2
(T_n-t)}{\bar{\sigma}_n\sqrt{T_n-t}}),
\end{equation}
where $\bar{\sigma}_n$ is the swaption's implied volatility.

\item Let $ESN_n(t;K)$ denote the value at time t of a European swaption
expiring at time $T_n$ with strike K on a swap with tenor
$[T_{n},T_{N+1}]$ and unit notional amount. It can be expressed
under the forward measure $Q^{n,N+1}$ by the martingale property,
\begin{eqnarray}
\nonumber ESN_n(t;K) & = &
P_n(t)\mathbb{E}_{t}^{n,N+1}[\frac{ESN_n(T_n;K)}{P_n(T_n)}] \\
\nonumber & = & P_n(t)\mathbb{E}_{t}^{n,N+1}[\frac{SV_n(t;K)
I_{\{\varphi (S_n(T_n)-K)>0 \}}}{P_n(T_n)}] \\
\nonumber & = & P_n(t)\mathbb{E}_{t}^{n,N+1}[\frac{P_{n}(T_n)\varphi
(S_n(T_n)-K)
I_{\{\varphi (S_n(T_n)-K)>0\}}}{P_n(T_n)}] \\
\nonumber & = & P_n(t)\mathbb{E}_{t}^{n,N+1}[\varphi(S_n(T_n)-K)
I_{\{\varphi (S_n(T_n)- K)>0\}}] \\
& = & P_n(t)\int_{-\infty}^{\infty}\varphi(y-K) I_{\{ \varphi(y - K)
>0\}}\phi(y) dy, \label{eq:ESN}
\end{eqnarray}
where $\phi$ denotes the probability density function of $S_n(T_n)$,
and $\varphi$ is 1 for a payer European swaption and -1 for a
receiver one. Assuming a lognormal distribution, we have the Black
formula\footnote{For derivation, we refer to Chapter 25 of Bjork
\cite{Bjork 2003}.} for a European swaption
\begin{equation} \label{eq:European}
ESN_n(t;K) = \varphi P_n(t)(S_n(t)\Phi(\varphi d_+)-K\Phi(\varphi
d_-)),
\end{equation}
where
\begin{equation}
d_{\pm} =
\frac{\log(\frac{S_{n}(t)}{K})\pm\frac{1}{2}\bar{\sigma}_n^2
(T_n-t)}{\bar{\sigma}_n\sqrt{T_n-t}}. \nonumber
\end{equation}

It should be noted that by differentiating Equation \ref{eq:ESN}
with respect to strike $K$, we get the Digital swaption's value in
Equation \ref{eq:DSN}, \emph{i.e.},
\begin{equation} \label{eq:first_dev}
DSN_n(t;K)=-\varphi\frac{\partial ESN_n(t;K)}{\partial K}.
\end{equation}
It's important to realize that we can only observe the implied
volatility quotes of European swaptions in the market. The values of
Digital swaptions are uniquely implied from their European
counterparts by the no-arbitrage principle, which is
model-independent.
\\ \\
If we further differentiate Equation \ref{eq:DSN} with respect to
$K$, we get the underlying's probability density function under
$Q^{n,N+1}$, \emph{i.e.},
\begin{equation} \label{eq:second_dev}
\phi(K)=\frac{\partial DSN_n(t;K)}{\partial K}.
\end{equation}
A useful idea is that rather than from the European options, we
may equivalently imply the underlying's density from their Digital
counterparts, which is one step less complicated.

\item Let $BSN(t;K)$ denote the value at time t of a co-terminal Bermudan
swaption with strike K on a swap with a unit notional amount. A
co-terminal Bermudan entitles the swaption holder to enter on
several predetermined dates into a swap that ends at a fixed
maturity date.
\end{itemize}

\section{Simplification of Notation} \label{sec:notation2}
As we are interested only in the discrete time points of tenor
structure $T_{1}, T_{2}, ...,T_{N+1}$, we apply some simplification
to make our math expressions more compact. We first simplify
$X(T_n)$ as follows.
\begin{equation}
X_n\triangleq X(T_n),
\end{equation}
where n=1, 2, \ldots, N+1. We denote a certain value of $X_n$ by
$x_n$. Moreover we omit the time parameter for a state variable, for
example,
\begin{equation}
D_k(X_n)\triangleq D_{k}(T_n,X_n)=D_{k}(T_n,X(T_n)),
\end{equation}
or for a sample point
\begin{equation}
D_k(x_n)\triangleq D_{k}(T_n,x_n),
\end{equation}
where $k\geq n$, since time is always a parameter of a state
variable by default. Likewise, we have
\begin{eqnarray}
L_n(X_n) & \triangleq & L_{n}(T_n,X(T_n))\nonumber\\
P_{n}(X_n) & \triangleq & P_{n}(T_n,X(T_n))\nonumber\\
S_{n}(X_n) & \triangleq & S_{n}(T_n,X(T_n))\\
SV_n(X_n;K) & \triangleq & SV_n(T_n,X(T_n);K)\nonumber\\
SV(X_n;K) & \triangleq & SV(T_n,X(T_n);K)\nonumber\\
DSN_n(X_n;K) & \triangleq & DSN_n(T_n,X(T_n);K)\nonumber\\
ESN_n(X_n;K) & \triangleq & ESN_n(T_n,X(T_n);K)\nonumber\\
BSN(X_n;K)& \triangleq & BSN(T_n,X(T_n);K).\nonumber
\end{eqnarray}

\section{The Greeks} \label{sec:greek}
If $V(t,S_t)$ is the value of a derivative, where $S_t$ denotes the
value of the underlying, the definition of some sensitivity ratios
of $V(t,S_t)$ are listed below. They are denoted by Greek letters
(except Vega) by convention.
\begin{eqnarray}
\mbox{Delta} :& \Delta \triangleq & \frac{\partial V(t,s)}{\partial s}\\
\mbox{Gamma} :& \Gamma \triangleq & \frac{\partial^2 V(t,s)}{\partial s^2}\\
\mbox{Theta} :& \Theta \triangleq & \frac{\partial V(t,S_t)}{\partial t}\\
\mbox{Vega}  :& \mathcal{V} \triangleq & \frac{\partial
V(t,S_t)}{\partial \sigma}. \label{eq:vega_def}
\end{eqnarray}
For Vega's definition in Equation \ref{eq:vega_def}, $\sigma$
denotes the volatility of the underlying.

\chapter{Integration of Polynomials against Gaussians} \label{chapter:Gaussians}

This appendix closely follows Pelsser \cite{Pelsser
1998-b}\cite{Pelsser 2000}. The numerical integration discussed here
is based on the following idea: \\
1. fit a polynomial to the payoff function defined on the grid; \\
2. calculate analytically the integral of the polynomial against the
Gaussian distribution.
\\ \\
{\bf Fitting a Polynomial.} Given a number of points $x_i$ and a set
of function values $f_i$, a polynomial that fits through these
values can be computed recursively using Neville's algorithm. Let
$P_{(i)\ldots(i+m)}$ denote the polynomial defined using the points
$x_i,\ldots,x_{i+m}$. Then the following relationship
holds\footnote{For more details of Neville's algorithm, please refer
to Section 3.1 of "Numerical Recipes in C++" \cite{NR 2002}.}
\begin{equation} \label{eq:Neville}
P_{(i)\ldots(i+m)} =
\begin{cases}
\frac{(x-x_{i+m})P_{(i)\ldots(i+m-1)}+(x_i-x)P_{(i+1)\ldots(i+m)}}{x_i-x_{i+m}}
& \mbox{if $m\geq 1$} \\
\;\;\;\;\; f_i & \mbox{if $m=0$}
\end{cases}
\end{equation}
where $m$ is the order for polynomial fitting. Each polynomial can
be expressed as $P_{(i)\ldots(i+m)}=\sum_{k=0}^m c_{i,k}x^k$. Using
Equation \ref{eq:Neville} we can then derive a recurrence formula
for the coefficients $c_{i,k}$ as follows
\begin{eqnarray}
c_{i,m}&=&\frac{c_{i,m-1}-c_{i+1,m-1}}{x_i-x_{i+m}} \nonumber\\
c_{i,k}&=&\frac{x_i c_{i+1,k} - x_{i+m}c_{i,k} + c_{i,k-1} -
c_{i+1,k-1}}{x_i-x_{i+m}}  \;\;\;\;\;\;\;   1 \leq k \leq m-1 \\
c_{i,0}&=&\frac{x_i c_{i+1,0}-x_{i+m} c_{i,0}}{x_i-x_{i+m}}
\nonumber
\end{eqnarray}
\\
{\bf Integrating against Gaussian.} The Markov process $x$ defined
in Equation \ref{eq:MF} has Gaussian density functions. Hence, the
calculation of integrals against a Gaussian density can be reduced
to evaluating for different powers $x^k$ of the polynomial $P$ the
following integral
\begin{equation}
G(k;h,\mu,\sigma)=\int_{-\infty}^h x^k
\frac{exp\{-\frac{1}{2}(\frac{x-\mu}{\sigma})^2\}}{\sigma \sqrt{2
\pi}}dx
\end{equation}
\\
Using partial integration, we derive the following recurrence
relation for $G$ in terms of $k$
\begin{eqnarray}
G(k)&=&\mu G(k-1) + (k-1)\sigma^2 G(k-2) - \sigma^2 h^{k-1}
\frac{exp\{-\frac{1}{2}(\frac{x-\mu}{\sigma})^2\}}{\sigma \sqrt{2
\pi}} \;\;\; k\geq 1\\
G(0)&=& N(\frac{h-\mu}{\sigma}) \;\;\;\; \mbox{and} \;\;\;\;
G(-1)\;\;=\;\; 0\nonumber
\end{eqnarray}
where $N(.)$ denotes the standard normal distribution function.

\begin{figure}[h!]
\centering \caption{Bad polynomial fit} \label{fig:bad_poly_fit}
\centering
\includegraphics[width=100mm,height=70mm]{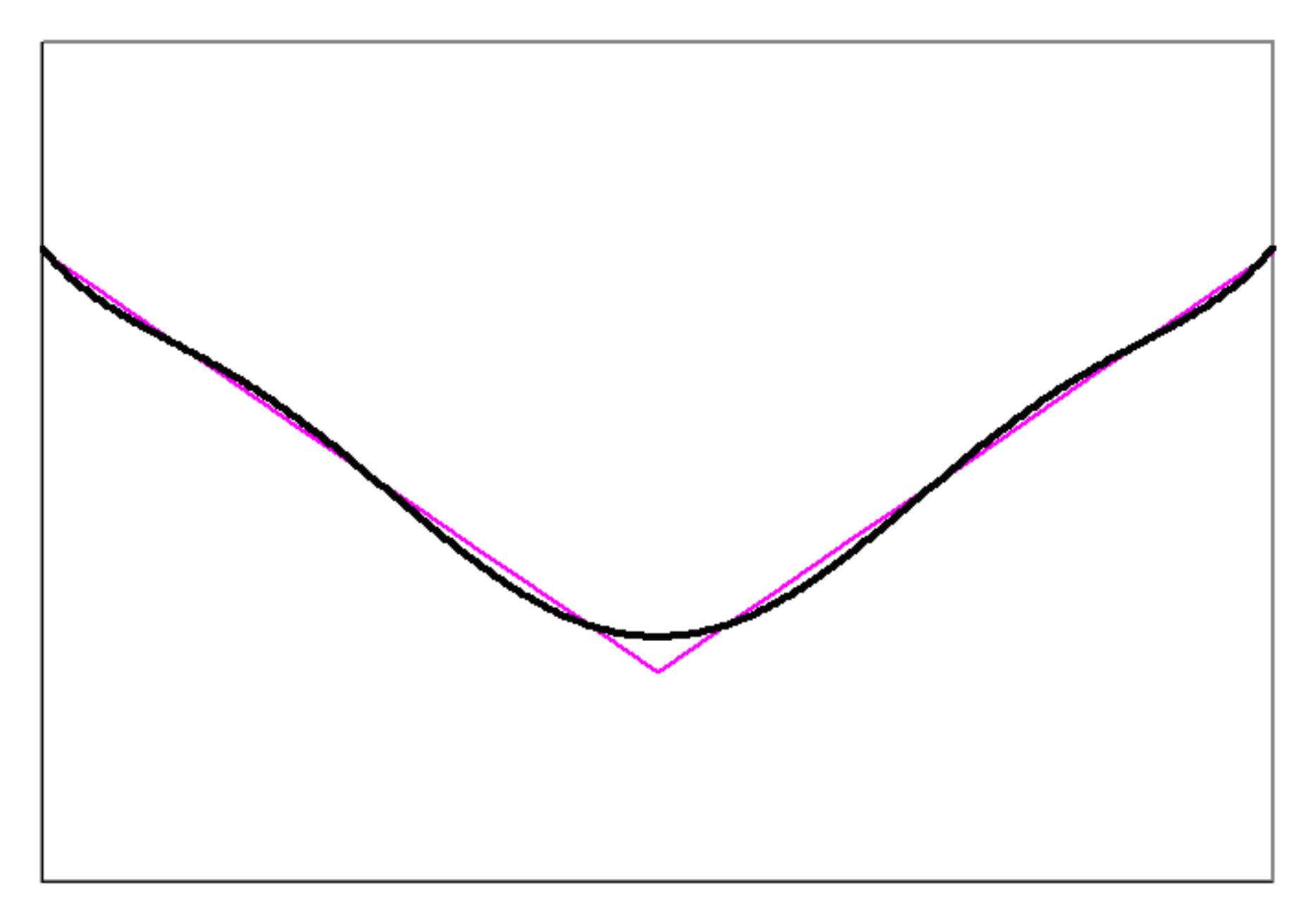}
\end{figure}

{\bf Calculating Expected Values.} Given a grid on which we are
working, option values are calculated by taking expectations of the
value function against the Gaussian density. Given that we have
calculated several option values at time $T_{n+1}$ at grid points
$x_j$, we want to calculate option vales at time $T_n$ for grid
points $x_i$. To do this we proceed as follows.
\begin{itemize}
\item Given an order $M$ the approximating polynomial $P(j-M/2)\ldots(j+1+M/2)$
for the interval $[x_j,x_{j+1}]$ is fitted through the points
$x_{j-M/2},\ldots,x_{j+1+M/2}$, where $M/2$ denotes integer division
($M$ div 2). The approximating polynomial has coefficients
$c_{j,k}$.
\item The integral over the approximating polynomial for the interval
$[x_j,x_{j+1}]$ can now be expressed as
\begin{eqnarray}
\sum_{k=0}^M
c_{j,k}\{\;G(\;k;x_{j+1},x_i,\sqrt{Var(X(T_{n+1})|X(T_n)=x_i)}\;) \nonumber\\
-\; G(\;k;x_j,x_i,\sqrt{Var(X(T_{n+1})|X(T_n)=x_i)}\;)\;\}
\end{eqnarray}
where
$Var(X(T_{n+1})|X(T_n)=x_i)=\int_{T_n}^{T_{n+1}}\tau^2(u)du$.\footnote{For
derivation, please refer to Appendix \ref{chapter:derivation}}
\item The option value in grid point $x_i$ will now be given by summing
over all intervals $[x_j,x_{j+1}]$:
\begin{eqnarray}
\sum_j\sum_{k=0}^M
c_{j,k}\{\;G(\;k;x_{j+1},x_i,\sqrt{Var(X(T_{n+1})|X(T_n)=x_i)}\;) \nonumber\\
-\; G(\;k;x_j,x_i,\sqrt{Var(X(T_{n+1})|X(T_n)=x_i)}\;)\;\}
\end{eqnarray}
\item Loop over all points $x_i$.
\end{itemize}

The fitting of the polynomials works well if the function that one
wants to approximate is smooth. However, many option payoffs are
determined as the maximum of two functions. This implies that the
payoff function will be smooth except at the crossover point where
the payoff function may have a kink. Since polynomials are "stiff"
they will fit a function with a kink very poorly. This is shown in
Figure \ref{fig:bad_poly_fit}.\footnote{Figure
\ref{fig:bad_poly_fit} is taken from Pelsser \cite{Pelsser 1998-c}.}
The way this problem can be solved is to fit polynomials to both
underlying functions, and to split the integration interval at the
crossover point, using the appropriate polynomial on either side of
the crossover point.

\chapter{Some Derivations} \label{chapter:derivation}

\section{Derivation of Equation \ref{eq:density} in Section
\ref{sec:assumption}} \label{sec:derivation1}

By Equation \ref{eq:MF} and setting $X_0=0$, we have
\begin{eqnarray}
X(s)
&=& \int_0^s \tau(u)dW^{N+1}_u \nonumber\\
&=& \int_0^t \tau(u)dW^{N+1}_u + \int_t^s
\tau(u)dW^{N+1}_u \nonumber \\
&=& X(t)+\int_t^s \tau(u)dW^{N+1}_u \label{eq:Xt},
\end{eqnarray}
$\tau(u)$ is a deterministic function, so $\int_t^s
\tau(u)dW^{N+1}_u$ is normally distributed. This results in that
conditional on $X(t)=x_{t}$, $X(s)$ is normally distributed with
mean equal to
\begin{eqnarray} \nonumber
\mathbb{E}(X(s)|X(t)=x_{t}) &=& \mathbb{E}(X(t)+\int_t^s \tau(u)dW^{N+1}_u|X(t)=x_{t}) \\
&=& x_{t}+\mathbb{E}(\int_t^s\tau(u)dW^{N+1}_u|X(t)=x_{t}) \nonumber\\
&=& x_{t}+ 0 =x_{t},
\end{eqnarray}
and variance equal to
\begin{eqnarray} \nonumber
Var(X(s)|X(t)=x_{t}) &=& Var(X_t+\int_t^s \tau(u)dW^{N+1}_u|X(t)=x_{t}) \\
\nonumber &=& 0 + Var(\int_t^s\tau(u)dW^{N+1}_u|X(t)=x_{t}) \\
\nonumber &=& \mathbb{E}[(\int_t^s\tau(u)dW^{N+1}_u)^2|X(t)=x_{t}] -
[\mathbb{E}(\int_t^s\tau(u)dW^{N+1}_u|X(t)=x_{t})]^2\\
\nonumber &=& \mathbb{E}[\int_t^s\tau^2(u)du|X(t)=x_{t}] -
0^2\\
&=& \int_t^s\tau^2(u)du.
\end{eqnarray}
Thus the probability density function of $X(s)$ given $X(t)=x_{t}$
is
\begin{equation}
\phi(X(s)|X(t)=x_{t})=\frac{exp(-\frac{1}{2}\frac{(X(s)-x_{t})^2}{\int_{t}^{s}\tau^{2}(u)du})}
{\sqrt{2\pi\int_{t}^{s}\tau^{2}(u)du}}.
\end{equation}
Besides, the Markov process $X(t)$ in this choice is
time-inhomogeneous as
\begin{eqnarray}
\nonumber \phi[X(s)|X(t)=x] & = &
\frac{exp(-\frac{1}{2}\frac{(X(s)-x)^2}{\int_{t}^{s}\tau^{2}(u)du})}
{\sqrt{2\pi\int_{t}^{s}\tau^{2}(u)du}}
\\
& \neq & \phi[X(s-t)|X_0=x]=
\frac{exp(-\frac{1}{2}\frac{(X(s-t)-x)^2}{\int_{0}^{s-t}\tau^{2}(u)du})}
{\sqrt{2\pi\int_{0}^{s-t}\tau^{2}(u)du}}.
\end{eqnarray}
\\
\section{Derivation of Equation \ref{eq:HW_MR} and \ref{eq:MR} in
Section \ref{sec:MR}} \label{sec:derivation2}

{\bf Derivation of Equation \ref{eq:HW_MR} in Section \ref{sec:MR}}
\\ \\
Multiplying both sides of Equation \ref{eq:HW} by $e^{at}$ and
re-arranging it, we can get
\begin{equation}
e^{at}\theta(t)dt+ e^{at}\sigma dW_t = e^{at} dr_t+a e^{at} r_t dt =
d(e^{at}r_t).
\end{equation}
Integrating both sides from $0$ to $t$, we have
\begin{equation}
e^{at}r_t - r_0 = \int_0^t e^{au}\theta(u)du+ \sigma\int_0^t
e^{au}dW_u,
\end{equation}
that is,
\begin{equation}
r_t = e^{-at}r_0 + e^{-at}\int_0^t e^{au}\theta(u)du+
e^{-at}\sigma\int_0^t e^{au} dW_u.
\end{equation}
Thus we have, for $t<s$,
\begin{eqnarray}
Corr(r(t),r(s))
&=& \frac{Cov(r(t),r(s))}{\sqrt{Var(r(t))Var(r(s))}}\nonumber\\
&=& \frac{E[(e^{-at}\sigma\int_0^t e^{au}dW_u)(e^{-as}\sigma\int_0^s
e^{au}dW_u)]}{\sqrt{E[(e^{-at}\sigma\int_0^t
e^{au}dW_u)^2]E[(e^{-as}\sigma\int_0^s e^{au}dW_u)^2]}}\nonumber\\
&=& \sqrt{\frac{E[(\int_0^t e^{au}dW_u)^2]}{E[(\int_0^s
e^{au}dW_u)^2]}} \nonumber\\
&=& \sqrt{\frac{\int_0^t (e^{au})^2 du}{\int_0^s (e^{au})^2 du}}\nonumber\\
&=& \begin{cases} \sqrt{\frac{t}{s}}
& \mbox{if $a=0$}\\
\sqrt{\frac{e^{2at}-1}{e^{2as}-1}} & \mbox{if $a\neq 0$}
\end{cases}.
\end{eqnarray}
\\
{\bf Derivation of Equation \ref{eq:MR} in Section
\ref{sec:MR}}
\\ \\
By Equation \ref{eq:Xt} and \ref{eq:MF_vol}, we have, for $t<s$,
\begin{eqnarray}
Corr(X(t),X(s))
&=& \frac{Cov(X(t),X(s))}{\sqrt{Var(X(t))Var(X(s))}}\nonumber\\
&=& \frac{Cov(\int_0^t e^{au}dW_u,\int_0^s e^{au}dW_u)}
{\sqrt{Var(\int_0^t e^{au}dW_u) Var(\int_0^s e^{au}dW_u)}} \nonumber\\
&=& \frac{E[\int_0^t e^{au}dW_u \int_0^t e^{au}dW_u]}
{\sqrt{E[(\int_0^t
e^{au}dW_u)^2] E[(\int_0^s e^{au}dW_u)^2]}} \nonumber\\
&=& \sqrt{\frac{E[(\int_0^t e^{au}dW_u)^2]}{E[(\int_0^s e^{au}dW_u)^2]}} \nonumber\\
&=& \sqrt{\frac{\int_0^t (e^{au})^2 du}{\int_0^s (e^{au})^2
du}}\nonumber\\
&=& \begin{cases} \sqrt{\frac{t}{s}}
& \mbox{if $a=0$}\\
\sqrt{\frac{e^{2at}-1}{e^{2as}-1}} & \mbox{if $a\neq 0$}
\end{cases}.
\end{eqnarray}

\chapter{Near-the-money Bermudan Swaption Prices Affected by More Pronounced Smiles}
\label{chapter:negative_vega}

\begin{figure}[hp!]
\label{fig:fatter_LIBOR1} \centering
\includegraphics[width=120mm,height=80mm]{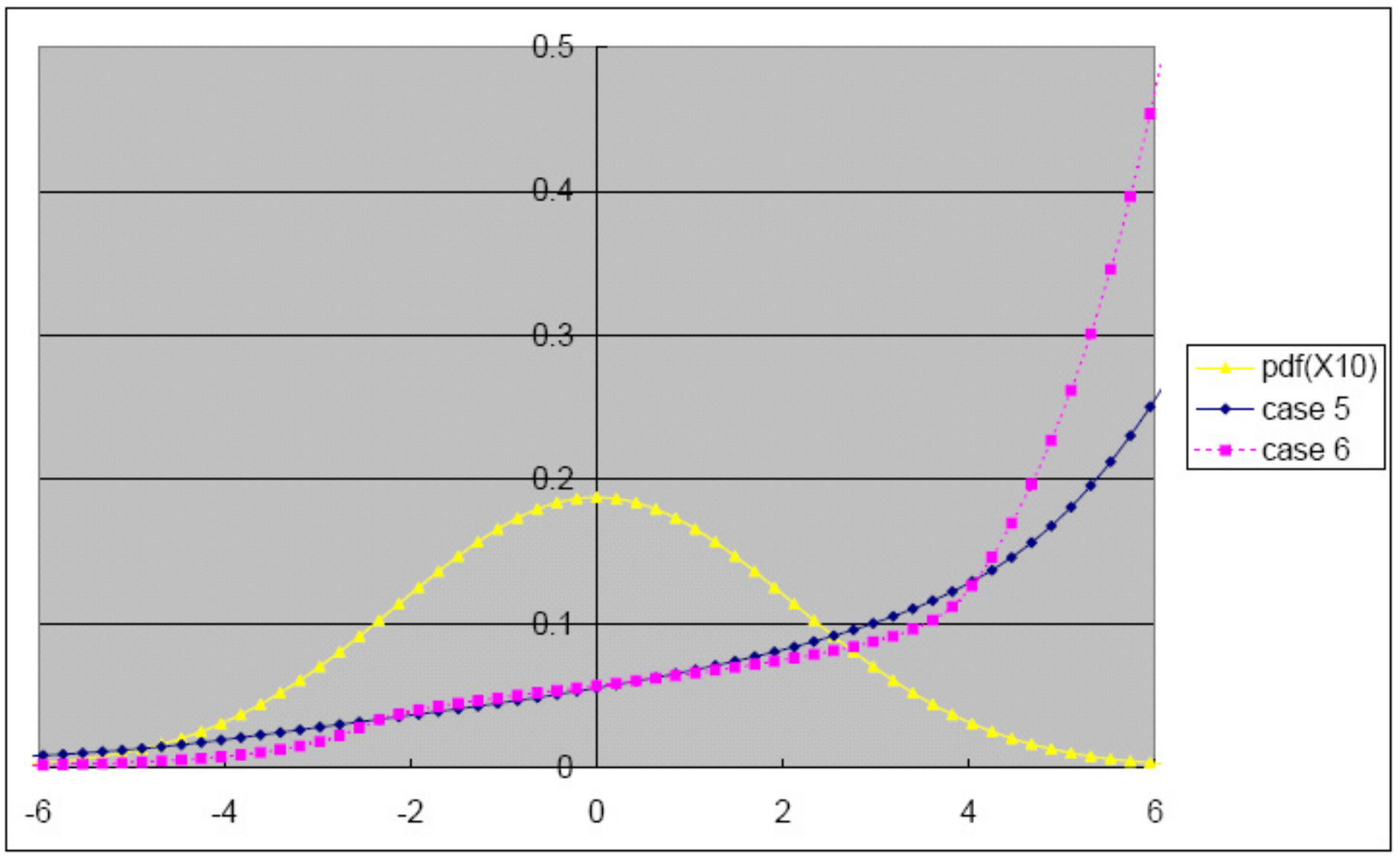}
\centering \caption{$L_{10}(X_{10})$ in case 5 and 6.}
\end{figure}

\begin{figure}[hp!]
\label{fig:fatter_LIBOR2} \centering
\includegraphics[width=120mm,height=80mm]{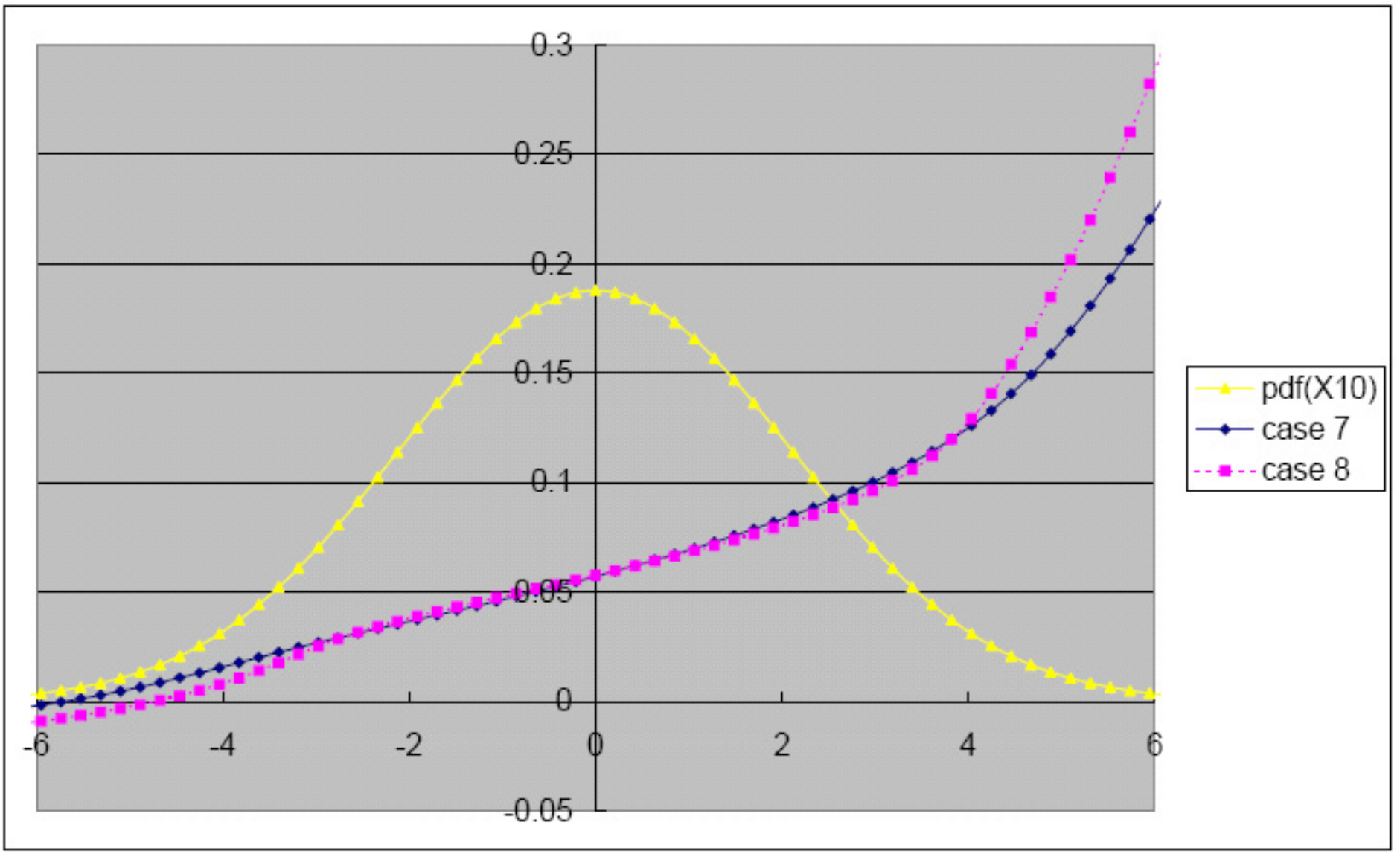}
\centering \caption{$L_{10}(X_{10})$ in case 7 and 8.}
\end{figure}

This appendix gives an explanation of the issue discussed in Section
\ref{sec:Bermudan_price} by considering an example. More precisely,
more pronounced smiles, \emph{i.e.}, distributions of the underlying
swap rates with fatter tails (both left side and right side), may
result in either a higher or lower near-the-money Bermudan value.
Recall Equation \ref{eq:terminal_swaprate} and that by assuming
$\alpha_k=1$, for $k=n,\cdots,N$, we have
\begin{eqnarray}
S_n(X_n) &=&\frac{1-D_{N+1}(X_n)}{\sum_{k=n+1}^{N+1}D_k(X_n)} \nonumber\\
&=&\frac{1-D_{N+1}(X_n)}{D_{n+1}(X_n)+\sum_{k=n+2}^{N+1}D_k(X_n)} \nonumber\\
&=&\frac{1-D_{N+1}(X_n)}{\frac{1}{1+L_n(X_n)}+\sum_{k=n+2}^{N+1}D_k(X_n)}.
\end{eqnarray}
Rearranging it we have
\begin{equation} \label{eq:LIBOR_swap}
L_n(X_n) =
\frac{1}{\frac{1-D_{n+1}(X_n)}{S_n(X_n)}-{\sum_{k=n+2}^{N+1}D_k(X_n)}}
-1.
\end{equation}
By observation of Equation \ref{eq:LIBOR_swap} we {\bf claim} that
in MF, distributions of swap rates $S_n(X_n)$ with fatter tails
\emph{may} result in fatter-tailed distributions of LIBOR rates
$L_n(X_n)$, for $n=1,\cdots,N$.
\\ \\
An evidence of this claim is shown in Figure \ref{fig:fatter_LIBOR1}
and \ref{fig:fatter_LIBOR2}. In Figure \ref{fig:fatter_LIBOR1}, we
plot $L_{10}(X_{10})$ in case 5 and 6 mentioned in Section
\ref{sec:Bermudan_price}, respectively, together with the
probability distribution function of $X_{10}$. Smiles in case 6 are
more pronounced than in case 5. We see from the figure that the
distribution of $L_{10}(X_{10})$ in case 6 has fatter tails than in
case 5. If we plot the similar graphs with respect to $L_n(X_n)$,
for $n=1,\cdots,9$, we would reach the same conclusion. Figure
\ref{fig:fatter_LIBOR2} is with respect to case 7 and 8 mentioned in
Section \ref{sec:Bermudan_price}, where we can reach exactly the
same conclusions as above.

\newpage

Based on the validity of the claim above, we make one extremely
simplified example. In this example, there are only two floating
reset dates $T_1$ and $T_2$. Each tenor's length $\alpha_n$, for
$n=1,2$, is one year. $T_1$ is one year ahead of "now", \emph{i.e.}, $T_0$.
At each $T_n$, for $n=1,2$, there are only three states, shown in
Table \ref{table:example_xgrid}. These three states represent the
deep ITM, near-the-money and deep OTM scenarios, respectively. The
transition probabilities are assigned below:
\begin{eqnarray}
\mathbb{P}(x0\rightarrow x_{1,j}) &=&\frac{1}{3},\;\;\;\mbox{for $j=-1,0,1$}\nonumber\\
\mathbb{P}(x_{1,i}\rightarrow x_{2,j}) &=& \frac{1}{3},\;\;\;
\mbox{for $i,j=-1,0,1$}. \nonumber
\end{eqnarray}
\\

\begin{table}[hp!]
\begin{center}
    \begin{tabular}{|c|c|c|}
        \hline
                $T_0$ & $T_1$ & $T_2$  \\
        \hline
                      & $x_{1,1}$  & $x_{2,1}$ \\
        \hline
                $x_0$ & $x_{1,0}$  & $x_{2,0}$ \\
        \hline
                      & $x_{1,-1}$ & $x_{2,-1}$ \\
        \hline
    \end{tabular}
\caption{Example - lattice of $X_n$.} \label{table:example_xgrid}
\end{center}
\end{table}

The corresponding LIBOR tree is shown in Table
\ref{table:example_LIBOR_tree}. We are dealing with a payer Bermudan
with the right to exercise at $T_1$ and $T_2$. The notional is
10000. The swap value tree is computed in Table
\ref{table:example_swap_tree}. In Table
\ref{table:example_option_tree}, we follow the principle in Section
\ref{sec:pricing_general} to calculate the Bermudan value with the
strike level $5.50\%$, which is near the money. The Bermudan price
is 122.49. With similar calculations, We also get the Bermudan
prices of 250.43 and 44.93 for the strike level $4.50\%$ (ITM) and
$6.50\%$ (OTM), respectively.

\begin{table}[hp!]
\begin{center}
    \begin{tabular}{|c|c|c|}
        \hline
                $T_0$ & $T_1$ & $T_2$  \\
        \hline
                                  & $L_1(x_{1,1})=7.00\%$  & $L_2(x_{2,1})=8.00\%$ \\
        \hline
                $L_0(x_0)=5.50\%$ & $L_1(x_{1,0})=5.50\%$  & $L_2(x_{2,0})=6.00\%$ \\
        \hline
                                  & $L_1(x_{1,-1})=4.00\%$ & $L_2(x_{2,-1})=4.00\%$ \\
        \hline
    \end{tabular}
\caption{Example - LIBOR tree.} \label{table:example_LIBOR_tree}
\end{center}
\end{table}

\begin{table}[hp!]
\begin{center}
    \begin{tabular}{|l | l|}
        \hline
       $SV(x_{1,1})$ =   & $SV(x_{2,1})$ =  \\
       $10000\times(7.00\%-5.50\%)+\frac{(250+50-150)}{3\times(1+7.00\%)}$
       & $10000\times(8.00\%-5.50\%)$        \\
                    = 196.73      & = 250.00\\

        \hline
       $SV(x_{1,0})$=  & $SV(x_{2,0})$ =  \\
       $10000\times(5.50\%-5.50\%)+\frac{(250+50-150)}{3\times(1+5.50\%)}$
       & $10000\times(6.00\%-5.50\%)$ \\
       =47.39    &= 50.00\\
        \hline
       $SV(x_{1,-1})$=  & $SV(x_{2,-1})$= \\
        $10000\times(4.00\%-5.50\%)+\frac{(250+50-150)}{3\times(1+4.00\%)}$
        & $10000\times(4.00\%-5.50\%)$ \\
        = -101.92 &= -150.00\\
        \hline
    \end{tabular}
\caption{Example - payer swap value tree for $K=5.50\%$.}
\label{table:example_swap_tree}
\end{center}
\end{table}

\begin{table}[h!]
\begin{center}
    \begin{tabular}{|l |l | l|}
        \hline
        &$BSN(x_{1,1})$= & $BSN(x_{2,1})$= \\
        &$Max(196.73,\frac{(250+50-150)}{3\times(1+7.00\%)})$ &
        Max(250.00,0)\\
        & = 196.73            &= 250.00\\
        \hline
    $BSN(x_0)$= &   $BSN(x_{1,0})$=  &$BSN(x_{2,0})$=  \\
    $\frac{(196.73+94.79-96.15)}{3\times(1+5.50\%)}$ &
    $Max(47.39,\frac{(250+50-150)}{3\times(1+5.50\%)})$ & Max(50.00,0)\\
             = 122.49        & = 94.79           &= 50.00\\
        \hline
       &$BSN(x_{1,-1})$=  & $BSN(x_{2,-1})$= \\
        &$Max(-101.92,\frac{(250+50-150)}{3\times(1+4.00\%)})$ &
        Max(-150.00,0)\\
       &               = 96.15       &= -150.00\\
        \hline
    \end{tabular}
\caption{Example - payer Bermudan swaption value tree for
$K=5.50\%$.} \label{table:example_option_tree}
\end{center}
\end{table}

\newpage

Now we have in Table \ref{table:example_LIBOR_tree2} a LIBOR tree
with fatter tails than the original one in Table
\ref{table:example_LIBOR_tree}. Following a similar line above, we
get the Bermudan prices of 252.52, 125.08 and 46.43 for the strike
level $4.50\%$ (ITM), $5.50\%$ (near-the-money) and $6.50\%$ (OTM),
respectively. In this case, more pronounced smiles result in a
higher near-the-money Bermudan price.
\begin{table}[h!]
\begin{center}
    \begin{tabular}{|c|c|c|}
        \hline
                $T_0$ & $T_1$ & $T_2$  \\
        \hline
                                  & $L_1(x_{1,1})=7.05\%$  & $L_2(x_{2,1})=8.05\%$ \\
        \hline
                $L_0(x_0)=5.50\%$ & $L_1(x_{1,0})=5.50\%$  & $L_2(x_{2,0})=6.00\%$ \\
        \hline
                                  & $L_1(x_{1,-1})=3.95\%$ & $L_2(x_{2,-1})=3.95\%$ \\
        \hline
    \end{tabular}
\caption{Example - LIBOR tree of fatter-tailed distribution (A).}
\label{table:example_LIBOR_tree2}
\end{center}
\end{table}

However we have in Table \ref{table:example_LIBOR_tree3} another
LIBOR tree which also has fatter tails than the original one. We
instead get the Bermudan prices of 252.65, 122.06 and 46.41 for the
strike level $4.50\%$ (ITM), $5.50\%$ (near-the-money) and $6.50\%$
(OTM), respectively. In this case, more pronounced smiles result in
a lower near-the-money Bermudan price.

\begin{table}[h!]
\begin{center}
    \begin{tabular}{|c|c|c|}
        \hline
                $T_0$ & $T_1$ & $T_2$  \\
        \hline
                                  & $L_1(x_{1,1})=7.05\%$  & $L_2(x_{2,1})=8.05\%$ \\
        \hline
                $L_0(x_0)=5.50\%$ & $L_1(x_{1,0})=5.60\%$  & $L_2(x_{2,0})=5.90\%$ \\
        \hline
                                  & $L_1(x_{1,-1})=3.95\%$ & $L_2(x_{2,-1})=3.95\%$ \\
        \hline
    \end{tabular}
\caption{Example - LIBOR tree of fatter-tailed distribution (B).}
\label{table:example_LIBOR_tree3}
\end{center}
\end{table}

From this example, we see that whether more pronounced smiles result
in a higher or lower near-the-money Bermudan price may be subject to
how the near-the-money LIBOR probability mass is influenced.

\chapter{Market Data and Specification of Test Trades}

\section{Market Data Used in the Numerical Tests}
All the data sets in this report are chosen arbitrarily.

\subsection{Data Set I} \label{sec:DataSet1}

The yield curve\footnote{The currency of the yield curve and
implied volatility surface in Data Set I is unknown. We append
this data set so that every result based on it can be reproduced.}
in this data set is listed in Table \ref{table:yield_curve1}. For
example, the fifth row of the table represents the bid/ask
discount factor of 367 days from today (July 9th, 2002). Any
required discount factor not available in the table was calculated
by linear interpolation.
\begin{table}[h!]
\begin{center}
    \begin{tabular}{c| c}
        \hline
            Days & Bid / Ask \\
        \hline
        34  & 0.998367115 / 0.998367115 \\
        94 & 0.995269154 / 0.995269154 \\
        188 & 0.990025493 / 0.990025493 \\
        367 & 0.977629093 / 0.977629093 \\
        735 & 0.938822503 / 0.938822503 \\
        1098 & 0.893023545 / 0.893023545 \\
        1463 & 0.84517874 / 0.84517874 \\
        1828 & 0.796865431 / 0.796865431 \\
        2562 & 0.703583273 / 0.703583273 \\
        3655 & 0.5784443 / 0.5784443 \\
        5481 & 0.40916987 / 0.40916987 \\
        10961 & 0.152839928 / 0.152839928 \\
        \hline
    \end{tabular}
\caption{Yield curve of July 9th, 2002.} \label{table:yield_curve1}
\end{center}
\end{table}

The implied volatility surface (tenor length, expiry) of ATM
European swaptions in this data set is listed in Table
\ref{table:volatility surface1}. The value of each volatility is
the average of the original bid and ask values. Any required
volatility not available in the table was calculated by linear
surface interpolation.

\begin{table}[h!]
\begin{center}
    \begin{tabular}{c| c c c c c c c c c}
    \hline
         & \multicolumn{9}{c}{Exp. (Days)} \\
         Tenor (Days) & 32 & 63 & 92 & 182 & 360 & 730 & 1095 & 1463 & 1827 \\
    \hline
         360 & 0.457 & 0.4455 & 0.434 & 0.379 & 0.333 & 0.261 & 0.239 & 0.219 & 0.204 \\
         720 & 0.39 & 0.3805 & 0.371 & 0.338 & 0.294 & 0.25 & 0.23 & 0.213 & 0.2 \\
        1080 & 0.32 & 0.32 & 0.32 & 0.301 & 0.271 & 0.237 & 0.22 & 0.206 & 0.193 \\
        1440 & 0.29 & 0.29 & 0.29 & 0.279 & 0.255 & 0.228 & 0.213 & 0.2 & 0.188 \\
        1800 & 0.271 & 0.2705 & 0.27 & 0.263 & 0.244 & 0.221 & 0.208 & 0.195 & 0.184 \\
        2520 & 0.25 & 0.2475 & 0.245 & 0.242 & 0.228 & 0.211 & 0.2 & 0.188 & 0.177 \\
        3600 & 0.23 & 0.2265 & 0.223 & 0.223 & 0.213 & 0.201 & 0.19 & 0.178 & 0.168 \\
        5400 & 0.19 & 0.1905 & 0.191 & 0.191 & 0.185 & 0.179 & 0.17 & 0.161 & 0.152 \\
        10800 & 0.19 & 0.173 & 0.156 & 0.155 & 0.152 & 0.152 & 0.143 & 0.136 & 0.129 \\
    \hline
    \end{tabular}
\caption{ATM volatility surface of July 9th, 2002.}
\label{table:volatility surface1}
\end{center}
\end{table}

\subsection{Data Set II} \label{sec:DataSet2}
In this data set, the yield curve of EURO is listed in Table
\ref{table:yield_curve2}. Any required discount factor not
available in the table was calculated by linear interpolation.

\begin{table}[h!]
\begin{center}
    \begin{tabular}{c| c| c| c}
        \hline
            Days & Bid / Ask & Days & Bid / Ask\\
        \hline
            4 & 0.999658 / 0.999658 & 1466 & 0.855901 / 0.855901 \\
            11 & 0.999057 / 0.999057 & 1830 & 0.821377 / 0.821377 \\
            35 & 0.996992 / 0.996992 & 2196 & 0.787538 / 0.787538 \\
            40 & 0.996575 / 0.996575 & 2561 & 0.754289 / 0.754289 \\
            66 & 0.994276 / 0.994276 & 2926 & 0.721887 / 0.721887 \\
            96 & 0.991506 / 0.991506 & 3293 & 0.690247 / 0.690247 \\
            131 & 0.988237 / 0.988237 & 3657 & 0.659733 / 0.659733 \\
            221 & 0.97932 / 0.97932 & 4022 & 0.62999 / 0.62999 \\
            222 & 0.979217 / 0.979217 & 4387 & 0.601379 / 0.601379 \\
            313 & 0.969993 / 0.969993 & 4752 & 0.573949 / 0.573949 \\
            314 & 0.969891 / 0.969891 & 5120 & 0.547413 / 0.547413 \\
            404 & 0.960723 / 0.960723 & 5484 & 0.522205 / 0.522205 \\
            405 & 0.960621 / 0.960621 & 7311 & 0.412457 / 0.412457 \\
            495 & 0.951481 / 0.951481 & 9135 & 0.327908 / 0.327908 \\
            586 & 0.942281 / 0.942281 & 10962 & 0.262923 / 0.262923\\
            677 & 0.933159 / 0.933159 & 14614 & 0.171647 / 0.171647 \\
            678 & 0.933059 / 0.933059 & 18267 & 0.113148 / 0.113148 \\
            769 & 0.923968 / 0.923968 & 21920 & 0.073214 / 0.073214 \\
            1102 & 0.891103 / 0.891103 \\
        \hline
    \end{tabular}
\caption{EURO yield curve of August 11th, 2006.}
\label{table:yield_curve2}
\end{center}
\end{table}

\begin{table}[h!]
\begin{center}
    \begin{tabular}{c| c c c c c c c c c c}
    \hline
         & \multicolumn{10}{c}{Expiry (Days)} \\
         Tenor (Days) & 31 & 94 & 185 & 273 & 367 & 731 & 1096 & 1461 & 1826 & 2194 \\
    \hline
         360 & 0.129 & 0.136 & 0.145 & 0.15 & 0.153 & 0.158 & 0.159 & 0.156 & 0.153 & 0.1485 \\
         720 & 0.137 & 0.141 & 0.147 & 0.151 & 0.153 & 0.156 & 0.156 & 0.154 & 0.15 & 0.1455 \\
         1080 & 0.143 & 0.147 & 0.15 & 0.153 & 0.153 & 0.154 & 0.154 & 0.151 & 0.147 & 0.1425 \\
         1440 & 0.146 & 0.15 & 0.151 & 0.152 & 0.152 & 0.152 & 0.151 & 0.148 & 0.144 & 0.1395 \\
         1800 & 0.146 & 0.151 & 0.151 & 0.151 & 0.15 & 0.15 & 0.148 & 0.145 & 0.141 & 0.1365  \\
         2160 & 0.142 & 0.148 & 0.149 & 0.149 & 0.148 & 0.148 & 0.145 & 0.142 & 0.138 & 0.134 \\
         2520 & 0.139 & 0.145 & 0.146 & 0.146 & 0.146 & 0.145 & 0.143 & 0.139 & 0.136 & 0.1325 \\
         2880 & 0.136 & 0.142 & 0.143 & 0.144 & 0.144 & 0.143 & 0.14 & 0.137 & 0.134 & 0.131 \\
         3240 & 0.132 & 0.139 & 0.14 & 0.141 & 0.141 & 0.141 & 0.139 & 0.136 & 0.132 & 0.129 \\
         3600 & 0.13 & 0.136 & 0.137 & 0.138 & 0.139 & 0.139 & 0.137 & 0.134 & 0.131 & 0.128 \\
         5400 & 0.122 & 0.127 & 0.129 & 0.13 & 0.131 & 0.131 & 0.129 & 0.127 & 0.124 & 0.1215 \\
         7200 & 0.117 & 0.122 & 0.124 & 0.126 & 0.126 & 0.126 & 0.125 & 0.122 & 0.12 & 0.1175 \\
         9000 & 0.114 & 0.118 & 0.121 & 0.122 & 0.123 & 0.123 & 0.122 & 0.12 & 0.117 & 0.115 \\
         10800 & 0.112 & 0.116 & 0.118 & 0.12 & 0.121 & 0.121 & 0.12 & 0.118 & 0.115 & 0.113 \\
         14400 & 0.112 & 0.116 & 0.118 & 0.12 & 0.121 & 0.121 & 0.12 & 0.118 & 0.115 & 0.113 \\
    \hline
    \hline
         & \multicolumn{10}{c}{Expiry (Days)} \\
         Tenor (Days)  & 2558 & 2922 & 3287 & 3653 & 5479 & 7035 & 9131 & 10958 & 14612\\
    \hline
        360 &  0.144 & 0.1397 & 0.1353 & 0.131 & 0.12 & 0.114 & 0.11 & 0.108 & 0.108 \\
        720 & 0.141 & 0.137 & 0.133 & 0.129 & 0.117 & 0.112 & 0.108 & 0.106 & 0.106 \\
        1080 & 0.138 & 0.1343 &0.1307 & 0.127 & 0.117 & 0.111 & 0.108 & 0.105 & 0.105 \\
        1440 & 0.135 & 0.1317 &0.1283 & 0.125 & 0.116 & 0.11 & 0.106 & 0.103 & 0.103 \\
        1800 & 0.132 & 0.129 & 0.126 & 0.123 & 0.114 & 0.109 & 0.105 & 0.102 & 0.102 \\
        2160 & 0.13 & 0.1273 & 0.1247 & 0.122 & 0.113 & 0.108 & 0.104 & 0.102 & 0.102 \\
        2520 & 0.129 & 0.1263 & 0.1237 & 0.121 & 0.113 & 0.108 & 0.104 & 0.102 & 0.102 \\
        2880 & 0.128 & 0.1253 & 0.1227 & 0.12 & 0.112 & 0.108 & 0.104 & 0.102 & 0.102 \\
        3240 & 0.126 & 0.124 & 0.122 & 0.12 & 0.112 & 0.108 & 0.104 & 0.102 & 0.102 \\
        3600 & 0.125 & 0.123 & 0.121 & 0.119 & 0.112 & 0.108 & 0.104 & 0.102 & 0.102 \\
        5400 & 0.119 & 0.117 & 0.115 & 0.113 & 0.106 & 0.101 & 0.099 & 0.097 & 0.097 \\
        7200 & 0.115 & 0.113 & 0.111 & 0.109 & 0.103 & 0.098 & 0.096 & 0.094 & 0.094 \\
        9000 & 0.113 & 0.1107 & 0.1083 & 0.106 & 0.1 & 0.096 & 0.094 & 0.093 & 0.093 \\
        10800 & 0.111 & 0.1087 & 0.1063 & 0.104 & 0.099 & 0.094 & 0.093 & 0.093 & 0.093 \\
        14400 & 0.111 & 0.1087 & 0.1063 & 0.104 & 0.099 & 0.094 & 0.093 & 0.093 & 0.093 \\
    \hline
    \end{tabular}
\caption{ATM volatility surface for EURIBOR of August 11th, 2006.}
\label{table:volatility surface2}
\end{center}
\end{table}

The implied volatility surface (tenor length, expiry) of ATM
European swaptions on EURIBOR is listed in Table
\ref{table:volatility surface2}. The value of each volatility is the
average of the original bid and ask values. For away-from-the-money
swaptions, we have the ratio data for strikes with offsets relative
to the ATM strike of -600bp, -500bp, -400bp, -300bp, -250bp, -200bp,
-150bp, -100bp, -75bp, -50bp, -25bp, 0, 25bp, 50bp, 75bp, 100bp,
150bp, 200bp, 250bp, 300bp, 400bp, 500bp, 600bp, 700bp, 800bp,
900bp, 1000bp, 1200bp, 1400bp, 1600bp, 1800bp, 2000bp, 2500bp,
3000bp, 3500bp and 4000bp.\footnote{The ratio data were generated by
the SABR model. For the SABR model, we refer to Hagan \cite{Hagan
2002}.} The ATM point has a ratio of 1.0. If a strike, let's say, at
ATM+50bp has a ratio of 1.1, this means the volatility of the strike
ATM+50bp is $1.1\times \mbox{ATM vol}$. Any required volatility not
available in Table \ref{table:volatility surface2} and ratio data
was calculated by linear interpolation of the volatility cube.

\newpage
\section{Specification of Test Trades} \label{sec:TestTrades}
\begin{table}[h!]
\begin{center}
    \begin{tabular}{l| l l}
    \hline
                                    & Trade I & Trade II          \\
    \hline
    Valuation Date                  & 09-07-2002 & 11-08-2006        \\
    Start Date                  & 12-07-2002 & 11-02-2007        \\
    Notional                        & 10000      & 10000             \\
    Exercise Type                   & Payer      & Payer             \\
    Number of Floating Periods      & 10         & 20                \\
    Floating Frequency              & 6 months   & 12 months         \\
    Floating Margin                 & 0          & 0                 \\
    Floating Margin Increment       & 0          & 0                 \\
    Floating Date-roll              & Modified Following& Modified Following\\
    Floating Day-count              & ACT/360    & ACT/360           \\
    Fixed Frequency                 & 6 months   & 12 months         \\
    Fixed Coupon Increment          & 0          & 0                 \\
    Fixed Date-roll                 & Modified Following& Modified Following\\
    Fixed Day-count                 & ACT/360    & ACT/360           \\
    Exercise Fee                    & 0          & 0                 \\
    Exercise Fee Increment          & 0          & 0                 \\
    \hline
    Steps per Deviation             &10          &10                \\
    Number of Deviations            &10          &10                \\
    Maximum Polynomial Order        &3           &3                \\
    \hline
    \end{tabular}
\caption{Test Trades.} \label{table:test_trades}
\end{center}
\end{table}

Some comments for the Test Trades:
\begin{itemize}
\item "Start Date" means the closest starting date among all the
co-terminal swaps.
\item In this report, we always set the "Fixed Frequency" equal
to the "Floating Frequency" for simplicity.
\item "Steps per Deviation", "Number of Deviations" and "Maximum Polynomial
Order" are actually grid specification for Gaussian numerical
integration. The "Steps per Deviation" is the "number of steps in
the interval length equal to one $\sigma_{X_n}$" in Section
\ref{sec:numerical}. The "Number of Deviations" is the "$m$" in
Section \ref{sec:numerical}, which applies to a single side
(positive side or negative side) of $X_n$.
\end{itemize}

\begin{table}[h!]
\begin{center}
    \begin{tabular}{l| c}
    \hline
    Trade Period                    & from 05-28-2004 to 07-29-2005 \\
    Start Date                      & 08-31-2005                    \\
    Notional                        & 10000                         \\
    Exercise Type                   & Payer                         \\
    Number of Floating Periods      & 10                            \\
    Floating Frequency              & 12 months                     \\
    Floating Margin                 & 0                             \\
    Floating Margin Increment       & 0                             \\
    Floating Date-roll              & Modified Following            \\
    Floating Day-count              & ACT/360                       \\
    Fixed Frequency                 & 12 months                     \\
    Fixed Coupon Increment          & 0                             \\
    Fixed Date-roll                 & Modified Following            \\
    Fixed Day-count                 & ACT/360                       \\
    Exercise Fee                    & 0                             \\
    Exercise Fee Increment          & 0                             \\
    \hline
    Steps per Deviation             &10                              \\
    Number of Deviations            &7                               \\
    Maximum Polynomial Order        &3                               \\
    \hline
    \end{tabular}
\caption{Trade specification of the hedge tests in Chapter
\ref{chapter:Hedging}.} \label{table:hedge_trade}
\end{center}
\end{table}

\begin{figure}[hp!]
\centering \nonumber
\includegraphics[width=\textwidth,height=90mm]{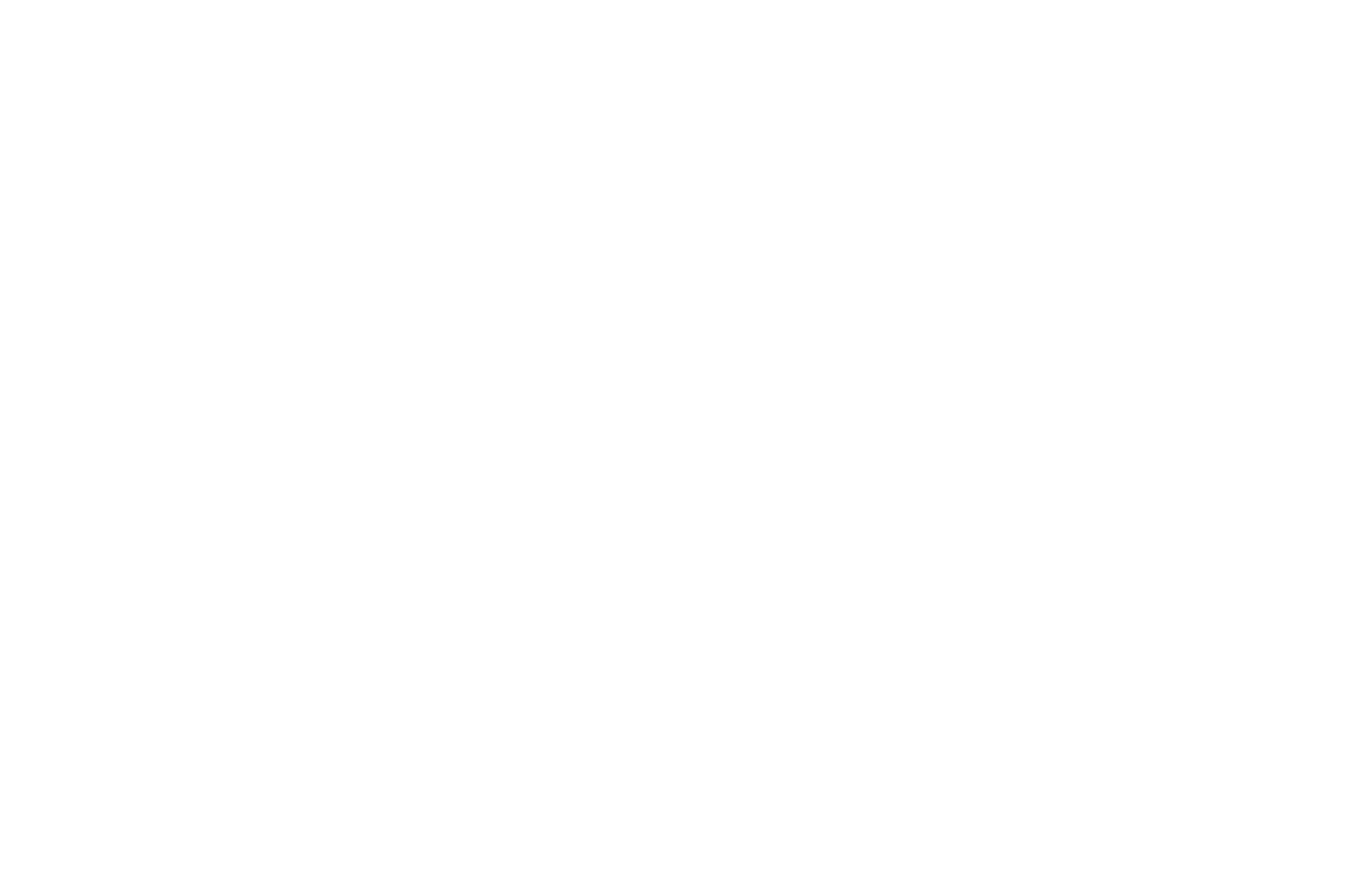}
\end{figure}


\bibliographystyle{plain}

\end{document}